\documentclass[12pt]{article}
\usepackage{times}
\usepackage{graphicx}
\usepackage{amssymb,amsmath,amsthm}
\usepackage{amsfonts}%,amscd}

\newtheorem{defn}{Definition}[section]

\newtheorem{lemma}[defn]{Lemma}

\newtheorem{thm}[defn]{Theorem}
\newtheorem{theorem}[defn]{Theorem}

\newcommand{\be}{\begin{equation}}
\newcommand{\ee}{\end{equation}}
\newcommand{\bea}{\begin{eqnarray}}
\newcommand{\eea}{\end{eqnarray}}
\newcommand{\beas}{\begin{eqnarray*}}
\newcommand{\eeas}{\end{eqnarray*}}

\newcommand{\goto}{\rightarrow}

\newcommand{\ink}{\rule{.5\baselineskip}{.55\baselineskip}}
\newcommand{\ds}{\displaystyle}
\newcommand{\ts}{\textstyle}

\newcommand{\noi}{\noindent}

\newcommand{\ve}{\varepsilon}

\newcommand{\skp}{\vspace{\baselineskip}}

\newcommand{\R}{{\mathbb R}}

\newcommand{\N}{\mathbb N}
\newcommand{\C}{\mathbb C}

\newcommand{\atwo}{a_2}
\newcommand{\afour}{a_4}
\newcommand{\asix}{a_6}

\newcommand{\mmbetak}{{\cal M}_{\beta,K}}

\newcommand{\bark}{\bar{K}}
\newcommand{\barx}{\bar{x}}
\newcommand{\bary}{\bar{y}}
\newcommand{\barz}{\bar{z}}

\newcommand{\bn}{\beta_n}

\newcommand{\kn}{K_n}
\newcommand{\mbnkn}{m(\beta_n,K_n)}
\newcommand{\mn}{m_n}
\newcommand{\mbar}{\bar{m}}
\newcommand{\mbarn}{\bar{m}_n}
\newcommand{\mbarnone}{\bar{m}_{n_1}}
\newcommand{\mbarntwo}{\bar{m}_{n_2}}
\newcommand{\mbarnthree}{\bar{m}_{n_3}}
\newcommand{\mbarnfour}{\bar{m}_{n_4}}
\newcommand{\mbarnsix}{\bar{m}_{n_6}}
\newcommand{\gbk}{G_{\beta,K}}

\newcommand{\tildegbk}{\tilde{G}_{\beta,K}}
\newcommand{\gbnkn}{G_{\bn,\kn}}
\newcommand{\gprimebnkn}{G_{\bn,\kn}'}
\newcommand{\gn}{G_n}
\newcommand{\gntwo}{G_{n_2}}
\newcommand{\gnthree}{G_{n_3}}
\newcommand{\half}{\ts\frac{1}{2}}
\newcommand{\bc}{\beta_c}
\newcommand{\kc}{K(\bc)}
\newcommand{\bzero}{\beta_0}
\newcommand{\kbzero}{K(\bzero)}

\newcommand{\kbc}{K(\bc)}
\newcommand{\kprimebc}{K'(\bc)}

\newcommand{\pnbnkn}{P_{n,\bn,\kn}}
\newcommand{\enbnkn}{E_{n,\bn,\kn}}
\newcommand{\ntou}{n^{u}}

\newcommand{\tll}{\tau}

\newcommand{\rng}{R n^\gamma}

\newcommand{\tell}{\tilde{\ell}}
\newcommand{\tall}{\left|\rule{0pt}{10pt}}
\newcounter{bean}
\newcommand{\benuma}{\setlength{\labelwidth}{.25in}
\begin{list}%
{(\alph{bean})}{\usecounter{bean}}}
\newcommand{\eenuma}{\end{list}}

\def\theequation{\thesection.\arabic{equation}}
\def\theequation{\arabic{section}.\arabic{equation}}
\def\thedefn{\arabic{section}.\arabic{defn}}
\newcommand{\beginsec}{\setcounter{equation}{0}}

\def\m{m}
\def\tbe{\tilde{\beta}}

\def\mus{\mu_1}
\def\mut{\mu_2}
\def\phit{\varphi_t}
\def\a{a}
\def\b{b}

\begin{document}

% \pagewiselinenumbers
% use with "\usepackage{lineno}"

\title{
Ginzburg-Landau Polynomials and the \\
Asymptotic Behavior of the Magnetization\\
Near Critical and Tricritical Points}
\author{Richard S.\ Ellis\normalsize{$\,^1$} \vspace{-.1in} \\
\small{rsellis@math.umass.edu} \vspace{-.125in} \\ \\
Jonathan Machta\normalsize{$\,^2$} \vspace{-.1in}\\  
\small{machta@physics.umass.edu} \vspace{-.125in} \\ \\
Peter Tak-Hun Otto\normalsize{$\,^3$} \vspace{-.1in}\\
\small{potto@willamette.edu} \vspace{-.125in}\\ \\
\normalsize{$^1$ Department of Mathematics and Statistics} \vspace{-.05in} \\ 
\normalsize{University of Massachusetts} \vspace{-.05in} \\ 
\normalsize{Amherst, MA 01003} \vspace{-.1in}\\ \\
\normalsize{$^2$ Department of Physics} \vspace{-.05in} \\ 
\normalsize{University of Massachusetts} \vspace{-.05in} \\ 
\normalsize{Amherst, MA 01003} \vspace{-.1in}\\ \\
\normalsize{$^3$ Department of Mathematics} \vspace{-.05in}\\ 
\normalsize{Willamette University} \vspace{-.05in}\\ 
\normalsize{Salem, OR 97301}}
\maketitle

\begin{abstract}
The purpose of this paper is to prove unexpected connections among
the asymptotic behavior of the magnetization, the structure of the phase transitions,
and a class of polynomials that we call the Ginzburg-Landau polynomials. 
The model under study is a mean-field version of an important
lattice-spin model due to Blume and Capel.  It 
is defined by a probability distribution that depends
on the parameters $\beta$ and $K$, which represent, respectively, the inverse temperature and the interaction strength.  
Our main focus is on the asymptotic behavior of the magnetization $m(\beta_n,K_n)$ 
for appropriate sequences $(\beta_n,K_n)$ that converge to a second-order point or to the tricritical point of the model
and that lie inside various subsets of the phase-coexistence
region.  The main result states that as $(\beta_n,K_n)$
converges to one of these points $(\beta,K)$, $m(\beta_n,K_n) \sim \bar{x}|\beta - \bn|^\gamma \rightarrow 0$.
In this formula $\gamma$ is a positive constant, and 
$\bar{x}$ is the unique positive, global minimum point of a certain polynomial $g$. We call
$g$ the Ginzburg-Landau polynomial because of its close connection with the Ginzburg-Landau
phenomenology of critical phenomena.  This polynomial arises as a limit of appropriately
scaled free-energy functionals, the global minimum points of which define
the phase-transition structure of the model.  In the asymptotic
formula $m(\bn,\kn) \sim \bar{x}|\beta - \bn|^\gamma$, 
both $\gamma$ and $\bar{x}$ depend on the sequence $(\beta_n,K_n)$. 
Six examples of such sequences are considered, each
leading to a different asymptotic behavior of $m(\bn,\kn)$.
Our approach to studying the asymptotic behavior of the magnetization
has three advantages. First, for each sequence $(\bn,\kn)$ under study, 
the structure of the global minimum points of the 
associated Ginzburg-Landau polynomial
mirrors the structure of the global minimum points of the free-energy functional
in the region through which $(\bn,\kn)$ passes and thus reflects
the phase-transition structure of the model in that region.
In this way the properties of the Ginzburg-Landau polynomials
make rigorous the predictions of the Ginzburg-Landau phenomenology of critical phenomena. 
Second, we use these properties to discover new features of the first-order curve
in a neighborhood of the tricritical point.  Third, the 
predictions of the heuristic scaling theory of the tricritical
point are made rigorous by the 
asymptotic formula $m(\beta_n,K_n) \sim \bar{x}|\beta - \bn|^\gamma$,
which is the main result in the paper.
\end{abstract}

\noi
{\it American Mathematical Society 2000 Subject Classifications.}  Primary 82B20
\skp

\noi
{\it Key words and phrases:} Ginzburg-Landau phenomenology, second-order phase transition, first-order phase transition, tricritical point, scaling theory, Blume-Capel model

\pagestyle{myheadings}
\markboth{R.~S.~Ellis, J.~Machta, and P.~T.~Otto}{Ellis, Machta, and Otto: 
Asymptotics for the Magnetization}
% Need next line for line numbers.
% \pagewiselinenumbers

\section{Introduction}
\beginsec
\label{section:intro}

In this paper we prove unexpected connections among
the asymptotic behavior of the magnetization, the structure of the phase transitions,
and a class of polynomials that we call the Ginzburg-Landau polynomials. 
The investigation is carried out for a mean-field version of an important
lattice-spin model due to Blume and Capel, to which we refer as the 
B-C model \cite{Blu, Cap1, Cap2, Cap3}. 
This mean-field model is equivalent to the B-C model on 
the complete graph on $n$ vertices. It is
certainly one of the simplest models that 
exhibit the following intricate phase-transition structure: a curve of second-order 
points; a curve of first-order points; and 
a tricritical point, which separates the two curves. 
A generalization of the B-C model is studied in \cite{BluEmeGri}. 

The main result in the present paper is Theorem \ref{thm:exactasymptotics}, a general theorem
that gives the asymptotic behavior of the magnetization in the mean-field B-C model for suitable sequences. With only changes
in notation, the theorem also applies to other mean-field models including the Curie-Weiss
model \cite{Ellis} and the Curie-Weiss-Potts model \cite{EllWan}.

\iffalse
Applications of the original Blume-Emery-Griffiths model to a wide 
range of physical systems are discussed in \cite[\S 1]{EllOttTou} 
and in \cite[\S 3.3]{NagBon}, where the model is called the Blume-Emery-Griffiths-Rys model.  
The latter reference also points out that the model studied in the present paper 
is actually a mean-field version of a precursor of the Blume-Emery-Griffiths-Rys 
model due to Blume \cite{Blu} and Capel \cite{Cap1,Cap2,Cap3}. 
With apologies to these authors, we follow the nomenclature of our earlier papers
\cite{CosEllOtt, EllOttTou} by referring to this mean-field version 
as the BEG model.  
\fi

The mean-field B-C model is defined by a canonical ensemble that we denote by $P_{N,\beta,K}$; 
$N$ equals the number of spins, $\beta$ is the inverse temperature, 
and $K$ is the interaction strength. $P_{N,\beta,K}$ is defined in terms of the Hamiltonian
\[
H_{N, K}(\omega)=\sum_{j=1}^N \omega_j^2 - \frac{K}{N}\!\left( \sum_{j=1}^N 
\omega_j \right)^2,
\]
in which $\omega_j$ represents the spin at site $j \in \{1,2,\ldots, N\}$ and 
takes values in $\Lambda=\{1,0,-1\}$. 
The configuration space for the model is the set $\Lambda^N$ containing all 
sequences $\omega = (\omega_1,\omega_2, \ldots, \omega_N)$ with each $\omega_j \in \Lambda$.  

Before introducing the results in this paper,
we summarize the phase-transition structure of the model.
For $\beta > 0$ and $K >0$ we denote by $\mmbetak$ the set of equilibrium values
of the magnetization. 
$\mmbetak$ coincides with the set of zeroes of the rate function
in a large deviation principle for the spin per site [Thm.\ \ref{thm:ldppnbetak}(a)] and with the 
set of global minimum points of the free-energy functional $\gbk$, which
is related to the rate function via a Legendre-Fenchel transform [see (\ref{eqn:introgbetak})].
It is known from heuristic arguments and is proved in
\cite{EllOttTou} that there exists a critical
inverse temperature $\beta_c = \log 4$ and that for $0 < \beta \leq \bc$
there exists a quantity $K(\beta)$ and for $\beta > \bc$ there exists
a quantity $K_1(\beta)$ having the following properties:
\begin{enumerate}
\item For $0 < \beta \leq \bc$ and $0 < K \leq K(\beta)$, $\mmbetak$ consists of the unique
pure phase 0.
\item For $0 < \beta \leq \bc$ and $K > K(\beta)$, $\mmbetak$ consists of two symmetric,
nonzero values of the magnetization $\pm m(\beta,K)$. 
\item For $0 < \beta \leq \beta_c$, $\mmbetak$
undergoes a continuous bifurcation at $K = K(\beta)$, changing continuously
from $\{0\}$ for $K \leq K(\beta)$ to $\{\pm m(\beta,K)\}$ for $K > K(\beta)$.
This continuous bifurcation corresponds to a second-order phase transition.
\item For $\beta > \bc$ and $0 < K < K_1(\beta)$, $\mmbetak$ consists of the unique
pure phase 0.
\item For $\beta > \bc$ and $K = K_1(\beta)$, $\mmbetak$ consists of 0 
and two symmetric, nonzero values of the magnetization $\pm m(\beta,K_1(\beta))$. 
\item For $\beta > \bc$ and $K > K_1(\beta)$, $\mmbetak$ consists of two symmetric,
nonzero values of the magnetization $\pm m(\beta,K)$. 
\item For $\beta > \beta_c$, $\mmbetak$
undergoes a discontinuous bifurcation at $K = K_1(\beta)$, 
changing discontinuously from $\{0\}$ for $K < K(\beta)$ to 
$\{0, \pm m(\beta,K)\}$ for $K = K_1(\beta)$ to 
$\{\pm m(\beta,K)\}$ for $K > K_1(\beta)$. This discontinuous bifurcation 
corresponds to a first-order phase transition.
\end{enumerate}

Because of items 3 and 7, we refer to the curve $\{(\beta,K(\beta)), 0 < \beta
< \bc\}$ as the second-order curve and to the curve 
$\{(\beta,K_1(\beta)), \beta > \bc\}$ as the first-order curve.
Points on the second-order curve are called second-order points, and points
on the first-order curve first-order points. 
The point $(\beta_c,K(\beta_c)) = (\log 4, 3/2 \log 4)$ separates the second-order curve
from the first-order curve and is called the tricritical point.  The phase-coexistence
region consists of all points in the positive $\beta$-$K$ quadrant for which 
$\mmbetak$ consists of more than one value.  Thus this region consists of all
$(\beta,K)$ above the second-order curve, above the tricritical point, on the first-order
curve, and above the first-order curve; i.e., all
$(\beta,K)$ satisfying $0 < \beta \leq \bc$ and $K > K(\beta)$ and
satisfying $\beta > \bc$ and $K \geq K_1(\beta)$.  The sets that describe the phase-transition structure
of the model are shown in Figure 1.

\begin{figure}[h]
\begin{center}
\includegraphics[width=12cm]{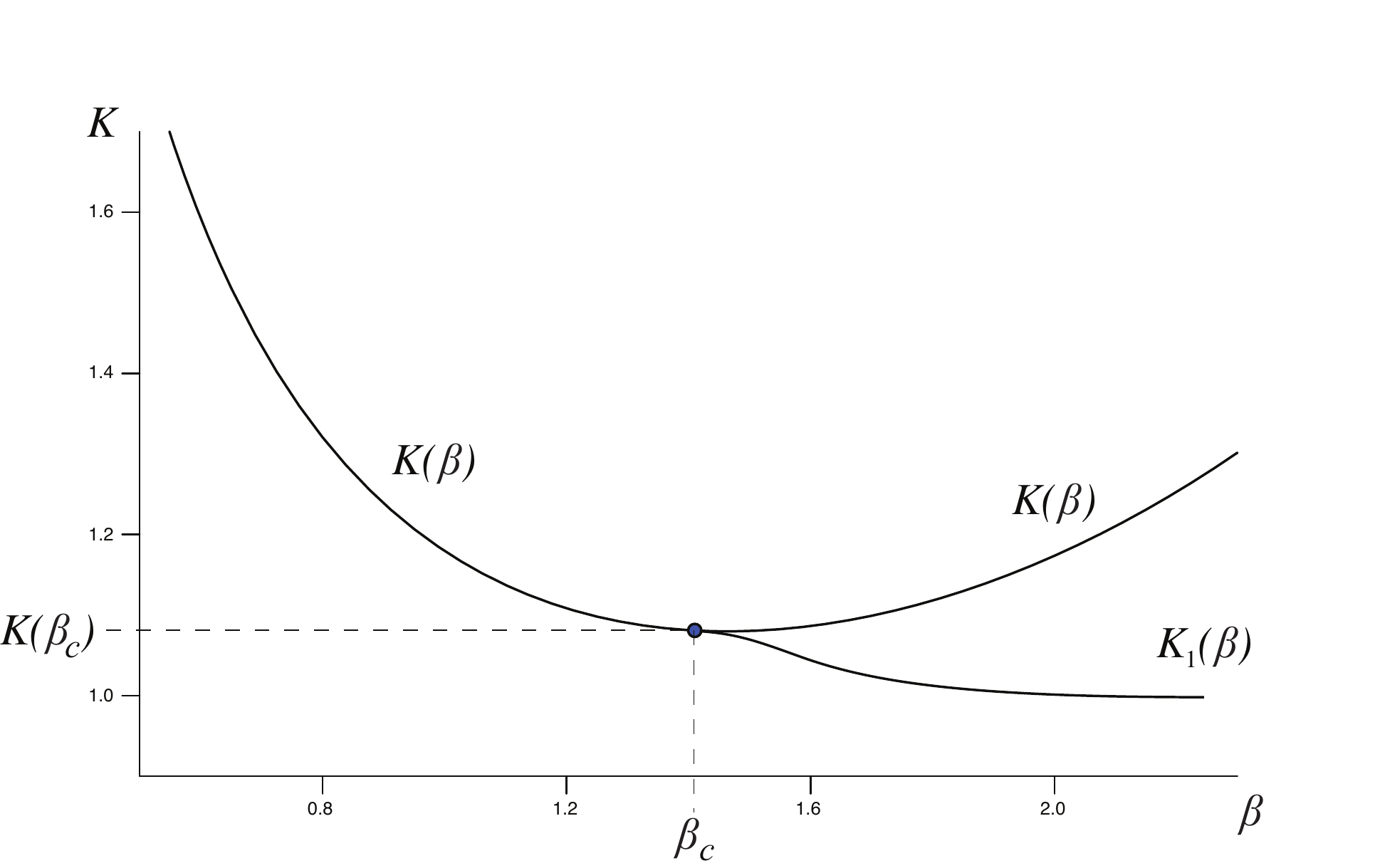}
\caption{\footnotesize The sets that describe the phase-transition structure of the BEG model:
the second-order curve $\{(\beta,K(\beta)), 0 < \beta < \bc\}$, the 
first-order curve $\{(\beta,K_1(\beta)), \beta > \bc\}$, and the
tricritical point $(\bc,\kc)$.  The phase-coexistence
region consists of all $(\beta,K)$ above the second-order curve, above the tricritical point, on the first-order
curve, and above the first-order curve. The extension of the second-order curve to $\beta > \bc$
is called the spinodal curve.}
\end{center}
\end{figure} 

We now turn to the main focus of this paper, which is 
the asymptotic behavior of the magnetization $m(\bn,\kn)$
for appropriate sequences $(\bn,\kn)$ that converge either to a second-order point or to the tricritical point from various subsets of the phase-coexistence
region.  In the case of second-order points we consider two
such sequences in Theorems \ref{thm:asymptoticsforb} and \ref{thm:asymptoticsforb2}, and in the case of the tricritical
point we consider four such sequences in Theorems 
\ref{thm:asymptoticsforc1}--\ref{thm:asymptoticsforc4}.  
Denoting the second-order point or the tricritical point by $(\beta,K)$, 
in each case we prove as a consequence of the general result
in Theorem \ref{thm:exactasymptotics} that $\mbnkn \goto 0$ according to the asymptotic formula
\be
\label{eqn:mainresult}
\mbnkn \sim \bar{x}|\beta - \bn|^\gamma; \ \mbox{ i.e. } \
\lim_{n \goto \infty} |\beta - \bn|^{-\gamma} \mbnkn = \bar{x}.
\ee
In this formula $\gamma$ is a positive
constant, and $\bar{x}$ is the unique positive, global minimum point of a certain polynomial $g$.
We call $g$ the Ginzburg-Landau polynomial because of its close connection
with the Ginzburg-Landau phenomenology of critical phenomena \cite{LanLif}.
Both $\gamma$ and $\bar{x}$ depend on the sequence $(\bn,\kn)$. The exponent $\gamma$ and the 
polynomial $g$ arise via the limit of suitably
scaled free-energy functionals; specifically, for appropriate choices of $u \in \R$ and $\gamma > 0$
and uniformly for $x$ in compact subsets of $\R$
\be
\label{eqn:gbnkng}
\lim_{n \goto \infty} n^{1-u} \gbnkn(x/n^\gamma) = g(x).
\ee

\iffalse
It is important to note that in the definition of the canonical ensemble
$\pnbnkn$, $n$ denotes the size of the system. By contrast, in the definition
of the sequence $(\bn,\kn)$ and in the expressions $\mbnkn$ and $\gbnkn$,
$n$ is simply an index and does not refer to system size.
\fi

The Ginzburg-Landau polynomials $g$ play several roles in this paper.
First, the structure of the set of global minimum points of each $g$ mirrors the structure
of the set of global minimum points of the free-energy functional in the subset
of the phase-coexistence region through which the corresponding sequence $(\bn,\kn)$ passes.
Details of this mirroring are given in the discussions leading up to Theorem \ref{thm:asymptoticsforb}
and Theorem \ref{thm:asymptoticsforc2}; of all the sequences that we consider, the sequence considered in the
latter theorem shows the most varied behavior.
Since the global minimum points of the free-energy functional determine
the phase-transition structure of the model, one can also investigate the
phase-transition structure using properties of the Ginzburg-Landau polynomials, 
which are polynomials of degree 4 or 6 and thus have a much simpler form than
the free-energy functional.  In this way, properties of the Ginzburg-Landau
polynomials make rigorous the predictions of the Ginzburg-Landau
phenomenology of critical phenomena, which replaces appropriate
thermodynamic quantities by the first few terms of their Taylor expansions
in an ad hoc manner.  An example of such an application of the Ginzburg-Landau
polynomials is given in section \ref{section:firstordercurve}, where we use
the polynomials to discover new features of the first-order curve.
The Ginzburg-Landau phenomenology is used in section 2 to motivate the phase-transition
structure of the model.

The Ginzburg-Landau polynomials are also intimately related to probabilistic
limit theorems for the total spin $S_N$ with respect to 
the canonical ensemble $P_{N,\beta,K}$.  These limit
theorems are studied in the sequel to the present paper  when $N = n$ \cite{EllMacOtt2}; 
i.e., when the system size $N$ coincides with the index $n$ parametrizing the sequence $(\bn,\kn)$.
For each sequence $(\bn,\kn)$ for
which the asymptotic behavior of $\mbnkn$ is studied, 
there exists $\gamma_0 \in (0,1/4]$ such that up to a multiplicative constant,
the exponential of the associated Ginzburg-Landau polynomial is 
the limiting density in a scaling limit for $S_n/n^{1-\gamma_0}$. 
This limit is supplemented by other scaling limits for $\gamma \not =
\gamma_0$. In addition, for all $\gamma \in (0,\gamma_0)$, up to an additive constant
the Ginzburg-Landau polynomial is the rate function
in a moderate deviation principle (MDP) for $S_n/n^{1-\gamma}$.,
After deriving the MDPs in \cite{EllMacOtt2}, we apply them to study refined asymptotics of the spin.  
This fascinating feature is discussed at the end of the introduction.

The connection with probabilistic limit theorems reveals a close relationship
between the present study 
and our previous paper \cite{CosEllOtt}. In that paper we reveal
the intricate structure of the phase transitions in the BEG model by proving 
a total of 18 scaling limits and 18 MDPs for $S_n/n^{1-\gamma}$.  
These results are obtained for appropriate sequences $(\bn,\kn)$
converging either to points in the single-phase region located under
the second-order curve (1 scaling limit and 1 MDP), to second-order points 
(4 scaling limits and 4 MDPs), and to the tricritical
point (13 scaling limits and 13 MDPs). Our goal in that paper 
was to obtain the maximal number of probabilistic limit theorems.
In order to achieve that goal, we chose sequences
$(\bn,\kn)$ for which certain terms in a Taylor expansion have appropriate
large-$n$ behavior.  However, the physical significance of those sequences is not obvious.

By contrast, in the present paper we focus, not on probabilistic limit theorems as in
\cite{CosEllOtt, EllMacOtt2}, but on the asymptotic behavior of the 
magnetization using physically significant sequences 
$(\bn,\kn)$. These sequences converge either to second-order points 
or to the tricritical point from specific subsets of the phase-coexistence
region.
Doing so allows us to use properties of the Ginzburg-Landau polynomials
in order to study the phase transitions in these various subsets.

This paper puts on a rigorous footing the idea, first introduced by Ginzburg and Landau, 
that low-order polynomial approximations to the free energy functional give correct asymptotic results 
near continuous phase transitions for mean-field models \cite{LanLif}. 
The use of sequences $(\beta_n, K_n)$  that approach 
second-order points or the tricritical point permits us to establish the validity of truncating the expansion of the free-energy functional at an appropriate low order.  The higher order terms are driven to zero by a power of $n$ and are shown to be asymptotically irrelevant.  While the renormalization group methodology also demonstrates the irrelevance of higher order terms in the expansion of the free-energy functional, it does so via a different route that depends on heuristics. By contrast, our approach is rigorous and shows in (\ref{eqn:gbnkng})
how to obtain the Ginzburg-Landau polynomial as a limit of suitably scaled free-energy functionals. No heuristic approximations appear.

Let $(\bn,\kn)$ be any particular sequence converging to a second-order point or the tricritical point from the phase-coexistence
region and denote the limiting point by $(\beta,K)$. It is not difficult
to obtain an asymptotic formula expressing the rate at which $\mbnkn$ converges to 0. Since $\mbnkn$ is the unique positive minimum
point of $\gbnkn$, it solves the equation $G_{\bn,\kn}'(\mbnkn) = 0$. 
As we illustrate in appendix \ref{section:moreasymptotics} in two examples,
expanding $G_{\bn,\kn}'$ in a Taylor series of appropriate order,
one obtains the formula $\mbnkn \sim \barx|\beta - \bn|^\gamma$ for some $\barx > 0$ and $\gamma > 0$. However, this method gives only the functional form for $\barx$, not associating it with the model
via the Ginzburg-Landau polynomial. This is in contrast to our general
result in Theorem \ref{thm:exactasymptotics}. That result identifies $\barx$ as the unique positive, global minimum point of the Ginzburg-Landau polynomial, using the uniform convergence in (\ref{eqn:gbnkng}).  The proof of that result completely avoids Taylor expansions, making use of
the fact that under appropriate conditions, the positive global minimum points of $n$-dependent minimization problems converge to the positive global minimum point of a limiting minimization problem when such a minimum point is unique. 

\iffalse
Another contribution of the present paper is to demonstrate how the structure
of the set of global minimum points of the associated 
Ginzburg-Landau polynomials $g$ mirrors the phase-transition structure
of the subsets through which $(\bn,\kn)$ passes.  In this way the properties of the Ginzburg-Landau polynomials
make rigorous the predictions of the Ginzburg-Landau phenomenology of critical phenomena.  
Details of this mirroring are given in the discussions leading up to Theorem \ref{thm:asymptoticsforb}
and Theorem \ref{thm:asymptoticsforc2}; of all the sequences that we consider, the sequence considered in the
latter theorem shows the most varied behavior.
\fi

Our work is also closely related to the scaling theory for critical and tricritical points.  By choosing sequences that approach second-order points or the tricritical point 
from various directions and at various rates, we are able to verify a number of predictions of scaling theory.  
The sequences that approach the tricritical point reveal the subtle geometry of the crossover between critical and tricritical behavior described in Riedel's tricritical scaling theory \cite{Rie}.   In section \ref{section:scalingtheory}
we will see that a proper application of scaling theory near the tricritical point requires that the scaling parameters be defined in a curvilinear coordinate system, an idea proposed in \cite{Rie} but, to our knowledge, not previously explored.

Some of the results proved here are contained in the work of Capel and
collaborators \cite{Cap1, Cap2, Cap3, CapOudPer, OudCapPer1, OudCapPer2, PerCapOud}.  These papers
introduce the mean-field B-C model and
provide a general framework for studying mean-field models and
obtaining the thermodynamic properties of these systems.  The work of Capel and his collaborators does not encompass the contributions of the present paper, which include 
a new and rigorous methodology involving Ginzburg-Landau polynomials for treating mean-field
models. We believe that this methodology will be valuable in future mathematical
investigations of related systems in statistical mechanics.

In order to highlight our new results, we summarize them for the six sequences considered
in Theorems \ref{thm:asymptoticsforb}--\ref{thm:asymptoticsforb2} and in Theorems
\ref{thm:asymptoticsforc1}--\ref{thm:asymptoticsforc4}.  Two of the six
cases involve the spinodal curve,
which is the extension of the second-order curve $\{(\beta,K(\beta)), 0 < \beta < \bc\}$
to values $\beta > \bc$.  In cases 1, 2, and 6 the limiting Ginzburg-Landau polynomial
has degree 4 while in cases 3, 4, and 5 the limiting Ginzburg-Landau polynomial 
has degree 6. In each case the quantity $\bar{x}$ equals the unique positive,
global minimum point of the associated Ginzburg-Landau polynomial.  
In the following six items the asymptotic behavior of $m(\bn,\kn) \goto 0$ is expressed as an appropriate function of $\beta - \bn$. 

\begin{enumerate}
\item In case 1 the sequence $(\bn,\kn)$ converges to a second-order point $(\beta,K(\beta))$
along a ray that lies in the phase-coexistence region. This ray is above
the tangent line to the second-order curve at that point. 
Given $0 < \beta < \bc$, $\alpha > 0$, $b \in \{1,0,-1\}$,
and $k \in \R$ satisfying $K'(\beta)b - k < 0$, the sequence is defined by
$\beta_n = \beta + b/n^\alpha$ and $K_n = K(\beta) + k/n^\alpha$.
As described in Theorem 
\ref{thm:asymptoticsforb}, $\mbnkn \sim \bar{x}/n^{\alpha/2}$.
If $b \not = 0$, then this becomes $\mbnkn \sim \bar{x}|\beta - \bn|^{1/2}$.
\item In case 2 the sequence $(\bn,\kn)$ converges to a second-order point $(\beta,K(\beta))$
along a curve that lies in the phase-coexistence region. It coincides with the second-order curve
to order $p-1$ in powers of $\beta_n - \beta$, where $p \geq 2$ is an integer. Hence the two
curves have the same tangent at $(\beta,K(\beta))$. Parametrized
by $\alpha > 0$, $b \in \{1,-1\}$, an integer $p \geq 2$, and a real number $\ell$
satisfying $(K^{(p)}(\beta) - \ell)b^p < 0$, the sequence is defined by
\be
\label{eqn:kjb}
\bn = \beta + {b}/{n^\alpha} \ \mbox{ and } \ 
\kn = K(\beta) + \sum_{j=1}^{p-1} {K^{(j)}(\beta) b^j}/(j! n^{j\alpha}) + 
{\ell b^p}/(p! n^{p\alpha}).
\ee
As described in Theorem \ref{thm:asymptoticsforb2}, $m(\bn,\kn) \sim \bar{x}/n^{p\alpha/2} = \barx|\beta - \bn|^{p/2}$. 
\item In case 3 the sequence $(\bn,\kn)$ converges to the tricritical point $(\bc,K(\bc))$ along a ray that 
lies in the phase-coexistence region. The ray is above
the tangent line to the phase-transition curve
at the tricritical point. The sequence is defined as in case 1
with $\beta$ replaced by $\bc$. As described in Theorem 
\ref{thm:asymptoticsforc1}, $\mbnkn \sim \bar{x}/n^{\alpha/4}$. 
If $b \not = 0$, then this becomes $\mbnkn \sim \bar{x}|\beta_c - \bn|^{1/4}$.
\item In case 4 the sequence $(\bn,\kn)$ converges to the tricritical point $(\bc,K(\bc))$
along a curve that lies in the phase-coexistence region and is tangent to the spinodal curve at the tricritical point.
Given Conjectures 1--3 in section \ref{section:firstordercurve},
in a neighborhood of the tricritical point
this curve either lies above the first-order
curve or coincides with that curve to order 2 in powers of $\beta - \bc$. 
The sequence is defined in (\ref{eqn:bnknell}) in terms of 
a curvature parameter $\ell$ and another parameter $\tell$, and four different cases are 
listed in items 4a--4d after the definition (\ref{eqn:bnknell}).
As described in Theorem \ref{thm:asymptoticsforc2}, in all four cases
$\mbnkn \sim \bar{x}/n^{\alpha/2} = \barx(\bn - \bc)^{1/2}$. 
\item In case 5 the sequence $(\bn,\kn)$ converges to the tricritical point $(\bc,K(\bc))$
along a curve that lies in the phase-coexistence region and coincides with the second-order
curve to order 2 in powers of $\beta - \bc$. Hence the two curves have the same tangent at the tricritical
point. The sequence is defined as 
in (\ref{eqn:bnknell}) with $\bn = \bc + 1/n^\alpha$ replaced by $\bn = \bc - 1/n^\alpha$
and with $\tell = 0$.
As described in Theorem \ref{thm:asymptoticsforc3}, for $\ell > K''(\bc)$,
$\mbnkn \sim \bar{x}/n^{\alpha/2} = \barx(\bc - \bn)^{1/2}$. 
\item In case 6 the sequence $(\bn,\kn)$ converges to the tricritical point $(\bc,K(\bc))$ 
along a curve that lies in the phase-coexistence region and coincides with the second-order curve 
to order $p-1$ in powers of $\beta - \bc$, where $p \geq 3$ is an
integer. Hence the two curves have the same tangent at the tricritical point. The sequence is defined as in (\ref{eqn:kjb})
with $b = -1$ and $\beta$ replaced by $\bc$.
As described in Theorem \ref{thm:asymptoticsforc4}, for appropriate
choices of $\ell \not = K^{(p)}(\bc)$, $\mbnkn \sim \bar{x}/n^{(p-1)\alpha/2} = \barx(\bc - \bn)^{(p-1)/2}$.
\end{enumerate}

Possible paths followed by the sequences in items 1--6 are shown
in Figure 2.  Two different paths are shown for each of the sequences in items 1, 2, and 3, four different
paths for the sequences in item 4, and one path for each of the sequences in items 5 and 6.
We believe that modulo uninteresting scale changes,
these are all the sequences of the form $\beta_n = \beta + b/n^{\alpha}$ 
and $K_n$ equal to $K(\beta)$ plus a polynomial in $1/n^\alpha$, where $(\beta,K(\beta))$ is 
either a second-order point or the tricritical point and for which $m(\bn,\kn) > 0$.

It is interesting to compare the sequences
in items 1 and 3 and the sequences in items 2 and 6. Although in both cases the sequences are defined similarly, the asymptotic formulas for $\mbnkn$ involve different powers of $n$. 
From the viewpoint of the scaling theory for critical phenomena, the discrepancies arise
because the sequences in items 1 and 2 converge to a second-order point while those in items 3 and 6 converge to the tricritical point; this is discussed in section \ref{section:scalingtheory}.

\begin{figure}[h]
\begin{center}
\includegraphics[width=12cm]{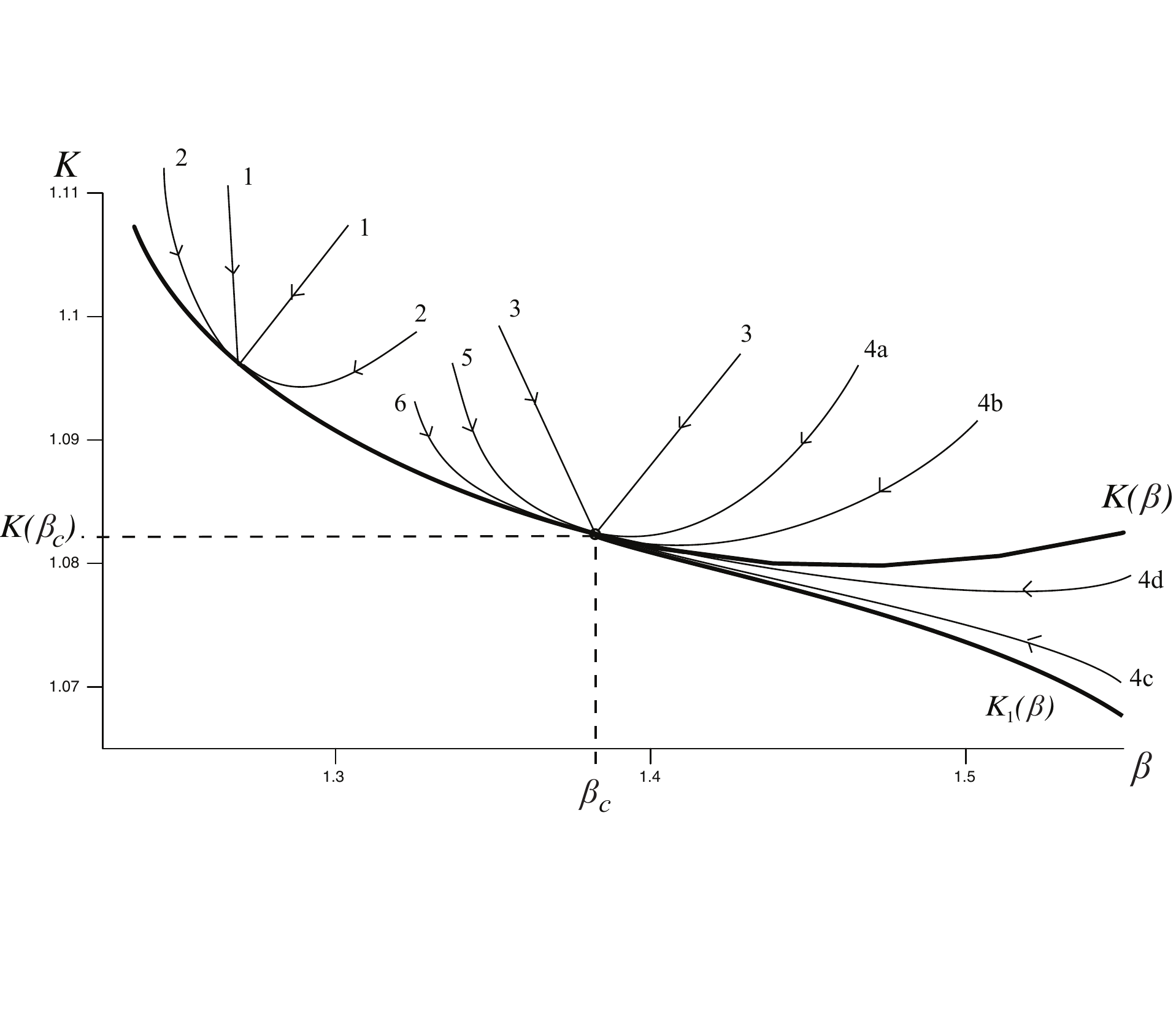}
\vspace{-.8in}
\caption{\footnotesize Possible paths for sequences converging to a 
second-order point and to the tricritical point. 
The curves labeled 1, 2, 3, 4a--4d, 5, and 6 are discussed in the respective items 1, 2, 3, 4, 5,
and 6. The sequences  on the curves labeled 4a--4d are defined in (\ref{eqn:bnknell}) and are discussed 
in more detail in the respective items 4a--4d appearing after (\ref{eqn:bnknell}).}
\end{center}
\end{figure} 

Table 1 summarizes the asymptotic behavior of $\mbnkn$ for the sequences
depicted in Figure 2 and indicates the theorem where the asymptotic behavior is proved.

\begin{center}
\begin{tabular}{||c|c|c|c||} \hline \hline
{{\bf sequence}} & {{\bf converges to}} & {{\bf theorem}} & {\bf asymptotic behavior } \\
& & & {\bf of} \boldmath $m(\bn,\kn)$ \unboldmath
\\ \hline \hline
1 & second-order point & Thm.\ \ref{thm:asymptoticsforb} & $m(\bn,\kn) \sim \bar{x}|\beta - \bn|^{1/2}$ \\ \hline
2 & second-order point & Thm.\ \ref{thm:asymptoticsforb2} & $m(\bn,\kn) \sim \bar{x}|\beta - \bn|^{p/2}$ \\ \hline
3 & tricritical point & Thm.\ \ref{thm:asymptoticsforc1} & $m(\bn,\kn) \sim \bar{x}|\bc - \bn|^{1/4}$ \\ \hline
4a--4d & tricritical point & Thm.\ \ref{thm:asymptoticsforc2} & $m(\bn,\kn) \sim \bar{x}(\bn - \bc)^{1/2}$ \\ \hline
5 & tricritical point & Thm.\ \ref{thm:asymptoticsforc3} & $m(\bn,\kn) \sim \bar{x}(\bc - \bn)^{1/2}$ \\ \hline
6 & tricritical point & Thm.\ \ref{thm:asymptoticsforc4} & $m(\bn,\kn) \sim \bar{x}(\bc - \bn)^{(p-1)/2}$ \\
\hline \hline
\end{tabular}
\vspace{.05in}
\end{center}
\vspace{-.15in}
\noi
Table 1: {\footnotesize Asymptotic behavior of $m(\bn,\kn) \goto 0$
for the sequences in Figure 2. For sequences 1 and 3 the asymptotic formula 
is valid provided $b \not = 0$ in the definition of the sequence.}

\skp
The sequences mentioned in item 4 and labeled 4a--4d in Figure 2
are particularly interesting. Parameterized by $\alpha > 0$,
a curvature parameter $\ell \in \R$, and another parameter $\tell \in \R$, these sequences are defined by
\be
\label{eqn:bnknell}
\beta_n = \bc + {1}/{n^\alpha} \ \mbox{ and } \ K_n = K(\bc) + K'(\bc)/{n^\alpha}
+ \ell/(2n^{2\alpha}) + \tell/(6n^{3\alpha}).
\ee
For appropriate choices of $\ell$ and $\tell$, these sequences converge
to the tricritical point while passing through the following
interesting subsets of the phase-coexistence region. 
\begin{enumerate}
\item[4a.] For $\ell > K''(\bc)$, $(\bn,\kn)$ passes through the phase-coexistence region
located above the spinodal curve.
\item[4b.]  For $\ell = K''(\bc)$,
$(\bn,\kn)$ converges to the tricritical point
along a curve that coincides with the spinodal curve to order 2
in powers of $\beta - \bc$.  
\item[4c.] For $\ell = \ell_c = K''(\bc) - 5/(4\bc)$,
$(\bn,\kn)$ converges to the tricritical point along a curve that coincides,
to order 2 in powers of $\beta - \bc$, with what we conjecture is the first-order curve.
\item[4d.] For $\ell$ in the open interval $(\ell_c,K''(\bc))$, $(\bn,\kn)$ converges to the tricritical point
along a curve that passes between what we conjecture is the first-order curve and the spinodal curve
in a neighborhood of the tricritical point.
\end{enumerate}  
In Figure 2 we do not show the curve along which $(\bn,\kn)$ converges to the tricritical point when 
$\ell < \ell_c$ and $\tell \in \R$. This curve is conjectured to lie in the single-phase region
under the first-order curve.

A number of examples are given in the paper of how the structure
of the set of global minimum points of the associated 
Ginzburg-Landau polynomials $g$ mirrors the phase-transition structure
of the subsets through which $(\bn,\kn)$ passes.  For example, for the sequence
defined in (\ref{eqn:bnknell}), we have the following picture. In all cases 
$g$ depends only on $\ell$, not on $\tell$. 
\begin{enumerate}
\item For $\ell > \ell_c$ and any $\tell \in \R$,
the global minimum points of $g$ are a symmetric, nonzero pair
$\pm \bar{x}(\ell)$, mirroring the fact that for $(\beta,K)$  above the
first-order curve the equilibrium values of the magnetization are the symmetric, nonzero
pair $\pm m(\beta,K)$.  
\item For $\ell = \ell_c$ and all sufficiently large $\tell$,
the global minimum points of $g$ are 0 and a symmetric,
nonzero pair $\pm \bar{x}(\ell_c)$, mirroring the fact that for
$(\beta,K) = (\beta,K_1(\beta))$  on the first-order curve the equilibrium
values of the magnetization are 0 and the symmetric, nonzero pair $\pm m(\beta,K_1(\beta))$.  
\item For $\ell < \ell_c$ and any $\tell \in \R$, $g$ has a unique global minimum point at 0,
mirroring the fact that for $(\beta,K)$  under the first-order curve
there is a unique pure phase at 0. 
\end{enumerate}
The values of the magnetization and the positive, global minimum points
$\bar{x}(\ell)$ are connected through the main result of
the paper, which is that as $(\bn,\kn)$ converges 
to the tricritical point $\mbnkn \sim \bar{x}(\ell)/n^{\alpha/4} = \barx(\ell)(\bc - \bn)^{1/4}$.

In section \ref{section:firstordercurve} we reverse this procedure, using properties of the appropriate
Ginzburg-Landau polynomials
not to mirror, but to predict features of the phase-transition structure. There we argue that at 
$\bc$ the first-order curve defined by
$(\beta,K_1(\beta))$ and the spinodal curve have the same right-hand tangent, that 
$K_1''(\bc) = \ell_c < K''(\bc)$, and that $K_1'''(\bc) > 0$ (see Conjectures 1, 2, and 3).  These conjectures are used to verify the asymptotic behavior of $\mbnkn \goto 0$ given in part (c) of Theorem \ref{thm:asymptoticsforc2}
when $\ell$ satisfies $\ell_c \leq \ell < K''(\bc)$.

We end the introduction by previewing our results on the refined asymptotics of $S_N$ when the system
size $N$ coincides with the index $n$ parametrizing the sequence $(\bn,\kn)$. 
These asymptotics are the main focus of the sequel to the present paper \cite{EllMacOtt2}. These refined asymptotics
reveal a fascinating relationship between the asymptotic formulas for $m(\bn,\kn)$ obtained here and 
the finite-size expectation $\enbnkn\{|S_n/n|\}$,
 where $\enbnkn$ denotes expectation with respect to $\pnbnkn$. In order
to illustrate this relationship, we focus on the sequence $(\bn,\kn)$ in 
Theorem \ref{thm:asymptoticsforb} that
converges to a second-order point $(\beta,K(\beta))$. A general result covering the other
five sequences considered in the present paper is given in \cite{EllMacOtt2}. 

According to part (c) of 
Theorem \ref{thm:asymptoticsforb}, for any $\alpha > 0$, $\mbnkn$ has the asymptotic behavior 
$\mbnkn \sim \bar{x}/n^{\alpha/2}$, where $\barx$ is the positive global minimum point of the 
associated Ginzburg-Landau polynomial.  When $\alpha \in (0,1/2)$, we prove in
\cite{EllMacOtt2} that $\mbnkn$ is asymptotic to the expectation of $|S_n/n|$; i.e., 
\[
\enbnkn\{|S_n/n|\} \sim \mbnkn \sim \bar{x}/n^{\alpha/2}.
\]  
A more refined statement is that when $\alpha \in (0,1/2)$, the 
probability distribution $\pnbnkn\{S_n/n \in dx\}$ is sharply peaked at $\pm \mbnkn$
as $n \goto \infty$. We prove this by considering the asymptotic behavior of the expectation 
\[
\enbnkn\{|\,|S_n/n| - \mbnkn| \},
\] 
showing that the fluctuations of $|S_n/n|$ around $\mbnkn$ as measured by this expectation are asymptotic
to $\barz/n^{1/2 - \alpha/2}$, where $\barz > 0$ is given explicitly. 
Since $\alpha \in (0,1/2)$, the rate $\barz/n^{1/2 - \alpha/2}$
at which this expectation converges to 0 is much faster than the rate $\barx/n^{\alpha/2}$
at which $\mbnkn$ converges to 0.  In this case $(\bn,\kn) \goto (\beta,K(\beta))$ slowly, and the system is in the phase-coexistence regime, where it is effectively infinite.  
Interestingly, the range $\alpha \in (0,1/2)$, for which $\mbnkn$ and $\enbnkn\{|S_n/n|\}$ 
have the same asymptotic behavior, is precisely the range of $\alpha$ for which we have a moderate deviation principle for $S_n/n^{1-\gamma}$
for appropriate $\gamma \in (0,1/4)$. The moderate deviation principle plays a key role in the proofs of the asymptotic
behaviors of the two expectations mentioned in this paragraph.

On the other hand, when $\alpha > 1/2$, $\mbnkn$  is not related to the finite-size 
expectation $\enbnkn\{|S_n/n|\}$.  In the $\alpha > 1/2$ regime, $\enbnkn\{|S_n/n|\}$ is asymptotic to $\bary/n^{1/4}$, where $\bary > 0$
is given explicitly.  In this case the fluctuations of $|S_n/n|$ 
as measured by this expectation are much larger than $\mbnkn$, which converges to 0 at the much faster rate
$\barx/n^{\alpha/2}$.  When $\alpha > 1/2$, $(\bn,\kn) \goto (\beta,K(\beta))$ quickly, and the system is in the
critical regime.  The theory of finite-size scaling predicts that when $\alpha > 1/2$,
 critical singularities are controlled by the size of the system rather than by the distance in parameter space from the phase transition
\cite{Barber,Bruce,Cardy1,RikKinGunKas}. 

\iffalse
The moderate deviation principle can be applied to yield other subsets $A_n \subset [-1,1]$
for which the conditional expectation $\enbnkn\{|S_n/n - \mbnkn| \tall \, S_n/n \in A_n\right.\}$
has the same asymptotic behavior as $\enbnkn\{|\,|S_n/n| - \mbnkn| \}$. These intriguing
issues and a number of related ones are explored in \cite{EllMacOtt2}.
\fi

The contents of the present paper are as follows. In section \ref{section:phasetr}
we use the Ginzburg-Landau phenomenology of critical phenomena to motivate the 
phase-transition structure of the model. We then present two theorems 
proved in \cite{EllOttTou} justifying the predictions of this phenomenology. 
In section \ref{section:asymptoticsforb} we illustrate the use of our main result
on the asymptotic behavior of the magnetization by applying it to two particular
sequence $(\bn,\kn)$ converging to second-order points. 
In section \ref{section:mbnkn} we prove our main result (\ref{eqn:mainresult})
on the asymptotic behavior of $\mbnkn \goto 0$ [Thm.\ \ref{thm:exactasymptotics}].  
In section \ref{section:asymptoticsforc}
that result is applied to four different sequences $(\bn,\kn)$
converging to the tricritical point from different subsets of the phase-coexistence
region.  Section \ref{section:firstordercurve}
is devoted to using the properties of appropriate Ginzburg-Landau polynomials
to discover new properties of the first-order curve. In section 7 we relate the results obtained earlier in this
paper to the scaling theory of critical phenomena. 

The paper also has two appendices. In appendix A we collect a number of results
on polynomials of degree 6 needed earlier in the paper.  In appendix B we illustrate another
technique for determining the asymptotic behavior of $\mbnkn \goto 0$
for two of the sequences considered earlier in the paper.  This technique 
is based on the fact that $\mbnkn$ is a positive global minimum point of the
associated free-energy functional and thus a positive zero of the derivative
of this function. Despite the naturalness of this characterization
of $\mbnkn$, the proofs of the asymptotic behavior of $\mbnkn$ given
in this appendix are much more complicated and the respective theorems
give less information than the proofs in the main part of the paper that
are based on properties of the Ginzburg-Landau polynomials.  This emphasizes
once again the elegance of the approach in Theorem \ref{thm:exactasymptotics},
where we use properties of
these polynomials to deduce the asymptotic behavior of $\mbnkn$. 

\skp
\noi
{\bf Acknowledgments.}  The research of Richard S.\ Ellis is supported 
in part by a grant from the National Science Foundation (NSF-DMS-0604071).

\section{Phase-Transition Structure of the BEG Model}
\beginsec
\label{section:phasetr}

After defining the mean-field B-C model, we introduce a function $\gbk$, called the free-energy functional.
The global minimum points of this function define the equilibrium values of the magnetization, and the minimum value of this function over $\R$ gives the canonical
free energy.  We then apply the Ginzburg-Landau
phenomenology to $\gbk$ in order to motivate the phase-transition structure of the model. 
The predictions of the Ginzburg-Landau
phenomenology are shown to be correct in Theorems
\ref{thm:secondorder} and \ref{thm:firstorder}.  

The mean-field B-C model is a lattice-spin model defined on the complete graph on $N$ 
vertices $1,2, \ldots, N$.  
The spin at site $j \in \{1,2,\ldots, N\}$ is denoted by $\omega_j$, a quantity taking values in $\Lambda=\{1,0,-1\}$. 
The configuration space for the model is the set $\Lambda^N$ containing all 
sequences $\omega = (\omega_1,\omega_2,
\ldots, \omega_N)$ with each $\omega_j \in \Lambda$.  In terms of a positive parameter $K$
representing the interaction strength, the Hamiltonian is defined by
\[
H_{N, K}(\omega)=\sum_{j=1}^N \omega_j^2 - \frac{K}{N}\!\left( \sum_{j=1}^N 
\omega_j \right)^2
\]
for each $\omega \in \Lambda^N$.  
Let $P_N$ be the product measure on $\Lambda^N$ with identical one-dimensional marginals 
$\rho = \ts \frac{1}{3}(\delta_{-1}+\delta_0+\delta_1)$.
Thus $P_N$ assigns the probability $3^{-N}$ to each $\omega \in \Lambda^N$.
For $N \in \N$, inverse temperature $\beta > 0$, and $K>0$,
the canonical ensemble for the BEG model is the sequence of probability measures that assign to each subset $B$ of $\Lambda^N$ the probability
\bea
\label{eqn:pnbetak}
P_{N, \beta, K}(B) & = & \frac{1}{Z_N(\beta, K)} \cdot \int_B \exp[-\beta H_{N,K}] \, dP_N \\
& = & \frac{1}{Z_N(\beta, K)} \cdot \sum_{\omega \in B} \exp[-\beta H_{N,K}(\omega)] 
\cdot 3^{-N}.
\nonumber
\eea
In this formula 
$Z_N(\beta, K)$ is the partition function equal to
\[
\int_{\Lambda^N} \exp[-\beta H_{N,K}] \, dP_N
= \sum_{\omega \in \Lambda^N} \exp[-\beta H_{N,K}(\omega)] \cdot 3^{-N}.
\]

The analysis of the 
canonical ensemble $P_{N, \beta, K}$ is facilitated by expressing it
 in the form of a Curie-Weiss-type model.  This is done by absorbing 
the noninteracting component of the Hamiltonian into the product measure $P_N$, obtaining
\be
\label{eqn:rewritecanon}
P_{N,\beta,K}(d\omega) = 
\frac{1}{\tilde{Z}_N(\beta,K)} \cdot \exp\!\left[ N \beta K 
\!\left(\frac{S_N(\omega)}{N} \right)^2 \right] P_{N,\beta}(d\omega).
\ee
In this formula $S_N(\omega)$ equals
the total spin $\sum_{j=1}^N \omega_j$,
$P_{N,\beta}$ is the product measure on $\Lambda^N$ with identical one-dimensional marginals
\be
\label{eqn:rhobeta}
\rho_\beta(d \omega_j) = 
\frac{1}{Z(\beta)} \cdot \exp(-\beta \omega_j^2) \, \rho(d \omega_j),
\ee
$Z(\beta)$ is the normalization equal to
$\int_\Lambda \exp(-\beta \omega_j^2) \rho(d \omega_j) = (1 + 2 e^{-\beta})/3$,
and $\tilde{Z}_N(\beta,K)$ is the normalization equal to
$[Z(\beta)]^N/Z_N(\beta,K)$. 

When rewritten as in (\ref{eqn:rewritecanon}), $P_{N,\beta,K}$ is reminiscent of the canonical ensemble 
for the Curie-Weiss model \cite[\S IV.4]{Ellis}. However, $P_{N,\beta,K}$
is much more complicated because of the $\beta$-dependent
product measure $P_{N,\beta}$ and the presence of the parameter $K$. 
As we will show in this section, the canonical ensemble $P_{N,\beta,K}$ for the mean-field B-C model
gives rise to a second-order phase transition, a first-order phase transition, and a tricritical
point, which separates the two phase transitions and is one of the main focuses of the present paper.

The starting point of the analysis of the 
phase-transition structure of the mean-field B-C model is the large deviation
principle (LDP) satisfied by the spin per site $S_N/N$ with respect to $P_{N,\beta,K}$.  
In order to state the form of the rate function, we introduce the cumulant generating function $c_\beta$ of the measure
$\rho_\beta$ defined in (\ref{eqn:rhobeta}); for $t \in \R$ this function is defined by 
\bea
\label{eqn:cbeta}
c_\beta(t) & = & \log \int_\Lambda \exp (t\omega_1) \, \rho_\beta (d\omega_1)
 \\ 
\nonumber & = &  
\log \!\left( \frac{1+e^{-\beta}(e^t+e^{-t})}{1+2e^{-\beta}} \right). \nonumber 
\eea
We also introduce the Legendre-Fenchel transform of $c_\beta$, which is defined for $x \in [-1,1]$ by
\[
J_\beta(x) = \sup_{t \in \R} \{tx - c_\beta(t)\};
\]
The smallest interval containing the range of $S_N/N$ is $[-1,1]$,
and $J_\beta(x)$ is finite for $x$ in that interval. $J_\beta$ is the rate function in Cram\'{e}r's theorem, which 
is the LDP for $S_n/n$ with respect to the product measures
$P_{N,\beta}$ \cite[Thm.\ II.4.1]{Ellis} and is one of the components of
the proof of the LDP for $S_N/N$ with respect to $P_{N,\beta,K}$.
This LDP and a related limit are stated in parts (a) and (b)
of the next theorem.  The theorem is proved like Theorem 3.1 in \cite{BET}
and Theorem 2.4 in \cite{EllHavTur}.

\begin{thm}
\label{thm:ldppnbetak}  For all $\beta > 0$ and $K > 0$ the following conclusions hold.

{\em (a)} With respect to the canonical ensemble $P_{N,\beta,K}$,
$S_N/N$ satisfies the LDP on $[-1,1]$ with rate function
\[
I_{\beta,K}(x) = J_\beta(x) - \beta K x^2 - \inf_{y \in [-1,1]}
\{J_\beta(y) - \beta K y^2\}. 
\]
In particular, for any closed set $F$ in $[-1,1]$ we have the large deviation
upper bound
\[
\limsup_{N \goto \infty} \frac{1}{N} \log P_{N,\beta,K}\{S_N/N \in F\} \leq
-I_{\beta,K}(F) = -\inf_{x \in F} I_{\beta,K}(x).
\]

{\em (b)}  We define the canonical free energy
\[
\varphi(\beta,K) = - \lim_{N \goto \infty} \frac{1}{N} \log Z_N(\beta,K),
\]
where $Z_N(\beta,K)$ is the partition function defined in {\em (\ref{eqn:pnbetak})}.
Then $\varphi(\beta,K) = \inf_{y \in \R}\{J_\beta(y) - \beta K y^2\}$.
\end{thm}

By definition, the infimum of $I_{\beta,K}$ over $[-1,1]$ equals 0.  
We define the set of equilibrium macrostates by
\[
\mmbetak = \{x \in [-1,1] : I_{\beta,K}(x) = 0\}.
\]
In order to justify this definition, let $F$ be any closed subset
of $[-1,1]$ that is disjoint from the closed set $\mmbetak$. 
Then $I_{\beta,K}(F) > 0$, and the large deviation upper bound implies
that for all sufficiently large $n$
\be
\label{eqn:upperld}
P_{N,\beta,K}\{S_N/N \in F\} \leq \exp[-N I_{\beta,K}(F)/2] \goto 0.
\ee
It follows that those $x \in [-1,1]$
satisfying $I_{\beta,K}(x) > 0$ have an exponentially small probability
of being observed in the canonical ensemble. 
Each $x \in \mmbetak$ is a global minimum point of $I_{\beta,K}$
and represents an equilibrium value of the magnetization.
In \cite{EllOttTou}, where Theorems \ref{thm:secondorder}
and \ref{thm:firstorder} are proved, the set $\mmbetak$
was denoted by $\tilde{\cal E}_{\beta,K}$.

For $x \in \R$ we define
\be
\label{eqn:introgbetak} 
G_{\beta,K}(x) = \beta K x^2 - c_\beta(2\beta K x).
\ee
The calculation of the zeroes of $I_{\beta,K}$ --- equivalently, the global minimum points
of $J_{\beta,K}(x) - \beta K x^2$ --- is greatly facilitated by the following observations
made in Proposition 3.4 in \cite{EllOttTou}: 
\begin{enumerate}
\item The global minimum points of 
$J_{\beta,K}(x) - \beta K x^2$ coincide with the global minimum points of $\gbk$,
which are much easier to calculate. 
\item The minimum values $\min_{x \in \R}\{J_{\beta,K}(x) - \beta K x^2\}$ 
and $\min_{x \in \R}G_{\beta,K}(x)$ coincide and both equal 
the canonical free energy $\varphi(\beta,K)$
defined in part (b) of Theorem \ref{thm:ldppnbetak}.
\end{enumerate}
Item 1 gives the alternate characterization 
\be
\label{eqn:ebetak}
\mmbetak = \{x \in [-1,1] : x \mbox{ is a global minimum point of } G_{\beta,K}(x)\}.
\ee
Because of item 2 we call $\gbk$ the free-energy functional of the mean-field B-C model.
In the context of Curie-Weiss-type models, the form of $G_{\beta,K}$ is explained
on page 2247 of \cite{EllOttTou}.

We next apply the Ginzburg-Landau phenomenology to $\gbk$ in order
to reveal the phase-transition structure of the model.  We then
state two theorems showing that the predictions of the Ginzburg-Landau
phenomenology are correct. As explained in \cite{LanLif}, 
the starting point is to represent
$\gbk$ in the positive quadrant of the $\beta$--$K$ plane 
by a polynomial consisting of the first few terms in its Taylor expansion.  The art 
in applying this phenomenology is to have a feel for how 
many terms should be kept.  The global minimum points
of this polynomial, which are easily determined, should indicate the structure
of the global minimum points of $\gbk$ and thus the phase-transition 
structure of the model.  The sets that describe the phase-transition structure
of the model are shown in Figure 1 in the introduction.

One additional aspect of the Ginzburg-Landau phenomenology is to correctly capture
the symmmetries of the order parameter. In the case of mean-field B-C model the order parameter
is the scalar magnetization $m(\beta,K)$, and the only symmetry is sign change, which 
rules out odd powers in the approximations (\ref{eqn:4degree}) and (\ref{eqn:gbk6}).
In more complicated models it becomes an important challenge to construct the correct
Ginzburg-Landau approximation to the free-energy functional that captures all the symmetries.

In the mean-field B-C model $\gbk$ is an even function vanishing at 0. Define $K(\beta) = (e^\beta + 2)/4\beta$.
For $\beta > 0$ and $K > 0$ the first three terms in the Taylor
expansion of $\gbk(x)$ at 0 are 
\be
\label{eqn:gbktaylor}
\frac{G_{\beta, K}^{(2)}(0)}{2!} x^2 + 
\frac{G_{\beta, K}^{(4)}(0)}{4!} x^4 + \frac{G_{\beta, K}^{(6)}(0)}{6!} x^6,
\ee 
where $G_{\beta, K}^{(2)}(0) = {2 \beta K (K(\beta) - K)}/{K(\beta)}$ and
$G_{\beta, K}^{(4)}(0) = {2 (2\beta K)^4 (4-e^{\beta})}/{(e^{\beta} + 2)^2}$.
We will comment on the sixth-order term later. 
The coefficient of $x^2$ in (\ref{eqn:gbktaylor}) is positive,
zero, or negative according to whether $K < K(\beta)$,
$K = K(\beta)$, or $K > K(\beta)$. 
Also, the coefficient of $x^4$ in (\ref{eqn:gbktaylor})
is positive, zero, or negative according to whether $\beta < \log 4$,
$\beta = \log 4$, or $\beta > \log 4$.  The value $\log 4$ defines
the critical inverse temperature $\bc$.  The coefficients
of $x^2$ and $x^4$ both vanish when $(\beta,K) = (\bc,\kc)
= (\log 4, 3/2 \log 4)$, which is the tricritical point.

For $0 < \beta < \bc$, the coefficient of $x^4$ in (\ref{eqn:gbktaylor})
is positive. We represent $\gbk$ by the first two terms in its Taylor
expansion, obtaining for $x$ near 0
\[
\gbk(x) \approx \frac{\beta K [K(\beta) - K]}{K(\beta)} x^2 +
\frac{4 (\beta K)^4 (4-e^{\beta})}{3(e^{\beta} + 2)^2} x^4.
\]
In order to simplify the calculation, we replace $K$ by $K(\beta)$ both in 
the term multiplying $K(\beta) - K$ in the coefficient of $x^2$ and in
the coefficient of $x^4$. 
After the substitution $\beta K(\beta) = (e^\beta + 2)/4$, the 
coefficient of $x^2$ becomes $\beta(K(\beta) - K)$, and the coefficient of $x^4$ 
becomes $c_4(\beta) = (e^\beta + 2)^2(4-e^\beta)/3 \cdot 4^3$. Thus for $0 < \beta < \bc$,
$K$ near $K(\beta)$, and $x$ near 0 we have the ad hoc approximation
\be
\label{eqn:4degree}
\gbk(x) \approx \tildegbk(x) = \beta(K(\beta) - K) x^2 + c_4(\beta) x^4.
\ee
In a different guise a polynomial having a similar form arises in
(\ref{eqn:herehere}) in the derivation
of the asymptotic behavior of $\mbnkn$ for appropriate sequences
$(\bn,\kn)$ converging to a point $(\beta,K(\beta))$ for $0 < \beta < \bc$.

We now describe the structure of the set of global minimum points of $\tildegbk$
for fixed $0 < \beta < \bc$ and variable $K$.
For $0 < K \leq K(\beta)$ the coefficient of $x^2$ in $\tildegbk$ is nonnegative, and $\tildegbk$
has a unique global minimum point at 0.  As $K$ increases through the value
$K(\beta)$, 0 becomes a local maximum point of $\tildegbk$, and $\tildegbk$
picks up two symmetric global minimum points, which we label
$\pm \bar{x}(\beta,K)$. The quantity $\bar{x}(\beta,K)$
is a positive, increasing, continuous function for 
$K > K(\beta)$, and as $K \goto (K(\beta))^+$,
$\bar{x}(\beta,K) \goto 0^+$.  Thus the set of global minimum points
of $\tildegbk$ exhibits a continuous bifurcation
at $K = K(\beta)$, changing continuously from $\{0\}$
for $0 < K \leq K(\beta)$ to $\{\pm \barx(\beta,K)\}$
for $K > K(\beta)$. This continuous bifurcation
corresponds to a second-order phase transition.
As we will see in Theorem 
\ref{thm:secondorder}, for $0 < \beta < \bc$ the behavior of the set of global minimum points 
of $\tildegbk$ has the same qualitative behavior as the behavior
of the set $\mmbetak$ of the set of global minimum points of the free-energy
functional $\gbk$; namely, a continuous
bifurcation corresponding to a second-order phase transition
as $K$ increases through the value $K(\beta)$.  For $0 < \beta < \bc$,
$K(\beta)$ describes the second-order curve.

The analysis for $\beta > \bc$ is much more complicated because of the 
much more intricate phase transition structure in the neighborhood
of the tricritical point.  
\iffalse
Later the point $(\bc,\kc)$ will be identified with the tricritical point of the model. 
By continuity, for all $(\beta,K)$ sufficiently near $(\bc,\kc)$, 
$G_{\beta, K}^{(6)}(0) > 0$.  Accordingly, the formula ({eqn:gbktaylor}) should
give information about the structure of the phase transitions for $(\beta,K)$
in a neighborhood of the tricritical point.
\fi
For $\beta > \bc$ and $(\beta,K)$ in a neighborhood of the tricritical point
we represent $\gbk$ by the first three terms in its Taylor
expansion, obtaining
\[
\gbk(x) \approx \frac{\beta K (K(\beta) - K)}{K(\beta)} x^2
+ \frac{4 (\beta K)^4 (4-e^{\beta})}{3(e^{\beta} + 2)^2} x^4
+ \frac{G_{\beta, K}^{(6)}(0)}{6!} x^6.
\]
In order to simplify the calculation, we replace 
the sixth-order coefficient $G_{\beta, K}^{(6)}(0)$ 
by the value of this derivative at 
the tricritical point $(\bc,\kc) = (\log 4, 3/2\bc)$.
This value is $G_{\bc, \kc}^{(6)}(0) = 2 \cdot 3^4$. 
We also replace $\beta$ and $K$ by $\bc$ and $\kc$ 
both in the terms multiplying $K(\beta) - K$ in the coefficient of $x^2$
and in the term multiplying $4 - e^\beta$ in the coefficient of $x^4$.
With these replacements the coefficient of $x^2$ becomes $\bc(K(\beta) - K)$,
and the coefficient of $x^4$ becomes $3(4 - e^\beta)/16$.
Thus for $\beta > \bc$, $(\beta,K)$ near the tricritical point, and $x$ near 0 
we have the ad hoc approximation
\be
\label{eqn:gbk6}
\gbk(x) \approx \tildegbk(x) =
\bc (K(\beta) - K) x^2 + c_4(4-e^{\beta}) x^4 + c_6 x^6,
\ee
where $c_4 = 3/16$ and $c_6 = 2 \cdot 3^4/6! = 9/40$.  
In a different guise a polynomial having a similar form arises in
(\ref{eqn:Taylorforc}) in the derivation
of the asymptotic behavior of $\mbnkn$ for appropriate sequences
$(\bn,\kn)$ converging to the tricritical point.

$\tildegbk$ is a polynomial of degree 6, the set of global minimum points of which
can be analyzed using Theorem \ref{thm:6order}. The details of this analysis are omitted.
The main point is that for $\beta > \bc$ the set of global minimum points
of $\tildegbk$ exhibits a discontinuous bifurcation
at a certain value $K = \tilde{K}_1(\beta)$, changing discontinuously from $\{0\}$ 
for $K < \tilde{K}_1(\beta)$ to $\{0, \pm \bar{x}(\beta,\tilde{K}_1(\beta))\}$ 
for $K = \tilde{K}_1(\beta)$
to $\{\pm \bar{x}(\beta,K)\}$ for $K > \tilde{K}_1(\beta)$.  In these formulas
$\bar{x}(\beta,K)$ is a positive quantity defined for $\beta > \bc$
and $K \geq \tilde{K}_1(\beta)$.  
This discontinuous bifurcation corresponds to a first-order phase transition.
As we will see in Theorem 
\ref{thm:firstorder}, for $\beta > \bc$ the behavior of the set of global minimum points 
of $\tildegbk$ has the same qualitative behavior as the behavior
of the set $\mmbetak$ of global minimum points of the free-energy
functional $\gbk$; namely, a discontinuous bifurcation corresponding
to a first-order phase transition as $K$ increases through a certain value $K_1(\beta)$. 

The next two theorems give the 
structure of $\mmbetak$ first for $0 < \beta < \bc = \log 4$
and then for $\beta > \bc$.  These theorems make rigorous
the discussion based on the structure of the set of global minimum points
of the approximating polynomials $\tildegbk$ defined in (\ref{eqn:4degree}) and (\ref{eqn:gbk6}).
The first theorem, proved in Theorem 3.6 in \cite{EllOttTou}, 
describes the continuous bifurcation in $\mmbetak$ for $0 < \beta < \bc$.  
This bifurcation corresponds to a second-order phase transition.
The quantity
$K(\beta)$ is denoted by $K^{(2)}_c(\beta)$ in \cite{EllOttTou}
and by $K_c(\beta)$ in \cite{CosEllOtt}.

\begin{thm}
\label{thm:secondorder} 
For $0 < \beta \leq \beta_c$, we define
\be
\label{eqn:kcbeta}
K(\beta) = {1}/[{2\beta c''_\beta(0)}] =
({e^\beta + 2})/({4\beta}).
\iffalse
\frac{1}{4\beta e^{-\beta}} + \frac{1}{2\beta}.
\fi
\ee
For these values of $\beta$, $\mmbetak$ has the following structure.

{\em(a)} For $0 < K \leq K(\beta)$,
${\mathcal{M}}_{\beta,K} = \{0\}$.

{\em(b)} For $K > K(\beta)$, there exists 
${m}(\beta,K) > 0$ such that
${\mathcal{M}}_{\beta,K} = \{\pm m(\beta,K) \}$.

{\em(c)} ${m}(\beta,K)$ is
a positive, increasing, continuous function for $K > K(\beta)$, and
as $K \goto (K(\beta))^+$, $m(\beta,K) \goto 0^+$. 
Therefore, ${\mathcal{M}}_{\beta,K}$ exhibits a continuous bifurcation
at $K(\beta)$.
\end{thm}

The next theorem, proved in Theorem 3.8 in \cite{EllOttTou},
describes the discontinuous bifurcation in $\mmbetak$ for $\beta > \bc$.
This bifurcation corresponds to a first-order phase transition. As shown in that theorem, for all $\beta > \bc$,
$K_1(\beta) < K(\beta)$. The quantity $K_1(\beta)$ is denoted by $K^{(1)}_c(\beta)$ in \cite{EllOttTou}
and by $K_c(\beta)$ in \cite{CosEllOtt}.  

\begin{thm}
\label{thm:firstorder} 
For $\beta > \beta_c$, $\mmbetak$ has
the following structure in terms of the quantity
$K_1(\beta)$, denoted by $K_c^{(1)}(\beta)$ in {\em \cite{EllOttTou}}
and defined implicitly for $\beta > \beta_c$ on page {\em 2231} of
{\em \cite{EllOttTou}}.

{\em(a)} For $0 < K < K_1(\beta)$,
${\mathcal{M}}_{\beta,K} = \{0\}$.

{\em(b)} For $K = K_1(\beta)$ there exists $m(\beta,K_1(\beta)) > 0$
such that ${\mathcal{M}}_{\beta,K_1(\beta)} =
\{0,\pm m(\beta,K_1(\beta))\}$.

{\em(c)} For $K > K_1(\beta)$ 
there exists $m(\beta,K) > 0$
such that ${\mathcal{M}}_{\beta,K} =
\{\pm m(\beta,K)\}$.

{\em(d)} $m(\beta,K)$
is a positive, increasing, continuous function for $K \geq K_1(\beta)$, and 
as $K \goto K_1(\beta)^+$, $m(\beta,K) \goto 
m(\beta,K_1(\beta)) > 0$.  Therefore,
${\mathcal{M}}_{\beta,K}$ exhibits a discontinuous bifurcation
at $K_1(\beta)$.
\end{thm}

We recall from (\ref{eqn:ebetak}) that $\mmbetak$ can be characterized as the set
of global minimum points of $\gbk$.  
In Figure 3 we exhibit the graphs of $\gbk$ for $0 < \beta \leq \bc$ and increasing
values of $K > 0$.  These graphs are based on the detailed analysis of the global
minimum points of a related function in section 3.2 of \cite{EllOttTou}.
The graph for $0 < K \leq K(\beta)$ is shown in (a), and the graph for
$K > K(\beta)$ is shown in (b).  

\begin{figure}[h]
\begin{center}
\includegraphics[width=10cm]{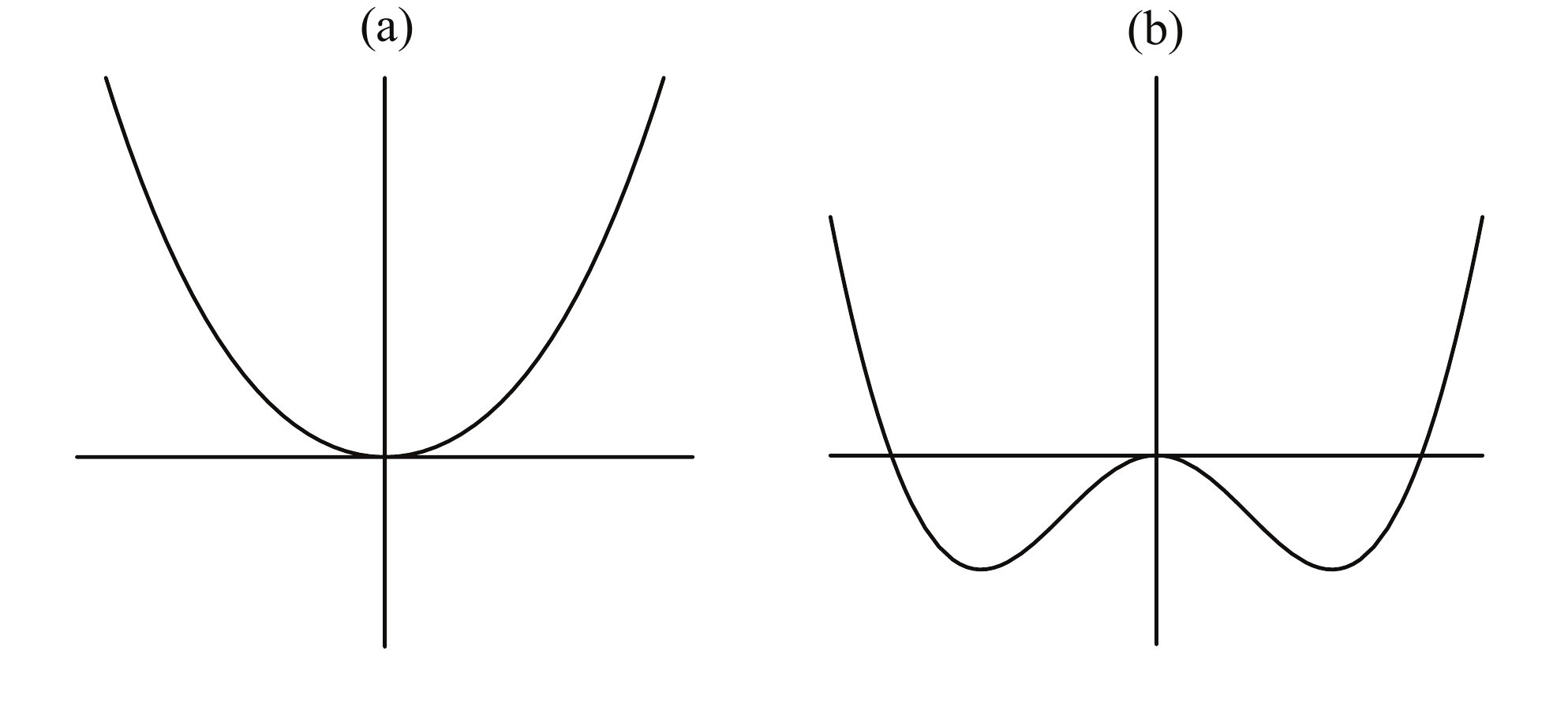}
\vspace{-.15in}
\caption{\footnotesize Graphs of $\gbk$ for $0 < \beta \leq \bc$. (a) $0 < K \leq K(\beta)$,
(b) $K > K(\beta)$.}
\end{center}
\end{figure} 

In Figures 4, 5, and 6, we exhibit the graphs of $\gbk$ for $\beta > \bc$ and 
increasing values of $K > 0$.
These graphs are based 
on the detailed analysis of the global minimum points of a related function
in section 3.3 of \cite{EllOttTou}.  The graphs for $0 < K < K_1(\beta)$
are shown in Figure 4, the graph for $K = K_1(\beta)$ in Figure 5,
and the graphs for $K > K_1(\beta)$ in Figure 6. These
three figures correspond, respectively, to parts (a), (b), and (c) of Theorem
\ref{thm:firstorder}. 

\begin{figure}[h]
\begin{center}
\includegraphics[width=10cm]{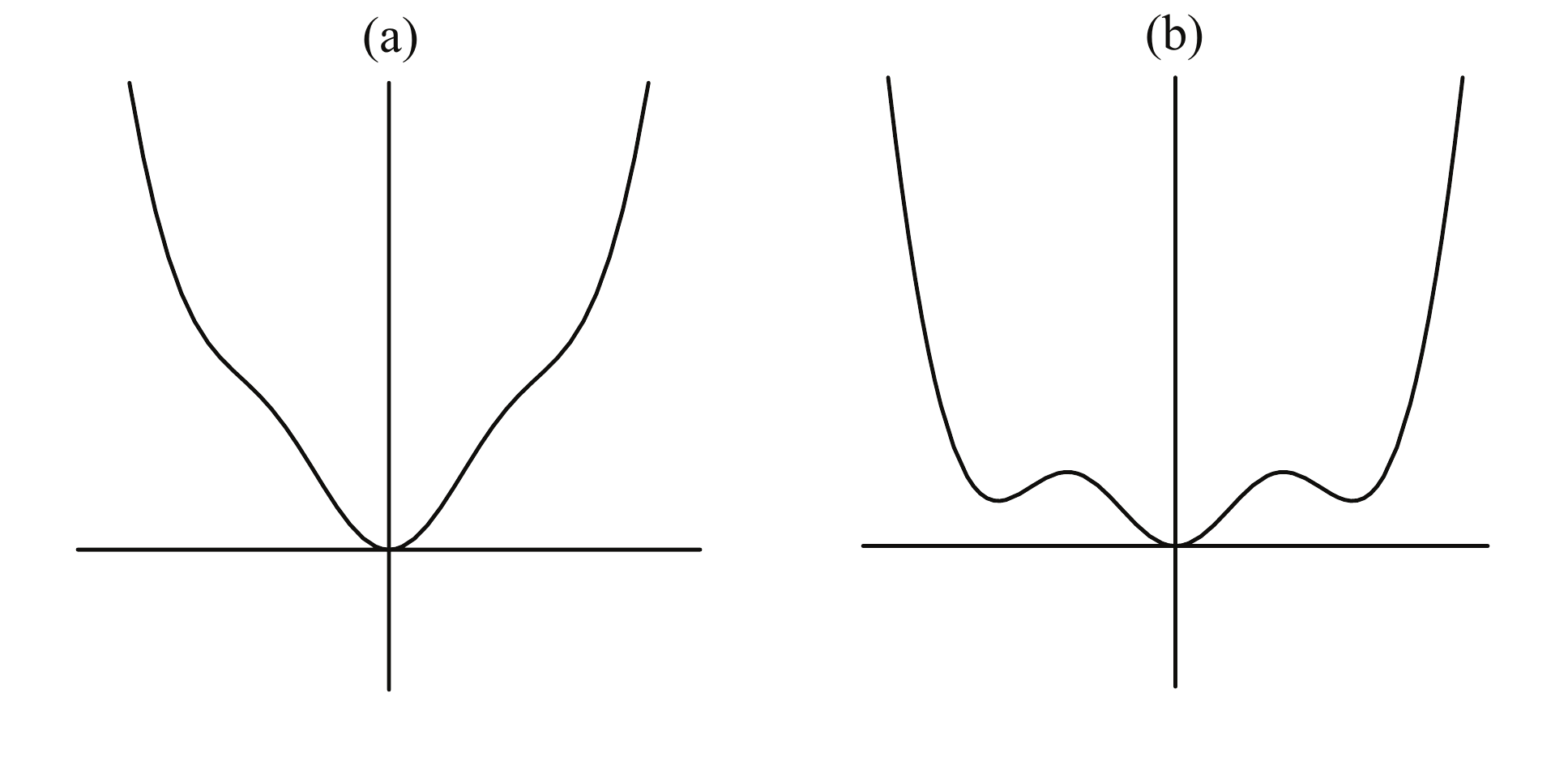}
\vspace{-.15in}
\caption{\footnotesize Graphs of $\gbk$ for $\beta > \bc$. (a) $0 < K < \kappa(\beta)$,
(b) $\kappa(\beta) < K < K_1(\beta)$. For $\beta > \bc$ the
set of minimum points of $\gbk$ undergoes the bifurcation shown in
graphs (a) and (b) as $K$ increases through the value $\kappa(\beta)$.}
\end{center}
\end{figure} 

\begin{figure}[h!]
\begin{center}
\includegraphics[width=5cm]{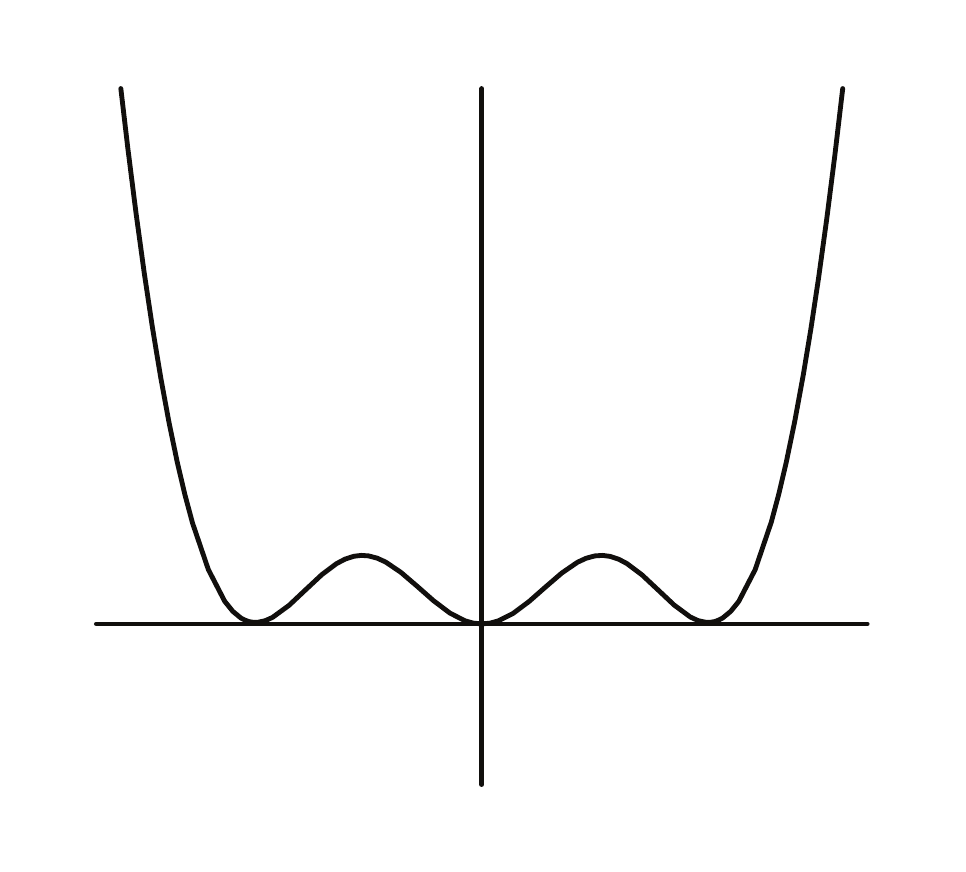}\vspace{-.15in}
\caption{\footnotesize Graph of $\gbk$ for $\beta > \bc$ and $K = K_1(\beta)$.}
\end{center}
\end{figure} 

\begin{figure}[h!]
\begin{center}
\includegraphics[width=12cm]{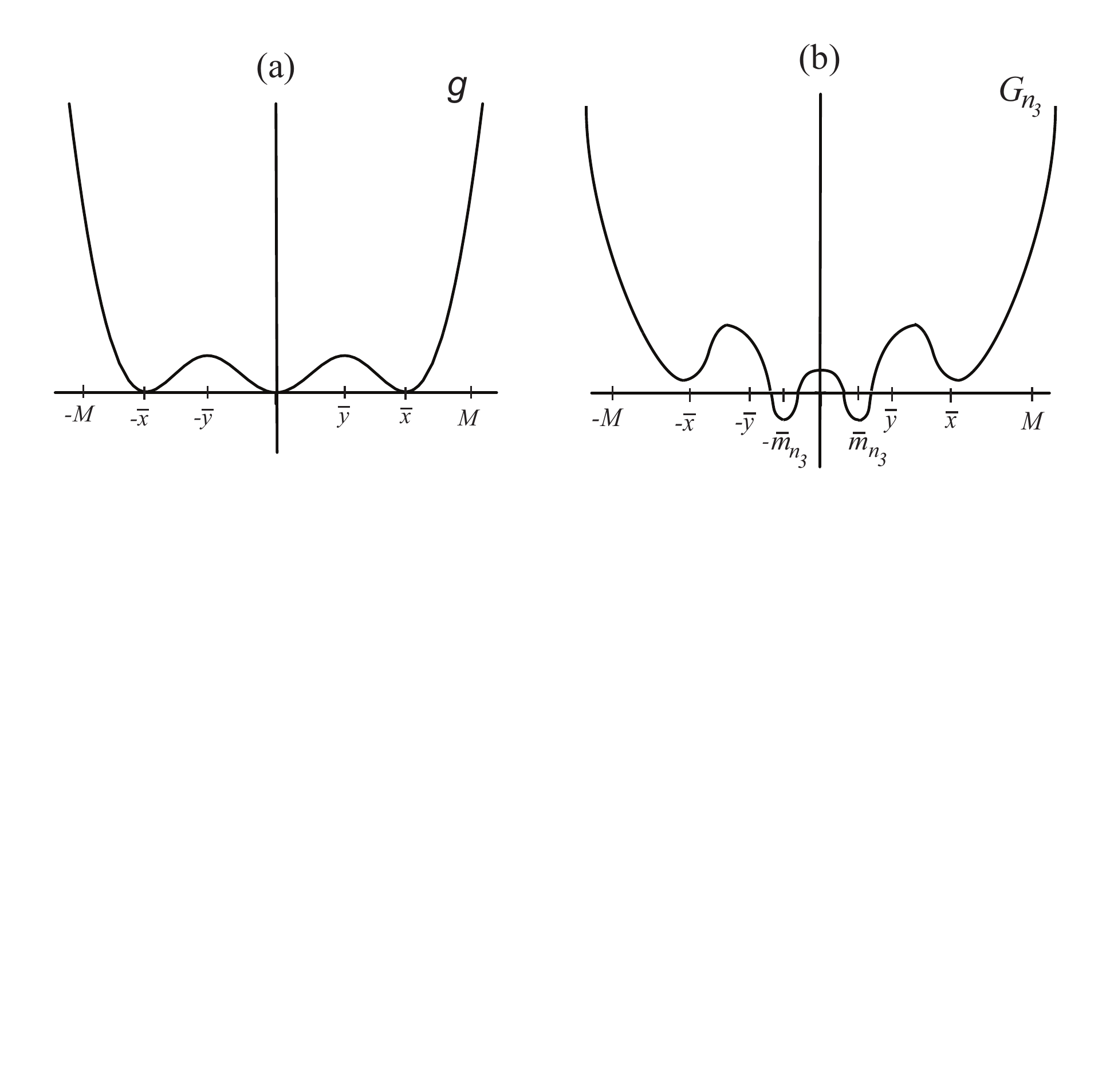}
\vspace{-.5in}
\caption{\footnotesize Graphs of $\gbk$ for $\beta > \bc$. (a) $K_1(\beta) < K < K(\beta)$,
(b) $K \geq K(\beta)$.}
\end{center}
\end{figure} 

In the next section we determine the asymptotic behavior of $m(\bn,\kn)$
for appropriate sequences $(\bn,\kn)$ converging from the phase-coexistence
region to a second-order point. In determining the asymptotic behavior, we will
see how to make rigorous the Ginzburg-Landau phenomenology by replacing, with
a well defined limit,
the approximation of the free-energy functional $\gbk$ by the fourth-degree
polynomial $\tildegbk$ in (\ref{eqn:4degree}).  

\section{Asymptotic Behavior of \boldmath $\mbnkn$ \unboldmath Near
a Second-Order Point}
\beginsec
\label{section:asymptoticsforb}

In this section we derive the asymptotic behavior of 
the magnetization $m(\bn,\kn)$ for two
sequences $(\beta_n,K_n)$.  For $0 < \beta < \bc$ each of these sequences converges
to a second-order point $(\beta,K(\beta))$ 
from the phase-coexistence region located above the second-order curve.
This section is a warm-up for section \ref{section:asymptoticsforc}, 
in which we analyze the much more
complicated asymptotic behavior of $\mbnkn$ in the neighborhood of the tricritical point.

By definition, when $(\bn,\kn)$ lies in the phase-coexistence region, $\mbnkn$ is the unique positive, global minimum point of the free-energy
functional $\gbnkn$. For each of the sequences considered in this section
the asymptotic behavior of $\mbnkn$ is expressed in terms
of the unique positive, global minimum point $\bar{x}$ of the limit of a suitable scaled version of $\gbnkn$. 
This limit is a fourth degree polynomial called the Ginzburg-Landau polynomial.
As we will see, properties of this polynomial reflect the phase-transition
structure of the mean-field B-C model, thus making rigorous the predictions of the Ginzburg-Landau
phenomenology of critical phenomena mentioned in section \ref{section:phasetr}.  

The two sequences $(\bn,\kn)$ to be considered in this section are defined in terms of a positive parameter
$\alpha$ that regulates the speed of approach of $(\bn,\kn)$ to a second-order point.
The asymptotic behavior of $\mbnkn$ for each of the two sequences is given in
Theorems \ref{thm:asymptoticsforb} and \ref{thm:asymptoticsforb2}.  This behavior
is derived from the general result in Theorem \ref{thm:exactasymptotics}.
 
For $0 < \beta < \bc$ let $(\bn,\kn)$ be an arbitrary positive sequence converging to a second-order
point $(\beta,K(\beta))$ and let
$\gamma > 0$ be given.
In the preceding section we motivated the phase-transition structure
for $0 < \beta < \bc$ by approximating $\gbk(x)$ in (\ref{eqn:4degree})
by a polynomial of degree 4 derived from the first two terms of its Taylor expansion.  
The starting point in determining the asymptotic behavior of $\mbnkn$ 
is to replace this two-term Taylor expansion for $\gbk(x)$ by 
the two-term Taylor expansion for $nG_{\bn,\kn}(x/n^\gamma)$ with an error term. 
According to Taylor's Theorem, for all $n \in \N$, any $R > 0$, and
all $x \in \R$
satisfying $|x/n^\gamma| < R$ there exists $\xi_n(x/n^\gamma)
\in [-x/n^\gamma,x/n^\gamma]$ such that
\bea
\label{eqn:Taylor}
\lefteqn{
nG_{\beta_n, K_n}(x/n^\gamma)} \\ \nonumber
&& = \frac{1}{n^{2\gamma-1}}\frac{G_{\beta_n, K_n}^{(2)}(0)}{2!} x^2 + 
\frac{1}{n^{4\gamma-1}}\frac{G_{\beta_n, K_n}^{(4)}(0)}{4!} x^4 + 
\frac{1}{n^{5\gamma - 1}} \frac{\gbnkn^{(5)}(\xi_n(x/n^\gamma))}{5!}  x^5.
\eea
In deriving this formula, we use the fact that $\gbnkn(0) = 0$
and that since $\gbnkn$ is an even function, $\gbnkn^{(1)}(0) = 0 = \gbnkn^{(3)}(0)$. 
Because the sequence
$(\bn,\kn)$ is positive and bounded, there exists $a \in (0,\infty)$ such that 
$0 < \bn \leq a$ and $0 < \kn \leq a$ for all $n$. As a continuous function 
of $(\beta,K,y)$ on the compact set $[0,a] \times [0,a] \times [-R,R]$, $G_{\beta,K}^{(5)}(y)$
is uniformly bounded.  It follows that the quantity $\gbnkn^{(5)}(\xi_n(x/n^\gamma))$
appearing in the error term in the Taylor expansion is 
uniformly bounded for $n \in \N$ and 
$x \in (-\rng,\rng)$.  We summarize this expansion by writing
\be
\label{eqn:TaylorG2again}
nG_{\beta_n, K_n}(x/n^\gamma) =
\frac{1}{n^{2\gamma-1}}\frac{G_{\beta_n, K_n}^{(2)}(0)}{2!} x^2 + 
\frac{1}{n^{4\gamma-1}}\frac{G_{\beta_n, K_n}^{(4)}(0)}{4!} x^4 
+ \mbox{O}\!\left(\frac{1}{n^{5\gamma-1}}\right) \! x^5,
\ee
where the big-oh term is uniform for $x \in (-Rn^\gamma,Rn^\gamma)$. 

In terms of the quantity $K(\beta) = (e^\beta + 2)/(4\beta)$,
the coefficients $G_{\beta_n, K_n}^{(2)}(0)$ and 
$G_{\beta_n, K_n}^{(4)}(0)$ in the Taylor expansion are given by
\[
G_{\beta_n, K_n}^{(2)}(0) = 
\frac{2 \beta_n K_n (K(\beta_n) - K_n)}{K(\beta_n)}
= 2\beta (K(\bn) - \kn) \cdot \frac{\bn\kn}{\beta K(\bn)}
\]
and 
\[
G_{\beta_n, K_n}^{(4)}(0) = \frac{2 (2\beta_n K_n)^4 (4-e^{\beta_n})}{(e^{\beta_n} + 2)^2}.
\]
In order to ease the notation, we let $\ve_n$ denote a sequence that converges to 0
and that represents the various error terms arising in the following calculation; 
we use the same notation $\ve_n$ to represent different error terms.
Since $(\bn,\kn)$ converges to $(\beta,K(\beta))$ 
and the function $K(\cdot)$ is continuous, we have 
$\beta_n K_n/K(\bn) \goto \beta$.  Thus 
\[
{G_{\beta_n, K_n}^{(2)}(0)}/{2!} = \beta (K(\beta_n) - K_n) (1 + \ve_n). 
\]
Define $c_4(\beta) = {(e^\beta + 2)^2 (4-e^\beta)}/({8 \cdot 4!})$.
Since $2\bn\kn \goto 2\beta K(\beta) = (e^\beta + 2)/2$, we also have
\be
\label{eqn:c4beta}
{G_{\beta_n, K_n}^{(4)}(0)}/{4!} = {(e^\beta + 2)^2 (4-e^\beta)}(1 + \ve_n)/({8 \cdot 4!})
= c_4(\beta) (1 + \ve_n);
\ee
$c_4(\beta) > 0$ since $4 - e^\beta = e^{\bc} - e^\beta > 0$.
Thus for all $n \in \N$, any $\gamma > 0$, any $R > 0$, and all $x \in \R$ satisfying $|x/n^\gamma| < R$
\bea
\label{eqn:herehere}
nG_{\beta_n, K_n}(x/n^\gamma) & = &
\frac{1}{n^{2\gamma-1}}\beta(K(\bn) - \kn) (1 + \ve_n) x^2 \\ 
\nonumber \hspace{.25in} && + 
\frac{1}{n^{4\gamma-1}}c_4(\beta) (1 + \ve_n) x^4 
+ \mbox{O}\!\left(\frac{1}{n^{5\gamma-1}}\right) \! x^5,
\eea
where the big-oh term is uniform for $x \in (-Rn^\gamma,Rn^\gamma)$.

For the moment, on the right side of the last display
let us replace $(\bn,\kn)$ by $(\beta,K)$, set $n = 1$, and drop
the big-oh term. Doing so,
we obtain the polynomial $\tildegbk$ that approximates the free energy
functional $\gbk$ in (\ref{eqn:4degree}) for $0 < \beta < \bc$, $K$
near $K(\beta)$, and $x$ near 0. Arising via the Ginzburg-Landau phenomenology, this 
polynomial is used in section \ref{section:phasetr} to motivate
the continuous bifurcation in the set of equilibrium values of the magnetization
that is described rigorously in Theorem \ref{thm:secondorder}. 
As we will soon see, by a suitable choice of $(\bn,\kn)$ and other parameters the
right side of the last display converges to a Ginzburg-Landau polynomial in terms of which
the asymptotic behavior of $\mbnkn$ is described.

We return to (\ref{eqn:herehere}), in which the term $(K(\beta_n) - K_n)$ converges
to 0 as $n \goto \infty$.  
The two different asymptotic behaviors of $\mbnkn$ to be considered in this section
each depends on the choice of the sequence $(\bn,\kn)$ converging to the second-order point $(\beta,K(\beta))$.
Each choice controls, in a different way, the rate at which $(K(\beta_n) - K_n) \goto 0$.
We analyze two different cases, each giving rise to a Ginzburg-Landau polynomial
having a unique positive, global minimum points at $\bar{x}$ for some $\bar{x} > 0$. This quantity enters
the respective asymptotic formula for $\mbnkn \goto 0$. 

Fix $0 < \beta < \bc$.  For the first choice of sequence
we take $\alpha > 0$, $b \in \{1,0,-1\}$,
and $k \in \R$ and define
\be
\label{eqn:newbetankn}
\beta_n = \beta + {b}/{n^\alpha} \ \mbox{ and } \ K_n = K(\beta) + {k}/{n^\alpha}.
\ee
If $b \not =0$, then $(\beta_n,K_n)$ converges to $(\beta,K(\beta))$ along a ray with slope $k/b$ (see path 1 in Figure 2).
We assume that $K'(\beta)b - k \not = 0$.  Since 
\[
K(\bn) = K(\beta + b/n^\alpha) = K(\beta) + K'(\beta) b/n^\alpha + \mbox{O}(1/n^{2\alpha}),
\]
we have
\be
\label{eqn:replace}
K(\beta_n) - K_n =  ({K'(\beta) b - k})/{n^\alpha} + \mbox{O}(1/n^{2\alpha}).
\ee
It follows from (\ref{eqn:herehere}) that for all $n \in \N$, 
any $\gamma > 0$, any $R > 0$, and all $x \in \R$ satisfying $|x/n^\gamma| < R$
\bea
\label{eqn:prelimgbnkn}
\lefteqn{
nG_{\beta_n, K_n}(x/n^\gamma) } \\ 
\nonumber && = \frac{1}{n^{2\gamma + \alpha - 1}}  \beta(K'(\beta) b - k) (1+\ve_n) x^2 \\
\nonumber && \hspace{.15in}  + \ \frac{1}{n^{4\gamma - 1}} c_4(\beta) (1+\ve_n) x^4 
+ \mbox{O}\!\left(\frac{1}{n^{2\gamma + 2\alpha -1}}\right) \!x^2
+ \mbox{O}\!\left(\frac{1}{n^{5\gamma -1 }}\right) \!x^5.
\eea
Because $K'(\beta) b - k \not = 0$ and $c_4(\beta) > 0$, the coefficients of $x^2$ and of $x^4$ are both nonzero.

The case where $K'(\beta)b - k = 0$ must be handled differently.
If this equality holds, then the asymptotic expression (\ref{eqn:replace}) for $(K(\bn) - \kn)$
is indeterminate.  In order to calculate the correct asymptotic expression
for $(K(\bn) - \kn)$ when $K'(\beta)b - k = 0$, one must consider the next term in
the Taylor expansion of $K(\beta + b/n^\alpha)$, obtaining (\ref{eqn:kbnknsim}) with
$p = 2$ and $\ell = 0$. 
We carry out the asymptotic analysis for this case in Theorem \ref{thm:asymptoticsforb2}.

We return to the sequence $(\bn,\kn)$ in (\ref{eqn:newbetankn}) when $K'(\bc)b - k \not = 0$.
In order to obtain the limit of $n\gbnkn(x/n^\gamma)$, we impose the condition
that the powers of $n$ appearing in the first two terms in 
(\ref{eqn:prelimgbnkn}) equal 0; i.e., $2\gamma + \alpha - 1 = 0 = 4 \gamma -1$,
which is equivalent to $\gamma = 1/4$ and $\alpha = 1-2\gamma = 1/2$.
With this choice of $\gamma$ and $\alpha$, the powers of $n$ appearing in the 
last two terms in (\ref{eqn:prelimgbnkn}) are positive, and so 
for all $x \in \R$ both terms
converge to 0 as $n \goto \infty$.  It follows that for 
$\gamma = 1/4$ and $\alpha = 1/2$, as $n \goto \infty$
we have for all $x \in \R$ 
\[
n\gbnkn(x/n^\gamma) \goto g(x) = \beta(K'(\beta) b - k) x^2 + c_4(\beta) x^4.
\]
We call $g$ the Ginzburg-Landau polynomial.  
Since the big-oh terms in (\ref{eqn:prelimgbnkn})
are uniform for $x \in (-Rn^\gamma, Rn^\gamma)$,
the convergence of $n\gbnkn(x/n^\gamma)$ to $g(x)$ is uniform for $x$ in compact subsets of $\R$. 

By a simple trick, we are able to derive a similar limit that is valid for all $\alpha > 0$.
Let $u$ be a real number that will be chosen momentarily.
Multiplying the numerator and denominator of the right side of (\ref{eqn:prelimgbnkn})
by $n^{u}$, we obtain $n \gbnkn(x/n^\gamma) = n^{u} G_n(x)$,
where for all $n \in \N$, any $\gamma > 0$, any $R > 0$, and all $x \in \R$ satisfying $|x/n^\gamma| < R$
\bea
\label{eqn:rewriteasympt}
\lefteqn{ G_n(x) } \\
\nonumber & & = 
\frac{1}{n^{2\gamma + \alpha - 1 + u}} \beta(K'(\beta) b - k) (1+\ve_n) x^2 \\
\nonumber && \hspace{.15in} + \ \frac{1}{n^{4\gamma - 1 + u}} c_4(\beta) (1+\ve_n) x^4
+ \mbox{O}\!\left(\frac{1}{n^{2\gamma + 2\alpha -1 + u}}\right) \!x^2
+ \mbox{O}\!\left(\frac{1}{n^{5\gamma -1 + u}}\right) \!x^5.
\eea
In this formula $\ve_n \goto 0$ and the big-oh terms are uniform
for $x \in (-Rn^\gamma,Rn^\gamma)$.
Again we impose the condition that the powers of $n$ appearing in the first two 
terms in (\ref{eqn:rewriteasympt}) equal 0; i.e., $2\gamma + \alpha - 1 + u = 0 = 4 \gamma -1 + u$. 
These two equalities are equivalent to $\gamma = \alpha/2$ and 
$u = 1 - 4\gamma = 1 - 2\alpha$.  
\iffalse
As $\alpha$ ranges through the interval $(0,\infty)$,
$\gamma$ takes all positive values and $u$ takes all values in $(-\infty,1)$. 
\fi
With this choice of $\gamma$ and $u$, the powers of $n$ appearing in the 
last two terms in (\ref{eqn:rewriteasympt}) are positive, and so 
for all $x \in \R$ both terms converge to 0 as $n \goto \infty$.  
It follows that as $n \goto \infty$, we have for all $x \in \R$
\be
\label{eqn:gnx}
G_n(x) = n^{1-u} \gbnkn(x/n^\gamma) \goto g(x) = \beta(K'(\beta) b - k) x^2 + c_4(\beta) x^4.
\ee
Again, since the big-oh terms in (\ref{eqn:rewriteasympt}) are uniform for $x \in (-Rn^\gamma,Rn^\gamma)$, the
convergence of $G_n(x)$ to $g(x)$ is uniform for $x$ in compact subsets of $\R$.

As we will see in Theorem \ref{thm:exactasymptotics}, by proving the convergence in (\ref{eqn:gnx}) for 
$u = 1 - 2\alpha \in (-\infty,1)$, we obtain the asymptotic behavior of $\mbnkn$ for any $\alpha > 0$.  If we worked
only with $u = 0$, then we would obtain the asymptotic behavior of $\mbnkn$ only for $\alpha = 1/2$.

The Ginzburg-Landau polynomial $g$ has two different behaviors depending on the sign of $\beta(K'(\beta) b - k)$, 
which is the coefficient of $x^2$.  We first consider the case where 
$\beta(K'(\beta) b - k) > 0$.  This corresponds to $(\beta_n,K_n)$
converging to $(\beta,K(\beta))$
\iffalse
either along the tangent line to $(\beta,K(\beta))$ with slope $k/b = K'(\beta)$ or 
\fi
along a ray lying beneath the tangent line to $(\beta,K(\beta))$.
Since $K(\beta)$ is a convex function for $0 < \beta < \bc$ [Lem.\ \ref{lem:kbeta}(c)], 
in this situation $(\beta_n,K_n)$ converges to $(\beta,K(\beta))$ from 
the single-phase region located beneath the second-order curve.
In this case, for all $n$ the free energy functional $\gbnkn$ 
has a unique global minimum point 0 [Thm.\ \ref{thm:secondorder}(a)],
a property reflected in the fact that the Ginzburg-Landau polynomial 
\[
g(x) = \beta(K'(\beta) b - k) x^2 + c_4(\beta) x^4
\]
also has a unique global minimum point 0. 

We now consider the case where $\beta(K'(\beta) b - k) < 0$,
which corresponds to $(\beta_n,K_n)$ converging to $(\beta,K(\beta))$ along a ray lying above the tangent line
to $(\beta,K(\beta))$.
If $b = 1$, then the slope of the ray satisfies $k/b > K'(\beta)$, which corresponds to 
$(\beta_n,K_n) \goto (\beta,K(\beta))$ from the northeast;
if $b = 0$, then $(\beta_n,K_n)$ converges to $(\beta,K(\beta))$ from the north; and if $b = -1$, 
then the slope of the ray satisfies $k/b < K'(\beta)$, which corresponds to 
$(\beta_n,K_n)$ converging to $(\beta,K(\beta))$ from the northwest.
In each of these situations $(\bn,\kn)$ lies in the phase-coexistence region for all sufficiently large 
$n$. For such values of $n$ the global minimum points of
the free energy functional $\gbnkn$ are $\pm \mbnkn$ [Thm.\ \ref{thm:secondorder}(b)],
a property reflected in the fact that the 
global minimum points of the Ginzburg-Landau polynomial 
$g$ are also a pair of symmetric nonzero points; these
points are $\pm \bar{x}$ for $\bar{x} > 0$ defined in (\ref{eqn:barxforb}).

\iffalse
In section \ref{section:asymptoticsforc} we will consider four examples of sequences $(\bn,\kn)$
converging to the tricritical point. For each of these sequences
the structure of the set of global minimum points of the 
corresponding Ginzburg-Landau polynomial
mirrors the structure of the set of global minimum points of the free-energy functional $G_{\bn,\kn}$.
In this sense the properties of the Ginzburg-Landau polynomial reflect
the phase-transition structure in the region through which $(\bn,\kn)$ passes.
This feature of the Ginzburg-Landau polynomials
makes rigorous the predictions of the Ginzburg-Landau
phenomenology of critical phenomena mentioned in section \ref{section:phasetr}.
\fi

In the next theorem we derive 
the asymptotic behavior of $m(\bn,\kn)$ when the sequence
$(\bn,\kn)$ defined in (\ref{eqn:newbetankn}) converges to a 
second-order point $(\beta,K(\beta))$ from the phase-coexistence region
located above the second-order curve; equivalently, when 
the coefficient $\beta(K'(\beta) b - k)$ of $x^2$ in the Ginzburg-Landau
polynomial $g$ is negative. According to Theorem \ref{thm:zngoto0},
in this case $\mbnkn \goto 0$. Theorem \ref{thm:exactasymptotics} shows that as a consequence of the 
uniform convergence of $G_n$ to $g$ and other hypotheses, the sequence $n^{\alpha/2}\mbnkn$
of positive global minimum points of $G_n$ converges to the positive global minimum point
$\barx$ of $g$.  This yields the asymptotic formula $\mbnkn \sim \barx/n^{\alpha/2} \goto 0$ given in part (c) of the next theorem.

\begin{thm}
\label{thm:asymptoticsforb}
For $\beta \in (0,\bc)$, $\alpha > 0$, $b \in \{1,0,-1\}$, and a real number $k \not = K'(\beta)b$, define
\[
\beta_n = \beta + {b}/{n^\alpha} \ \mbox{ and } \ K_n = K(\beta) + {k}/{n^\alpha}
\]
as well as $c_4(\beta) = (e^\beta + 2)^2 (4-e^\beta)/(8 \cdot 4!)$.
Then $(\bn,\kn)$ converges to the second-order point $(\beta,K(\beta))$.
The following conclusions hold.

{\em (a)} For any $\alpha > 0$, $u = 1-2\alpha$, and $\gamma = \alpha/2$
\[
G_n(x) = n^{1-u}\gbnkn(x/n^\gamma) \goto 
g(x) =  \beta(K'(\beta) b - k) x^2 + c_4(\beta) x^4
\]
uniformly for $x$ in compact subsets of $\R$.

{\em (b)} The Ginzburg-Landau polynomial $g$ has nonzero 
global minimum points if and only if $K'(\beta)b - k < 0$. If this inequality holds, then the global minimum points of
$g$ are $\pm \bar{x}$, where
\be
\label{eqn:barxforb}
\bar{x} = \left({\beta (k - K'(\beta) b)}/[2 c_4(\beta)]\right)^{1/2}
\ee

{\em (c)} Assume that $K'(\beta)b - k < 0$. Then for any $\alpha > 0$, $m(\bn,\kn) \goto 0$ and has the asymptotic behavior 
\[
\mbnkn \sim {\bar{x}}/{n^{\alpha/2}}; \ \mbox{ i.e., }
\lim_{n \goto \infty} n^{\alpha/2} \mbnkn = \bar{x}.
\]
If $b \not = 0$, then this becomes $\mbnkn \sim \barx|\beta - \bn|^{1/2}$.
\end{thm}

\noindent
{\bf Proof.} Part (a) follows from the discussion leading up to the
statement of the theorem.  The first assertion in part (b) is elementary. 
If $K'(\beta) b - k < 0$, then the equation $g'(x) = 2 \beta(K'(\beta) b - k) x + 4 c_4(\beta) x^3 = 0$ has solutions
at $\pm \bar{x}$ and at 0, where $\bar{x}$ is defined in (\ref{eqn:barxforb}).   One easily 
checks that $\pm \bar{x}$ are global minimum points and 0 a local maximum point.

We now verify the asymptotic behavior of $\mbnkn$ in part (c).  The convergence
$m(\bn,\kn) \goto 0$ is proved in Theorem \ref{thm:zngoto0}. 
The asymptotic behavior $\mbnkn \sim \bar{x}/n^{\alpha/2}$
is a consequence of Theorem \ref{thm:exactasymptotics}, a general result
covering the present case, the sequence considered in Theorem \ref{thm:asymptoticsforb2},
and the sequences $(\bn,\kn)$ converging to the tricritical point
that are considered in Theorems \ref{thm:asymptoticsforc1}--\ref{thm:asymptoticsforc4}.  
We now verify the four hypotheses of that theorem
for the sequence $(\bn,\kn)$ in Theorem \ref{thm:asymptoticsforb}, which converges
to the second-order point $(\beta,K(\beta))$. Hypothesis 
(i) is valid since $K'(\beta)b - k < 0$ is equivalent to 
$K_n > K(\bn)$ for all sufficiently $n$ [see
(\ref{eqn:replace})]. Hence the inequality $K'(\beta)b - k < 0$ guarantees that for all sufficiently large $n$,
$(\bn,\kn)$ lies in the phase-coexistence region above the second-order curve. 
Hypothesis (ii) is valid since $n^\alpha(\bn - \beta) = b$, 
$n^\alpha (\kn - K(\beta)) = k$, and either $b$ or $k$ is nonzero.
Hypothesis (iii) involves the Ginzburg-Landau polynomial
$g$ in part (a), which is an even polynomial of degree 4
satisfying $g(x) \goto \infty$ as $|x| \goto \infty$.
Hypothesis (iii)(a) states that there exist $\alpha_0 > 0$ 
and $\theta > 0$ such that for any $\alpha > 0$,  
if $u = 1 - \alpha/\alpha_0$ and $\gamma = \theta \alpha$, then 
\[
\lim_{n \goto \infty} n^{1-u} \gbnkn(x/n^\gamma) = g(x) =
\beta(K'(\beta) b - k) x^2 + c_4(\beta) x^4 
\]
uniformly for $x$ in compact subsets of $\R$.
As verified by the calculation leading up to the limit 
(\ref{eqn:gnx}) and as stated in part (a) of Theorem \ref{thm:asymptoticsforb},
hypothesis (iii)(a) is valid with $\alpha_0 = 1/2$ and $\theta = 1/2$. 
Hypothesis (iii)(b) is satisfied  
since $g$ has a unique positive, global minimum point $\bar{x}$,
Hypothesis (iv), the most technical of the four, is verified in the paragraph after the next one.
Since $\theta = 1/2$, the general result in Theorem \ref{thm:exactasymptotics}
takes the form $\mbnkn \sim \bar{x}/n^{\theta \alpha} = \bar{x}/n^{\alpha/2}$.
This is the conclusion of part (c) of Theorem \ref{thm:asymptoticsforb}.

We next consider hypothesis (iv) in Theorem \ref{thm:exactasymptotics}, which
states that there exists a polynomial $H(x)$ satisfying $H(x) \goto \infty$ as $|x| \goto \infty$
together with the following property: for any $\alpha > 0$ there exists $R > 0$ such that, if 
$u = 1 - \alpha/\alpha_0 = 1 - 2\alpha$
and $\gamma = \theta \alpha = 2 \alpha$, then
for all sufficiently large $n \in \N$ and for all $x \in \R$
satisfying $|x/n^\gamma| < R$, $n^{1-u} \gbnkn(x/n^\gamma) \geq H(x)$.  
This hypothesis is used in the proof
of the theorem to verify that the sequence $n^\gamma \mbnkn$ is bounded,
a key component in the derivation of the asymptotic behavior of $\mbnkn$.

In order to verify hypothesis (iv) in Theorem \ref{thm:exactasymptotics}, we 
fix $\alpha > 0$ and substitute $u = 1 - 2\alpha$ and $\gamma = \alpha/2$ in the expansion
(\ref{eqn:rewriteasympt}) for $G_n(x) = n^{1-u}\gbnkn(x/n^\gamma)$. It follows that for 
any $\alpha > 0$, any $R > 0$, all $n \in \N$, 
and all $x \in \R$ satisfying $|x/n^\gamma| < R$ 
\beas
n^{1-u}\gbnkn(x/n^\gamma) & = & \beta(K'(\beta) b - k) (1+\ve_n) x^2 \\
&& + \, c_4(\beta) (1+\ve_n) x^4
+ \mbox{O}({1}/{n^\alpha}) x^2
+ \mbox{O}({1}/{n^\gamma}) x^5 \\
& = & \left[\beta(K'(\beta) b - k) (1+\ve_n) + \mbox{O}(1/n^\alpha)\right] x^2  \\
&& + \left[c_4(\beta)(1+\ve_n) + \mbox{O}(x/n^\gamma)\right] x^4.
\eeas
Since the big-oh terms are uniform for
$x \in (-Rn^\gamma,Rn^\gamma)$, we conclude that for any $\alpha > 0$ there exists
$R > 0$ such that for all sufficiently large $n \in \N$ and all $x \in \R$ satisfying
$|x/n^\gamma| < R$
\be
\label{eqn:gh}
n^{1-u}\gbnkn(x/n^\gamma) \geq H(x) = 
-2\beta|K'(\beta) b - k| x^2 + \ts \frac{1}{2} c_4(\beta) x^4.
\ee
Since $H(x) \goto \infty$ as $|x| \goto \infty$, hypothesis (iv) in Theorem 
\ref{thm:exactasymptotics} is satisfied. 
This completes the verification that the four hypotheses of Theorem
\ref{thm:exactasymptotics} are valid in the context of Theorem \ref{thm:asymptoticsforb}
and so completes the proof of the latter theorem. \ \ink

\skp
We now consider the second choice of the sequence $(\bn,\kn)$ converging to a second-order point. This sequence
gives a different asymptotic behavior of $m(\bn,\kn) \goto 0$ from the sequence considered
in Theorem \ref{thm:asymptoticsforb}.
Let $(\bzero,\kbzero)$ be a second-order point corresponding to $0 < \bzero < \bc$.
Given $\alpha > 0$, $b \in \{1,-1\}$, an integer $p \geq 2$, 
and $\ell \in \R$ we define
\be
\label{eqn:bnknb2}
\bn = \bzero + {b}/{n^\alpha} \ \mbox{ and } \ 
\kn = \kbzero + \sum_{j=1}^{p-1} { K^{(j)}(\bzero)b^j}/({j! n^{j\alpha}}) + {\ell b^{p}}/({p! n^{p\alpha}}).
\ee
In order to simplify the analysis, we assume that $\ell \not = K^{(p)}(\bzero)$.  
The choice $\ell = K^{(p)}(\bzero)$ will be discussed after Theorem \ref{thm:asymptoticsforb2}.
\iffalse
COMMENT: THIS IS NOT SO.
The choice $p = 2$ and $\ell = 0$ reduces to the sequence $(\bn,\kn)$ 
in (\ref{eqn:newbetankn}). However, as pointed out in the paragraph after (\ref{eqn:prelimgbnkn}),
the asymptotic analysis given there is valid only when $K'(\bzero)b - k \not = 0$. 
\fi

Since $\bn - \bzero = b/n^\alpha$, we can write
\[
\kn = \kbzero + \sum_{j=1}^{p-1} {K^{(j)}(\bzero)(\bn - \bzero)^j}/{j!} + {\ell (\bn - \bzero)^p}/{p!}.
\]
Thus $(\bn,\kn)$ converges to $(\bzero,\kbzero)$ along the curve $(\beta,\tilde{K}(\beta))$, where for
$0 < \beta < \bc$
\be
\label{eqn:tildekbeta}
\tilde{K}(\beta) = \kbzero + \sum_{j=1}^{p-1} {K^{(j)}(\bzero)(\beta - \bzero)^j}/{j!} + 
{\ell (\beta - \bzero)^p}/{p!}.
\ee
This curve coincides with the second-order curve to order $p-1$ in powers of $\beta - \bzero$
in a neighborhood of $(\bzero,\kbzero)$ (see path 2 in Figure 2).  Thus the two curves have the same tangent
at $(\bzero,K(\bzero))$.

The relationship of the sequence $(\bn,\kn)$ 
to the second-order curve depends on the sign of $b$. 
We first assume that $b = 1$.
For all sufficiently large $n$, $\ell > K^{(p)}(\bc)$ corresponds to $(\bn,\kn)$ lying
in the phase-coexistence region located above the second-order curve 
and thus to the free energy functional $\gbnkn$ having its global minimum points at $\pm \mbnkn \not = 0$.
On the other hand, for all sufficiently large $n$,
$\ell < K^{(p)}(\bc)$ corresponds to $(\bn,\kn)$ 
lying in the single-phase region located under the second-order curve and thus to $\gbnkn$
having a unique global minimum point at 0.

When $b = -1$, we must take into account the parity of $p$.
If $p$ is even, then the situation is as in the last paragraph.
If $p$ is odd, then the situation is reversed.  
As we will see, in all cases the structure of the set of global minimum points
of the associated Ginzburg-Landau polynomial mirrors the structure
of the set of global minimum points of $\gbnkn$. 

In order to calculate the asymptotic behavior of $m(\bn,\kn)$ when 
$(\bn,\kn)$ is the sequence in (\ref{eqn:bnknb2}), we follow the pattern of proof of Theorem
\ref{thm:asymptoticsforb}.  Let $\alpha > 0$ be given. The first step is 
to calculate the appropriate expansion of $n\gbnkn(x/n^\gamma)$ in (\ref{eqn:herehere}). 
In order to ease the notation,
we indicate the second-order point by $(\beta,K(\beta))$ instead of by $(\bzero,\kbzero)$.
Since
\[
K(\bn) = K(\beta + b/n^\alpha) = 
K(\beta) + \sum_{j=1}^p {K^{(j)}(\beta)b^j}/({j! n^{j\alpha}})
+ \mbox{O}(1/n^{(p+1)\alpha)}),
\] 
we have 
\be
\label{eqn:kbnknsim}
K(\beta_n) - K_n = 
{(K^{(p)}(\beta) - \ell)b^p}/({p!n^{p\alpha}}) + \mbox{O}(1/n^{(p+1)\alpha)}).
\ee
Substituting this expression into (\ref{eqn:herehere}), we see
that for all $n \in \N$, any $\gamma > 0$, 
any $R > 0$, and all $x \in \R$ satisfying $|x/n^\gamma| < R$
\bea
\label{eqn:gbnkn32}
\lefteqn{ n\gbnkn(x/n^\gamma)} \\
\nonumber & & = 
\frac{1}{n^{2\gamma + p\alpha - 1}} \frac{1}{p!} \beta(K^{(p)}(\beta) - \ell)b^p (1+\ve_n) x^2 \\
\nonumber && \hspace{.15in} \ + \frac{1}{n^{4\gamma - 1}} c_4(\beta) (1+\ve_n)x^4 +
\mbox{O}\!\left(\frac{1}{n^{2\gamma + (p+1)\alpha - 1}}\right) \! x^2
+ \mbox{O}\!\left(\frac{1}{n^{5\gamma - 1}}\right) \! x^5.
\eea
In this formula $c_4(\beta) = (e^\beta + 2)^2 (4-e^\beta)/(8 \cdot 4!)$, $\ve_n \goto 0$,
and the big-oh terms are uniform for $x \in (-Rn^\gamma,Rn^\gamma)$.
Given $u \in \R$, we multiply the numerator and denominator of the right side 
of the last display by $n^u$,
obtaining $n\gbnkn(x/n^\gamma) = \ntou G_n(x)$,
where for any $R > 0$ and all $x \in \R$ satisfying $|x/n^\gamma| < R$
\bea
\label{eqn:whattocallit}
\lefteqn{ G_n(x) } \\
\nonumber & & = 
\frac{1}{n^{2\gamma + p\alpha - 1 + u}} \frac{1}{p!} \beta(K^{(p)}(\beta) - \ell)b^p (1+\ve_n) x^2 \\
\nonumber && \hspace{.15in} + \ \frac{1}{n^{4\gamma - 1 + u}} c_4(\beta) (1+\ve_n)x^4 +
\mbox{O}\!\left(\frac{1}{n^{2\gamma + (p+1)\alpha - 1 + u}}\right) \! x^2
+ \mbox{O}\!\left(\frac{1}{n^{5\gamma - 1 + u}}\right) \! x^5.
\eea
Since $\ell \not = K^{(p)}(\beta)$ and $c_4(\beta) > 0$, the coefficients of $x^2$ and $x^4$ are both nonzero.

The next step in calculating the asymptotic behavior of $\mbnkn$ is to obtain the limit of $G_n$.
In order to carry this out, we impose the condition that the two powers of $n$
appearing in the first two terms in (\ref{eqn:whattocallit}) equal 0;
i.e., $2\gamma + p\alpha - 1 + u = 0 = 4\gamma -1 + u $.  
These two equalities are equivalent to $\gamma = p\alpha/2$
and $u = 1 - 4\gamma = 1 - 2p\alpha$.
With this choice of $\gamma$ and $u$, the powers of $n$
appearing in the last two terms in (\ref{eqn:whattocallit}) are positive, and so for all $x \in \R$ both 
terms converge to 0 as $n \goto \infty$.  It follows that 
as $n \goto \infty$, we have for all $x \in \R$ 
\[
G_n(x) = n^{1-u} \gbnkn(x/n^\gamma) \goto g(x) = \ts \frac{1}{p!}
\beta(K^{(p)}(\beta) - \ell)b^p x^2 + c_4(\beta) x^4.
\]
Since the big-oh terms in (\ref{eqn:whattocallit}) are uniform for $x \in (-Rn^\gamma,Rn^\gamma)$, the convergence 
of $G_n(x)$ to $g(x)$ is uniform for $x$ in compact subsets of $\R$. 

The structure of the set of global minimum points of the Ginzburg-Landau polynomial $g$ mirrors
precisely the structure of the set of global minimum points of $\gbnkn$ in the region 
through which $(\bn,\kn)$ passes.  The structure of the set of global minimum points of
$\gbnkn$ is noted in the two paragraphs after (\ref{eqn:tildekbeta}).  We first assume that $b = 1$. 
The choice $\ell > K^{(p)}(\beta)$ yields a polynomial $g$ for which the global minimum points are a symmetric
nonzero pair $\pm \bar{x}$, where $\bar{x}$ is defined in (\ref{eqn:barx2b}).  On the other hand, 
the choice $\ell < K^{(p)}(\beta)$ yields a polynomial $g$ having a unique global minimum point at 0.
When $b = -1$, we must take into account the parity of $p$.  If $p$ is even, then the situation
is the same as for $b = 1$. If $p$ is odd, then the situation is reversed.  The choice
$\ell < K^{(p)}(\beta)$ yields a polynomial $g$ for which the global minimum points are the symmetric
nonzero pair $\pm \bar{x}$, while the choice $\ell > K^{(p)}(\beta)$ yields a polynomial $g$ having a unique global minimum point at 0.

We are now ready to state the asymptotic behavior of $\mbnkn$ for the sequence $(\bn,\kn)$ defined
in (\ref{eqn:bnknb2}) and for the choices of $\ell$ for which the Ginzburg-Landau polynomial
has a unique positive, global minimum point $\bar{x}$. The relationships between 
$\ell$ and $K^{(p)}(\beta)$ in parts (b)(i) and (b)(ii) guarantee that for all sufficiently
large $n$, $(\bn,\kn)$
lies in the phase-coexistence region above the second-order curve.

\begin{thm}
\label{thm:asymptoticsforb2}
For $\beta \in (0,\bc)$, $\alpha > 0$, $b \in \{1,-1\}$, an integer $p \geq 2$, 
and a real number $\ell \not = K^{(p)}(\beta)$, define
\[
\bn = \beta + b/{n^\alpha} \ \mbox{ and } \
\kn = K(\beta) + \sum_{j=1}^{p-1} K^{(j)}(\beta)b^j/(j! n^{j\alpha}) +  \ell b^p/(p! n^{p\alpha})
\]
as well as $c_4(\beta) = (e^\beta + 2)^2 (4-e^\beta)/(8 \cdot 4!)$.  
Then $(\bn,\kn)$ converges to the second-order point $(\beta,K(\beta))$.
The following conclusions hold.

{\em (a)}  For any $\alpha> 0$, $u = 1 - 2p\alpha$, and $\gamma = p\alpha/2$
\[
G_n(x) = n^{1-u}\gbnkn(x/n^\gamma) \goto g(x) = 
\ts \frac{1}{p!} \beta (K^{(p)}(\beta) - \ell)b^p x^2 + c_4(\beta) x^4 
\]
uniformly for $x$ in compact subsets of $\R$.

{\em (b)} The Ginzburg-Landau polynomial $g$ has nonzero global minimum points if and
only if $(K^{(p)}(\beta) - \ell)b^p < 0$. If this inequality holds, then 
the global minimum points of $g$ are $\pm \bar{x}$, where 
\be
\label{eqn:barx2b}
\bar{x} = \left({\beta(\ell - K^{(p)}(\beta))b^p}/[2 c_4(\beta) p!]\right)^{1/2}.
\ee

{\em (c)} Assume that $(K^{(p)}(\beta) - \ell)b^p < 0$.
Then for any $\alpha > 0$, $m(\bn,\kn) \goto 0$ and has the asymptotic behavior 
\[
\mbnkn \sim {\bar{x}}/{n^{p\alpha/2}} = \barx|\beta - \bn|^{p/2}; \ \mbox{ i.e., }
\lim_{n \goto \infty} n^{p\alpha/2} \mbnkn = \bar{x}.
\]
\end{thm}

\noi
{\bf Proof.}  Part (a) follows from
the discussion leading up to the statement of the
theorem. The assertions in part (b) are elementary.  
If $(K^{(p)}(\beta) - \ell)b^p < 0$, then the 
equation $g'(x) = 2\beta (K^{(p)}(\beta) - \ell) b^p x/p!+ 4 c_4(\beta) x^3  = 0$
has the three real solutions 0 and $\pm \bar{x}$, where $\bar{x}$ 
is defined in (\ref{eqn:barx2b}).
One easily checks that $\pm \bar{x}$ are global minimum points and 0 a local maximum point.

We now verify the asymptotic behavior of $\mbnkn$ in part (c). According to Theorem \ref{thm:zngoto0}, $\mbnkn \goto 0$.  
The validity of hypotheses (i) and (ii) of Theorem \ref{thm:exactasymptotics} 
follows from the definition of the sequence $(\bn,\kn)$ and 
the inequality $(K^{(p)}(\beta) - \ell)b^p < 0$, which by (\ref{eqn:kbnknsim}) is equivalent 
to $K_n > K(\bn)$ for all sufficiently large $n$.  Thus if $(K^{(p)}(\beta) - \ell)b^p < 0$,
then for all sufficiently large $n$, $(\bn,\kn)$ lies in the phase-coexistence region above
the second-order curve.
Hypothesis (iii) of Theorem \ref{thm:exactasymptotics}
is parts (a) and (b) of the present theorem. We now verify hypothesis (iv) of Theorem \ref{thm:exactasymptotics}.
Using (\ref{eqn:whattocallit}) with $u = 1-2\alpha$ and $\gamma = p\alpha/2$, one easily
proves that for any $\alpha > 0$ there exists
$R > 0$ such that for all sufficiently large $n \in \N$ and all $x \in \R$ satisfying
$|x/n^\gamma| < R$
\[
% \label{eqn:xsquared}
G_n(x) = n^{1-u}\gbnkn(x/n^\gamma) \geq H(x) = \ts -\frac{2}{p!}\beta |K^{(p)}(\beta) - \ell| x^2 
+ \ts \frac{1}{2} c_4(\beta) x^4.
\]
Since $H(x) \goto \infty$ as $|x| \goto \infty$, hypothesis (iv) of Theorem
\ref{thm:exactasymptotics} is satisfied.  This completes the verification
of the four hypotheses of Theorem \ref{thm:exactasymptotics}. We now apply the theorem
to conclude that for any $\alpha > 0$, $\mbnkn \sim
\bar{x}/n^\gamma = \barx/n^{p \alpha/2}$. Part (c) 
of the present theorem is proved. \ \ink

\skp
In order to derive the asymptotic behavior of $\mbnkn \goto 0$ for the sequence $(\bn,\kn)$
in the last theorem, we choose $\ell \not = K^{(p)}(\beta)$.  
The choice $\ell = K^{(p)}(\beta)$ corresponds to the sequence $(\bn,\kn)$ lying 
on a curve that coincides with the second-order curve to order $p$ in powers of $(\beta - \bc)$.
In order to analyze this case, we must know the sign of $K^{(p+1)}(\beta)$.
Because we are unable to determine this sign analytically for arbitrary
$\beta \in (0,\bc)$, the discussion of this case is omitted.

We have seen several examples in which the structure of the set of the global
minimum points of the Ginzburg-Landau polynomial
mirrors the structure of the set of global minimum points of the free energy functional,
and thus the phase-transition structure,
in the region through which the associated sequence $(\bn,\kn)$ passes.
We now reverse this procedure 
by using the structure of the global minimum points of the Ginzburg-Landau polynomial $g$ 
in the last theorem not to mirror, but to explore the phase-transition
structure in the region through which the sequence
$(\bn,\kn)$ passes.  If $p = 2$ and $\ell = 0$, then the Ginzburg-Landau
polynomial takes the form
\[
g(x) = \ts \frac{1}{2}\beta K''(\beta) b^2 x^2 + c_4(\beta) x^4.
\]
According to part (c) of Lemma \ref{lem:kbeta}, $K''(\beta) > 0$.
With this choice of parameters, $g$ has a unique global minimum point
at 0, and the sequence $(\bn,\kn)$ converges to
$(\beta,K(\beta))$ along the tangent line to $(\beta,K(\beta))$.
This suggests that for each $0 < \beta < \bc$ the points on the tangent line 
sufficiently close to $(\beta,K(\beta))$ lie in the single-phase region located beneath the second-order curve. In turn, this suggests that for each $0 < \beta < \bc$ the second-order curve lies
above the tangent line at $(\beta, K(\beta))$ except at the point of tangency.
This feature of the second-order curve is equivalent to the strict convexity 
of the function $K(\beta)$ that defines this curve, a property that can be 
verified directly [Lem.\ \ref{lem:kbeta}(c)].

This completes our analysis of the asymptotic behavior of $\mbnkn \goto 0$ for the sequences considered
in Theorems \ref{thm:asymptoticsforb} and \ref{thm:asymptoticsforb2}.  In each case the asymptotic
behavior of $\mbnkn$ is expressed in terms
of the unique positive, global minimum point of the Ginzburg-Landau polynomial appearing
in the statement of the theorem.  This is a consequence of the general asymptotic
result given in Theorem \ref{thm:exactasymptotics}, which we derive in the next section.

\section{Asymptotic Behavior of \boldmath $m(\bn,\kn)$ \unboldmath in Terms of Ginzburg-Landau Polynomials}
\beginsec
\label{section:mbnkn}
\skp

Theorem \ref{thm:exactasymptotics} is the main result in this paper. 
It gives the asymptotic behavior of $m(\bn,\kn)$
for appropriate sequences $(\beta_n,K_n)$ lying in the phase-coexistence region
and converging either to a second-order point or to the tricritical point.
The asymptotic behavior is expressed in terms of the unique positive, global minimum point
of the associated Ginzburg-Landau polynomial.  We already
illustrated the use of this theorem in the previous section, where we considered 
$(\beta_n,K_n)$ converging to a second-order point along a ray [Thm.\ \ref{thm:asymptoticsforb}]
and along a curve [Thm.\ \ref{thm:asymptoticsforb2}]. 
The theorem will be applied again in the next section,
where we study the much more complicated asymptotic behavior of $m(\bn,\kn)$ in the neighborhood 
of the tricritical point.

The phase-coexistence region is defined to be all $(\beta,K)$ satisfying $0 < \beta \leq \bc$ and 
$K > K(\beta)$ and all $(\beta,K)$ satisfying $\beta > \bc$ and $K \geq K_1(\beta)$.  
Thus for $0 < \beta \leq \bc$, the phase-coexistence region consists of the region
located above the second-order curve and above the tricritical point. For $\beta > \bc$, the phase-coexistence region consists of the first-order curve $(\beta,K_1(\beta))$ and the region 
located above that curve. For all $(\beta,K)$ in the phase-coexistence region 
there exists $m(\beta,K) > 0$ such that
$\{\pm m(\beta,K)\} \subset \mmbetak$.  This is an equality for all $(\beta,K)$
in the phase-coexistence region except for $\beta > \bc$ and $K = K_1(\beta)$, in which case
$\mmbetak = \{0, \pm m(\beta,K)\}$. 

The first theorem in this section
shows that for any sequence $(\bn,\kn)$ converging either to a second-order point
or to the tricritical point, $m(\bn,\kn) \goto 0$. 
\iffalse
This is certainly plausible since $\mmbetak = \{0\}$ for all such points.
\fi
For appropriate sequences $(\bn,\kn)$ lying in the phase-coexistence
region and converging either to a second-order point or to the tricritical point,
the exact asymptotic behavior of $m(\bn,\kn) \goto 0$ is expressed in Theorem \ref{thm:exactasymptotics}. The next theorem is an 
essential component in the proof of the asymptotic behavior of $m(\bn,\kn)$ 
given in Theorem \ref{thm:exactasymptotics}.

We are able to prove an extension of Theorem \ref{thm:zngoto0} valid for first-order points.  Given $\beta > \bc$, let
$(\beta,K_1(\beta))$ be a point on the first-order curve.  If $(\bn,\kn)$ is a positive
sequence converging to $(\beta,K_1(\beta))$ from the phase-coexistence region located
above the first-order curve, then $\lim_{n \goto \infty}\mbnkn = m(\beta,K_1(\beta)) > 0$.
Because this extension of the theorem is not used in the paper, the proof is omitted.

\begin{thm}
\label{thm:zngoto0}
Let $(\bn,\kn)$ be an arbitrary positive sequence converging either to a second-order
point $(\beta,K(\beta)), 0 < \beta < \bc$, or to the tricritical point $(\beta,K(\beta)) = (\bc,\kc)$.
Then $\lim_{n \goto \infty} m(\bn,\kn) = 0$. 
\end{thm}

\noi
{\bf Proof.}  Since $\gbnkn$ is a real analytic function, $\gbnkn(\mbnkn) \leq 0$, and
$\gbnkn(x) \goto \infty$ as $|x| \goto \infty$, $\gbnkn$ has a largest positive zero, which
we denote by $x_n$. We have the inequality $0 < \mbnkn < x_n$. 
For any $t \in \R$, $c_\beta(t) \leq \log(4 e^{|t|}) = \log 4 + |t|$.
Because the sequence $(\bn,\kn)$ is bounded
and remains a positive distance from the origin and the 
coordinate axes, there exist numbers $0 < b_1 < b_2 < \infty$
such that $b_1 \leq \beta_n \leq b_2$ and $b_1 \leq K_n \leq b_2$ for all $n \in \N$.
Hence
\beas
G_{\beta_n,K_n}(x) & = & \beta_n K_n x^2 - c_{\beta_n}(2\beta_n K_n x) \\
& \geq & \beta_n K_n x^2 - 2\beta_n K_n |x| - \log 4 \: \geq \:
b_1^2 (|x|-1)^2 - b_2^2 - \log 4.
\eeas
Therefore, if $x^*$ denotes the positive zero of the quadratic
$b_1^2 (|x|-1)^2 - b_2^2 - \log 4$, then 
\[
0 < \sup_{n \in \N} \mbnkn \leq \sup_{n \in \N} x_n \leq x^*.
\] 
It follows that $\mbnkn$ is a bounded sequence.  
Thus given any subsequence $m(\beta_{n_1},K_{n_1})$, there exists
a further subsequence $m(\beta_{n_2},K_{n_2})$ and $\tilde{x} \in \R$ 
such that $m(\beta_{n_2},K_{n_2}) \goto \tilde{x}$
as $n_2 \goto \infty$.  We complete the proof by showing that 
independently of the subsequence chosen, $\tilde{x} = 0$.  To prove this, we use the fact that
\[
G_{\beta_{n_2},K_{n_2}}(m(\beta_{n_2},K_{n_2})) = \inf_{y \in \R} G_{\beta_{n_2},K_{n_2}}(y).
\]
Hence for any $y \in \R$, $G_{\beta_{n_2},K_{n_2}}(m(\beta_{n_2},K_{n_2})) \leq
G_{\beta_{n_2},K_{n_2}}(y)$.  Since $G_{\beta_{n_2},K_{n_2}}(x) \goto
G_{\beta,K(\beta)}(x)$ uniformly for $x$ in compact subsets of $\R$, it follows that for all $y \in \R$
\[ 
G_{\beta,K(\beta)}(\tilde{x}) = \lim_{n_2 \goto \infty} G_{\beta_{n_2},K_{n_2}}(m(\beta_{n_2},K_{n_2}))
\leq \lim_{n_2 \goto \infty} G_{\beta_{n_2},K_{n_2}}(y) = G_{\beta,K(\beta)}(y).
\]
Therefore $\tilde{x}$ is a minimum point of $G_{\beta,K(\beta)}$.  Because $(\beta,K(\beta))$
is either a second-order point or the tricritical point,
$\tilde{x}$ must coincide with the unique positive, global
minimum point of $G_{\beta,K(\beta)}$ at 0 [Thm.\ \ref{thm:secondorder}(a), Thm.\ \ref{thm:firstorder}(a)].  
We have proved that any subsequence 
$m(\beta_{n_1},K_{n_1})$ of $\mbnkn$ has a further subsequence 
$m(\beta_{n_2},K_{n_2})$ such that $m(\beta_{n_2},K_{n_2}) \goto 0$
as $n_2 \goto \infty$. The conclusion is that $\lim_{n \goto \infty}
\mbnkn = 0$, as claimed. \ \ink

\skp
In sections \ref{section:asymptoticsforb} and \ref{section:asymptoticsforc} we consider six different
sequences $(\bn,\kn)$ converging either to a second-order point or to the tricritical point.
The fact that each of these sequences lies
in the phase-coexistence region for all sufficiently large $n$ is the first hypothesis of Theorem \ref{thm:exactasymptotics}; this
property implies that $\mbnkn > 0$ for all sufficiently large $n$ and $\mbnkn \goto 0$ [Thm.\ \ref{thm:zngoto0}].
Under three additional hypotheses Theorem \ref{thm:exactasymptotics} describes the exact asymptotic behavior of $\mbnkn \goto 0$.   
Examples of sequences for which 
the hypotheses of the theorem are valid are given in 
Theorems \ref{thm:asymptoticsforb} and \ref{thm:asymptoticsforb2}
for sequences converging to a second-order point 
and in Theorems \ref{thm:asymptoticsforc1}--\ref{thm:asymptoticsforc4}
for sequences converging to the tricritical point.

\begin{thm}
\label{thm:exactasymptotics}
Let $(\bn,\kn)$ be a positive sequence that converges either to a second-order point
$(\beta,K(\beta))$, $0 < \beta < \bc$, or to the tricritical point $(\beta,K(\beta)) = (\bc,\kc)$. 
We assume that $(\bn,\kn)$ satisfies the following four hypotheses{\em :}
\begin{itemize}
\item[\em (i)] $(\beta_n,K_n)$ lies in the phase-coexistence region for all sufficiently large $n$. 
\item [\em (ii)] The sequence $(\bn,\kn)$ is parametrized by $\alpha > 0$. This parameter regulates the speed of approach of 
$(\bn,\kn)$ to the second-order point or the tricritical point in the following sense:
\[
b = \lim_{n \goto \infty} n^\alpha (\bn - \beta) \ \mbox{ and } \
k = \lim_{n \goto \infty} n^\alpha (\kn - K(\beta))
\]
both exist, and $b$ and $k$ are not both zero; if $b \not = 0$, then $b$ equals {\em 1} or $-${\em 1}.
\item[\em (iii)] There exists an even 
polynomial $g$ of degree $4$ or $6$ satisfying $g(x) \goto \infty$ as $|x| \goto \infty$
together with the following two properties{\em ;} $g$ is called the Ginzburg-Landau polynomial.

\begin{itemize}
  \vspace{-.1in}
  \item[\em (a)]  $\exists\alpha_0 > 0$ and $\exists\theta > 0$ such that $\forall \alpha > 0$,  
if $u = 1 - \alpha/\alpha_0$ and $\gamma = \theta \alpha$, then 
\[
\lim_{n \goto \infty} n^{1-u} \gbnkn(x/n^\gamma) = g(x)
\]
uniformly for $x$ in compact subsets of $\R$. 

\item [\em (b)] $g$ has a unique positive, global minimum point $\bar{x}$; thus the set 
of global minimum points of $g$ equals $\{\pm\barx\}$ or $\{0,\pm\barx\}$.
\end{itemize}

\item [\em (iv)] There exists a polynomial $H$ satisfying $H(x) \goto \infty$ as $|x| \goto \infty$
together with the following property{\em :} $\forall\alpha > 0$ $\exists R > 0$ such that, 
if $u = 1 - \alpha/\alpha_0$ and $\gamma = \theta\alpha$, then
$\forall n \in \N$ sufficiently large and $\forall x \in \R$
satisfying $|x/n^\gamma| < R$, $n^{1-u} \gbnkn(x/n^\gamma) \geq H(x)$.  
\end{itemize}
Under hypotheses {\em (i)}--{\em (iv)}, for any $\alpha > 0$
\[
\mbnkn \sim {\bar{x}}/{n^{\theta \alpha}}; \ \mbox{ i.e., }
\lim_{n \goto \infty} n^{\theta \alpha} \mbnkn = \bar{x}.
\]
If $b \not = 0$, then this becomes $\mbnkn \sim \barx|\beta - \bn|^\theta$.
\end{thm}  

\noi
{\bf Proof.}  Since $\gbnkn(0) = 0$ and $\gbnkn$ is even, by hypotheses (iii) $g$ is an even polynomial
of degree 4 or 6 satisfying $g(0) = 0$. Hence the global minimum points of $g$ are either
$\pm \bar{x}$ for some $\bar{x} > 0$ or 0 and $\pm \bar{x}$ for some $\bar{x} > 0$.
The proof of the asymptotic relationship
$\mbnkn \sim \bar{x}/n^{\theta \alpha}$ is much easier in the case where the global minimum points of 
$g$ are $\pm \bar{x}$ for some $\bar{x} > 0$.  After a number of preliminary steps, we will prove the theorem 
for such polynomials $g$. We will then turn to the case where the 
global minimum points of $g$ are 0 and $\pm \bar{x}$ for some $\bar{x} > 0$.

As in hypothesis (iii)(a), let $\alpha > 0$ be given and define
$u = 1 - \alpha/\alpha_0$ and $\gamma = \theta \alpha$. In order to ease the notation, we write $\mbarn = n^\gamma \mbnkn$
and $G_n(x) = n^{1-u}\gbnkn(x/n^\gamma)$. For all sufficiently large $n$, 
since $(\bn,\kn)$ lies in the phase-coexistence region, we have $\mbnkn > 0$ and 
\[
\gbnkn(\mbnkn) = \inf_{y \in \R} \gbnkn(y).
\]
It follows that for all sufficiently large $n$
\bea
\label{eqn:infgn}
\gn(\mbarn) & = & n^{1-u} \gbnkn(\mbnkn) \\
\nonumber
& = & \inf_{y \in \R} [n^{1-u} \gbnkn(y)] = \inf_{y \in \R} \gn(y);
\eea
i.e., $G_n$ attains its minimum over $\R$ at $\mbarn > 0$.  This fact will be used several times in the proof.

We first prove that the sequence $\{\mbarn, n \in \N\}$ is bounded.  
If the sequence $\mbarn$ is not 
bounded, then there exists a subsequence $\mbarnone$ of $\mbarn$ such that
$\mbarnone \goto \infty$ as $n_1 \goto \infty$. 
Let $R$ be the quantity in hypothesis (iv). Since $m(\beta_{n_1},K_{n_1}) > 0$
and $m(\beta_{n_1},K_{n_1})  \goto 0$ [Thm.\ \ref{thm:zngoto0}], we have
$0 < \mbarnone/n^\gamma = m(\beta_{n_1},K_{n_1}) < R$ for all sufficiently large $n_1$, 
and so by hypothesis (iv)
\[
G_{n_1}(\mbarnone) \geq H(\mbarnone) \goto \infty \ \mbox{ as } n_1 \goto \infty.
\]
However, this contradicts the inequality
\[
G_n(\mbarn) = \inf_{y \in \R} \gn(y) \leq G_n(0) = 0,
\]
which is valid for all $n$.  This contradiction proves that the sequence $\mbarn$ is bounded.

We now prove that $\mbnkn \sim \bar{x}/n^\gamma = \barx/n^{\theta \alpha}$ in the case where the global minimum points of
$g$  are $\pm \bar{x}$ for some $\bar{x} > 0$. 
Let $\mbarnone$ be any subsequence of $\mbarn$.  Since the sequence
$\mbarnone$ is bounded, there exists a further subsequence $\mbarntwo$ and $\hat{x} 
\geq 0$ such that $\mbarntwo \goto \hat{x}$ as $n_2 \goto \infty$. 
According to (\ref{eqn:infgn}), for any $y \in \R$, $G_{n_2}(\mbarntwo)
\leq G_{n_2}(y)$.  Since $G_n(x) \goto g(x)$ uniformly for $x$ 
in compact subsets of $\R$, it follows that
\[
g(\hat{x}) = \lim_{n_2 \goto \infty} G_{n_2}(\mbarntwo) \leq 
\lim_{n_2 \goto \infty} G_{n_2}(y) = g(y).
\]
Hence $\hat{x}$ is 
a nonnegative global minimum point of $g$. Because $g$ has a unique 
nonnegative, global minimum point $\bar{x}$, which
is positive, $\hat{x}$ coincides with $\bar{x}$. We have proved
that any subsequence $\mbarnone$ of $\mbarn$ has a further subsequence $\mbarntwo$ such that 
$\mbarntwo \goto \bar{x}$ as $n_2 \goto \infty$.  The conclusion is that 
$\lim_{n \goto \infty} \mbarn = \bar{x}$, which implies that $\mbnkn \sim \bar{x}/n^\gamma 
= \bar{x}/n^{\theta \alpha}$.

We now prove that $\mbnkn \sim \bar{x}/n^\gamma = \barx/n^{\theta \alpha}$ in the case where the global minimum points of $g$ 
are 0 and $\pm \bar{x}$ for some $\bar{x} > 0$. 
In this case $g$ is a polynomial of degree 6.
There are two subcases to consider: (1) there exists an infinite
subsequence ${n_1}$ in $\N$ such that the global minimum points of $G_{n_1}$ are $\pm \mbarnone$; 
(2) there exists an infinite
subsequence ${n_4}$ in $\N$ such that the global minimum points of $G_{n_4}$ are 0 and $\pm \mbar_{n_4}$.  Examples of sequences for which both subcases hold are given
in the second paragraph before Theorem \ref{thm:asymptoticsforc2}.

In subcase 1 we will prove that any subsequence ${n_2}$ of ${n_1}$
has a further subsequence ${n_3}$ for which $\mbar_{n_3} \goto \barx$.  This implies
that $\mbar_{n_1} \goto \barx$.  In subcase 2 a similar proof shows that 
any subsequence ${n_5}$ of ${n_4}$
has a further subsequence ${n_6}$ for which $\mbar_{n_6} \goto \barx$.  This implies
that $\mbar_{n_4} \goto \barx$.  Now let $n_7$ be an arbitrary subsequence in $\N$. 
Then $n_7$ contains either infinitely many elements of the subsequence $n_1$ or 
infinitely many elements of the subsequence $n_4$.  In either case $n_7$ contains a further
subsequence $n_8$ for which $\mbar_{n_8} \goto \barx$.  The conclusion is
that $\mbar_n \goto \barx$, which yields the desired conclusion,
namely, $\mbnkn \sim \bar{x}/n^\gamma = \bar{x}/n^{\theta \alpha}$. 

We focus on subcase 1; subcase 2 is handled similarly. 
In order to understand the subtlety of the proof, we return to the argument just given in the case
where the global minimum points of $g$  are $\pm \bar{x}$ for some $\bar{x} > 0$.
Let $n_1$ be the subsequence in subcase 1 and let $n_2$ be any further subsequence.
Since the sequence $\mbar_{n_2}$ is bounded, the same argument shows that there exists
a further subsequence $n_3$ such that $\mbar_{n_3} \goto \hat{x}$ as $n_2 \goto \infty$, where $\hat{x}$
is a nonnegative global minimum point of $g$.  When the global minimum points of $g$  are $\pm \bar{x}$ for some $\bar{x} > 0$,
we are able to conclude in fact that $\hat{x}$ equals $\bar{x}$.  However, in the present case where the 
global minimum points of $g$ are 0 and $\pm \bar{x}$ for some $\bar{x} > 0$, it might turn out 
that $\hat{x}$ equals the global minimum point of $g$ at 0. In this situation we would conclude that 
$\mbar_{n_3} \goto 0$, which is not the asymptotic relationship that we want.

As this discussion shows, in subcase 1 it suffices to prove that there exists no subsequence ${n_3}$ of ${n_2}$ for which 
$\mbarnthree \goto 0$ as $n_3 \goto \infty$.  Under the assumption that there exists such a subsequence, we will reach a contradiction of the fact that $\mbarnthree$ is the largest critical point of $G_{n_3}$.  This fact is not directly stated in \cite{EllOttTou}, but it is a straightforward consequence of Lemmas 3.9 and 3.10(a) and Theorem 3.5 in that paper.  From these three results it follows that when $(\beta, K)$ lies in the two-phase region, the positive global minimum point of the function $F_{\beta, K}(z) = G_{\beta,K}(z/2\beta K)$ is also its largest critical point.  From the definition of $G_n$ in terms of $G_{\beta_n, K_n}$, it then 
follows that $\bar{m}_n$ is the largest critical point of $G_n$ for all $n$.   

Since the global minimum points of $g$ are 0 and $\pm \bar{x}$ for some $\bar{x} > 0$,  
there exists $\bar{y} \in (0,\bar{x})$
such that $g$ attains its maximum on the interval $[0,\bar{x}]$ at $\bar{y}$ and attains
its maximum on the interval $[-\bar{x},0]$
at $-\bar{y}$.  In addition, $g(\pm \bar{y}) > 0 = g(0) =
g(\pm \bar{x})$.  The graph of $g$ is shown in graph (a) in Figure 7.  The graph of $G_{n_3}$ under the assumption that $\mbarnthree \goto 0$ is shown in graph (b) in Figure 7.  Referring to these graphs should help the reader follow the proof.

\begin{figure}[h]
\begin{center}
\includegraphics[width=12cm]{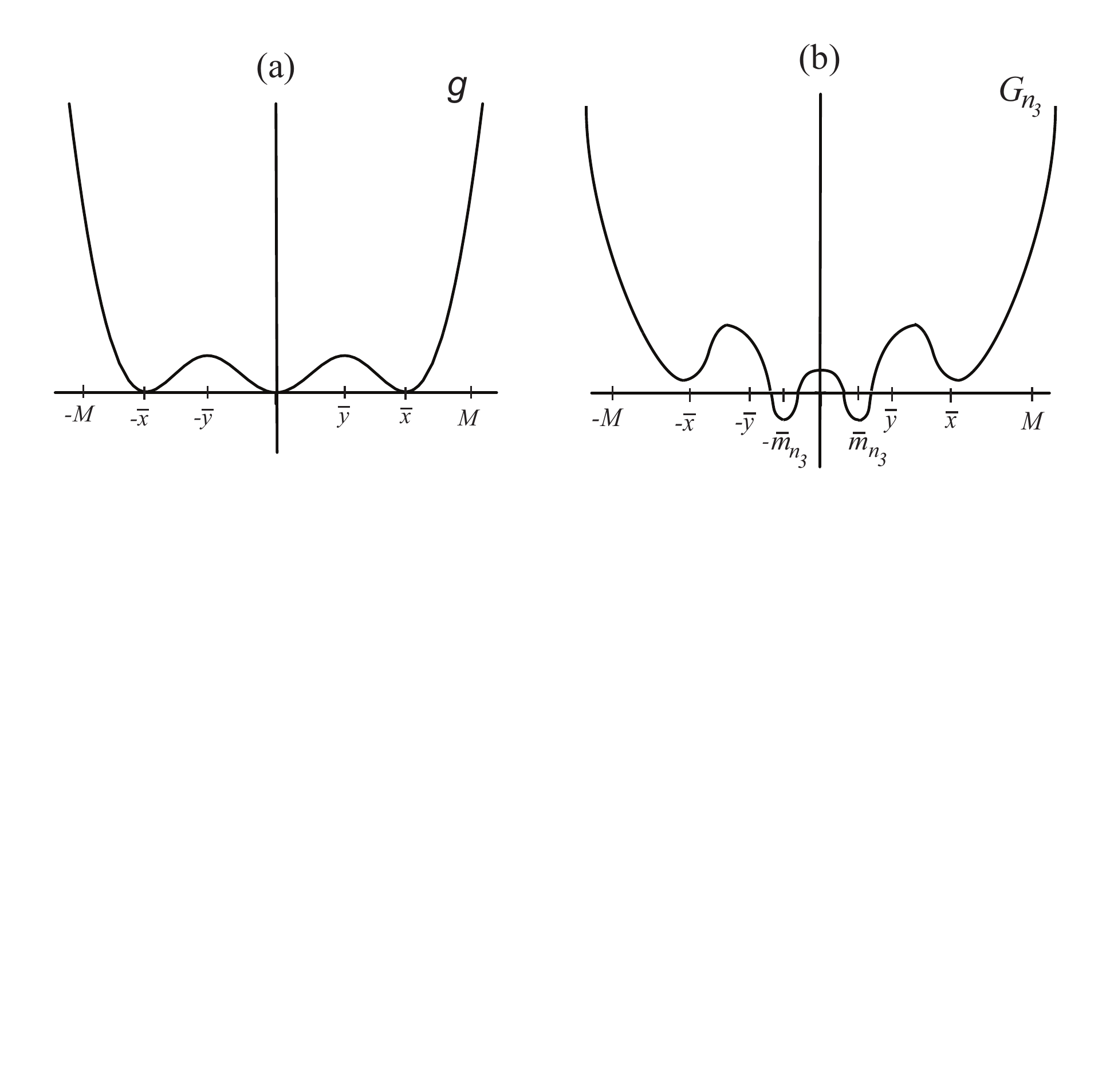}
\vspace{-2.5in}
\caption{\footnotesize Proof of Theorem \ref{thm:exactasymptotics} in subcase 1. (a) Graph of Ginzburg-Landau
polynomial $g$ having three global minimum points, (b) graph of $G_{n_3}$ showing $\bar{m}_{n_3} \goto 0$.}
\end{center}
\end{figure} 

By hypothesis (iii)(a), as $n_3 \goto \infty$, $\gnthree(z) \goto g(z)$ 
uniformly on compact subsets of $\R$. Thus for all sufficiently large $n_3$ and each choice of sign
\be
\label{eqn:gbary}
\gnthree(\pm \bar{y}) \geq 2 g(\bar{y})/3 > 0, \
\gnthree(\pm \bar{x}) \leq g(\bar{y})/3.
\ee
By definition of subcase 1 the global minimum points of $\gnthree$ are $\pm \mbarnthree$, and 
by assumption $\mbarnthree \goto 0$ as $n_3 \goto \infty$.  For all sufficiently large $n_3$, the inequality $G_{n_3}(\mbarnthree) < G_{n_3}(0) = 0$ and 
the two inequalities in (\ref{eqn:gbary}) imply
that there exists $\bar{y}_{n_3} \in (\mbarnthree,\bar{x})$ such 
that $\gnthree$ attains its maximum on the interval $[\mbarnthree,\bar{x}]$ at $\bar{y}_{n_3}$ 
and attains its maximum on the interval $[-\bar{x}, -\mbarnthree]$ at $-\bar{y}_{n_3}$.  Therefore, $\bar{y}_{n_3}$ is a critical point of 
$G_{n_3}$ greater than $\mbarnthree$. This contradicts the fact that $\mbarnthree$ is the largest critical point of $G_{n_3}$.
The proof of subcase 1 is complete.

Subcase 2 is handled similarly. Let $n_4$ be the subsequence in subcase 2 and let
$n_5$ be any further subsequence.
If there exists a subsequence $n_6$ of $n_5$ for which 
$\mbar_{n_6} \goto 0$ as $n_6 \goto \infty$, then for all sufficiently large $n_6$, 
there exists $\bar{y}_{n_6} \in (\mbarnsix,\bar{x})$ such 
that $G_{n_6}$ attains its maximum on the interval $[\mbarnsix,\bar{x}]$ at $\bar{y}_{n_6}$ 
and attains its maximum on the interval $[-\bar{x}, -\mbarnsix]$ at $-\bar{y}_{n_6}$.  
Again this contradicts the fact that $\mbarnsix$ is the largest critical point of $G_{n_6}$.
This completes the proof of the theorem. \ \ink

\skp
In the next section we use Theorem \ref{thm:exactasymptotics}
to derive the asymptotic behavior of $m(\bn,\kn)$ for
appropriate sequences $(\beta_n,K_n)$ converging to the tricritical point
$(\bc,K(\bc))$ from various subsets of the phase-coexistence region.
A number of new phenomena arise in this case that are not observed in 
the cases studied in section \ref{section:asymptoticsforb}.

\section{Asymptotic Behavior of \boldmath $\mbnkn$ \unboldmath Near the Tricritical Point}
\beginsec
\label{section:asymptoticsforc}

In this section we derive the asymptotic behavior of 
the magnetization $m(\bn,\kn)$ for appropriate
sequences $(\beta_n,K_n)$ converging to the tricritical point
$(\bc,K(\bc))$
from various subsets of the phase-coexistence region.
The situation is much more complicated than in section \ref{section:asymptoticsforb},
in which we studied the asymptotic behavior of 
$m(\bn,\kn)$ for two different
sequences $(\beta_n,K_n)$ converging to points $(\beta,K(\beta))$ on the second-order curve from the phase-coexistence
region located above that curve.  For each of these sequences there is a different
asymptotic behavior.  

By contrast, in the present section there are four distinct asymptotic behaviors of $\mbnkn$
corresponding to four different choices of the sequences $(\bn,\kn)$.  
These are treated in Theorems \ref{thm:asymptoticsforc1}--\ref{thm:asymptoticsforc4}.
In the first three cases the limiting Ginzburg-Landau polynomial has degree 6, and in the fourth case
it has degree 4. The most interesting example is treated in Theorem \ref{thm:asymptoticsforc2}. 
The sequence ($\bn,\kn)$ in that theorem converges to the tricritical point 
for $\bn > \bc$ along a curve that is tangent to the spinodal
curve at the tricritical point and depends on a curvature parameter.
For those sequences that lie in the phase-coexistence region,
Theorem \ref{thm:asymptoticsforc2} shows that $\mbnkn \sim \barx(\bn - \bc)^{1/2}$.  

As in section \ref{section:asymptoticsforb}, properties of the
Ginzburg-Landau polynomials in these four cases reflect the phase-transition
structure of the mean-field B-C model in the region through which the associated sequence
$(\bn,\kn)$ passes. This again makes rigorous the predictions of the Ginzburg-Landau
phenomenology of critical phenomena mentioned in section \ref{section:phasetr}.  

Let $(\bn,\kn)$ be an arbitrary positive sequence converging to the tricritical
point $(\bc,\kbc) = (\log 4, 3/2 \log 4)$ and let $\gamma > 0$ be given.
In section \ref{section:phasetr} we motivated the phase-transition
structure for $\beta > \bc$ by approximating $\gbk(x)$ in (\ref{eqn:gbk6})
by a polynomial of degree 6 derived from the first three terms in its Taylor expansion.  
The starting point in determining the asymptotic behavior of $\mbnkn$ 
is to replace this three-term Taylor expansion for $\gbk(x)$ by
the three-term Taylor expansion for $nG_{\bn,\kn}(x/n^\gamma)$ with an error term.
According to Taylor's Theorem, for any $R > 0$ and all $x \in \R$
satisfying $|x/n^\gamma| < R$ there exists $\xi_n(x/n^\gamma) \in [-x/n^\gamma,x/n^\gamma]$ such that
\bea
\label{eqn:TaylorG6}
\lefteqn{
nG_{\beta_n, K_n}(x/n^\gamma)} \\ \nonumber
&& = \frac{1}{n^{2\gamma-1}}\frac{G_{\beta_n, K_n}^{(2)}(0)}{2!} x^2 + 
\frac{1}{n^{4\gamma-1}}\frac{G_{\beta_n, K_n}^{(4)}(0)}{4!} x^4 \\ \nonumber 
&& \hspace{.2in} + \frac{1}{n^{6\gamma-1}}\frac{G_{\beta_n, K_n}^{(6)}(0)}{6!} x^6 + 
\frac{1}{n^{7\gamma - 1}} \frac{\gbnkn^{(7)}(\xi_n(x/n^\gamma))}{7!} x^7.
\eea
In deriving this formula, we use the fact that $\gbnkn(0) = 0$
and that since $\gbnkn$ is an even function, $\gbnkn^{(1)}(0) = 0 = \gbnkn^{(3)}(0)
= \gbnkn^{(5)}(0)$. Because the sequence
$(\bn,\kn)$ is positive and bounded, there exists $a \in (0,\infty)$ such that 
$0 < \bn \leq a$ and $0 < \kn \leq a$ for all $n$. As a continuous function 
of $(\beta,K,y)$ on the compact set $[0,a] \times [0,a] \times [-R,R]$, $G_{\beta,K}^{(7)}(y)$
is uniformly bounded.  It follows that the quantity $\gbnkn^{(7)}(\xi_n(x/n^\gamma))$
appearing in the error term in the Taylor expansion is 
uniformly bounded for $x \in (-\rng,\rng)$.  We summarize this expansion by writing
\bea
\label{eqn:TaylorG6again}
nG_{\beta_n, K_n}(x/n^\gamma)& = &
\frac{1}{n^{2\gamma-1}}\frac{G_{\beta_n, K_n}^{(2)}(0)}{2!} x^2 + 
\frac{1}{n^{4\gamma-1}}\frac{G_{\beta_n, K_n}^{(4)}(0)}{4!} x^4 \\
\nonumber && + \ \frac{1}{n^{6\gamma-1}}\frac{G_{\beta_n, K_n}^{(6)}(0)}{6!} x^6
+ \mbox{O}\!\left(\frac{1}{n^{7\gamma-1}}\right) \!x^7,
\eea
where the big-oh term is uniform for $x \in (-Rn^\gamma,Rn^\gamma)$.

In terms of the quantity $K(\beta) = (e^\beta + 2)/4\beta$,
the coefficients $G_{\beta_n, K_n}^{(2)}(0)$ and $G_{\beta_n, K_n}^{(4)}(0)$ in
the Taylor expansion are given by
\[
G_{\beta_n, K_n}^{(2)}(0) = 
\frac{2 \beta_n K_n (K(\beta_n) - K_n)}{K(\beta_n)}
= 2\beta (K(\bn) - \kn) \cdot \frac{\bn\kn}{\beta K(\bn)}
\]
and 
\[
G_{\beta_n, K_n}^{(4)}(0) = \frac{2 (2\beta_n K_n)^4 (4-e^{\beta_n})}{(e^{\beta_n} + 2)^2}.
\]
In order to ease the notation, we let $\ve_n$ denote a sequence that converges to 0
and that represents the various error terms arising in the following calculation; 
we use the same notation $\ve_n$ to represent different error terms.
Since $(\bn,\kn)$ converges to $(\bc,K(\bc))$ and the function $K(\cdot)$ is continuous, we have 
$\beta_n K_n/K(\bn) \goto \bc$.  Thus
\[
{G_{\beta_n, K_n}^{(2)}(0)}/{2!} = \bc (K(\beta_n) - K_n) (1+\ve_n). 
\]
Let $c_4 = 3/16$.  
Since $2\bn\kn \goto 2\bc K(\bc) = (e^{\bc} + 2)/2 = 3$ and $e^{\beta_n} + 2 \goto e^{\bc} + 2 = 6$, we also have
\be
\label{eqn:c4betanew}
{G_{\beta_n, K_n}^{(4)}(0)}/{4!} = {2 \cdot 3^4} (4-e^{\bn})(1+\ve_n)/{6^2 \cdot 4!} = c_4 (4-e^{\bn})(1+\ve_n).
\ee
In section \ref{section:asymptoticsforb} we have $\bn \goto \beta \in (0,\bc)$ and thus
$4 - e^{\bn} = (4 - e^\beta)(1+\ve_n) > 0$. In the present section, however, $4 - e^{\bn} \goto 0$
as $\bn \goto \bc$, and so we must keep this term in the last display.
Finally, let $c_6 = 9/40$.  Since $G_{\beta_n, K_n}^{(6)}(0) \goto G_{\bc, \kbc}^{(6)}(0) = 2\cdot3^4$,
we have
\[
{G_{\beta_n, K_n}^{(6)}(0)}/{6!} = {2 \cdot 3^4}(1+\ve_n)/{6!} = c_6 (1+\ve_n).
\]
Substituting these expressions into the Taylor expansion (\ref{eqn:TaylorG6again}), we obtain
for all $n \in \N$, any $\gamma > 0$, any $R > 0$, and all $x \in \R$ satisfying $|x/n^\gamma| < R$
\bea
\label{eqn:Taylorforc}
\lefteqn{
nG_{\beta_n, K_n}(x/n^\gamma) } \\ \nonumber
&& = \frac{1}{n^{2\gamma-1}} \bc (K(\beta_n) - K_n) (1+\ve_n) x^2
+ \frac{1}{n^{4\gamma-1}} c_4 (4 - e^{\bn})(1+\ve_n) x^4 \\
\nonumber && + \ \frac{1}{n^{6\gamma -1}} c_6 (1+\ve_n) x^6 + \mbox{O}\!\left(\frac{1}{n^{7\gamma-1}}\right)\!x^7,
\eea
where $c_4 = 3/16$ and $c_6 = 9/40$.

For the moment, in the polynomial on the right side of the last display,
let us replace $(\bn,\kn)$ by $(\beta,K)$ and set $n = 1$. Doing so,
we obtain the polynomial $\tildegbk$ that approximates the free energy
functional $\gbk$ in (\ref{eqn:gbk6}) for $\beta > \bc$, $(\beta,K)$
near the tricritical point, and $x$ near 0. Arising via the Ginzburg-Landau phenomenology, this 
polynomial is used in section \ref{section:phasetr} to motivate
the discontinuous bifurcation in the set of equilibrium values of the magnetization
that is described rigorously in Theorem \ref{thm:firstorder}. 
As we will soon see, by suitable choices of $(\bn,\kn)$ and other parameters the
polynomial on the right side of the last display converges to a Ginzburg-Landau polynomial
in terms of which the convergence of $\mbnkn$ to 0 is described.

We return to (\ref{eqn:Taylorforc}), in which the terms 
$K(\bn) - \kn$ and $4 - e^{\bn}$ both converge to 0 as $n \goto \infty$. This formula
is the seed from which will blossom the various asymptotic behaviors of $\mbnkn$,
each depending on the choice of the sequence $(\bn,\kn)$ converging to the tricritical point. 
Each choice controls, in a different way, the rate at
which $K(\beta_n) - K_n \goto 0$ and $e^{\bn} - 4 \goto 0$.
We analyze four separate cases, each giving rise to a Ginzburg-Landau polynomial
having a unique positive, global minimum points at $\bar{x}$ for some $\bar{x} > 0$. This quantity enters
the respective asymptotic formula for $\mbnkn \goto 0$. 

For the first choice we take $\alpha > 0$, $b \in \{1,0,-1\}$, and $k \in \R$ and define
\be
\label{eqn:newbetanknbc1}
\beta_n = \bc + {b}/{n^\alpha} \ \mbox{ and } \ K_n = \kbc + {k}/{n^\alpha}.
\ee
If $b \not = 0$, then $(\bn,\kn)$ converges to the tricritical point along a ray with slope $k/b$
(see path 3 in Figure 2).  We assume that $K'(\bc)b - k \not = 0$.  
Since 
\[
K(\bn) = K(\bc + b/n^\alpha) = K(\bc) + K'(\bc) b/n^\alpha + \mbox{O}(1/n^{2\alpha}),
\]
we have
\be 
\label{eqn:kbetankn}
K(\beta_n) - K_n = (K'(\bc) b - k)/{n^\alpha} + \mbox{O}(1/n^{2\alpha})
\ee
and
\be 
\label{eqn:ebn}
4 - e^{\bn} = e^{\bc}(1 - e^{b/n^\alpha}) = -4b/n^\alpha + \mbox{O}(1/n^{2\alpha}).
\ee

The case where $K'(\bc)b - k = 0$ must be handled differently.
If this equality holds, then the expression for $K(\bn) - \kn$
given here is indeterminate.  In order to calculate the correct behavior
of $K(\bn) - \kn$ when $K'(\bc)b - k = 0$, one must consider the next term in
the Taylor expansion of $K(\bc + b/n^\alpha)$, obtaining (\ref{eqn:K''ell}) with $\ell = 0 = \tell$. 
We carry out the analysis for this case after Theorem \ref{thm:asymptoticsforc1}.

We return to the sequence $(\bn,\kn)$ in (\ref{eqn:newbetanknbc1}) when $K'(\bc)b - k \not = 0$.
Substituting (\ref{eqn:kbetankn}) and (\ref{eqn:ebn})
into (\ref{eqn:Taylorforc}), we obtain for all $n \in \N$, any $\gamma > 0$, 
any $R > 0$, and all $x \in \R$ satisfying $|x/n^\gamma| < R$
\bea
\label{eqn:return}
\lefteqn{
nG_{\beta_n, K_n}(x/n^\gamma)} \\ \nonumber
&&  = 
\frac{1}{n^{2\gamma + \alpha - 1}}  \bc(K'(\bc) b - k)(1+\ve_n) x^2
- \frac{1}{n^{4\gamma + \alpha - 1}} 4  c_4 b (1+\ve_n) x^4 \\
\nonumber && \hspace{.2in} + \ \frac{1}{n^{6\gamma -1}} c_6 (1+\ve_n) x^6
+ \mbox{O}\!\left(\frac{1}{n^{2\gamma + 2\alpha - 1}}\right)\!x^2 \\
\nonumber && \hspace{.2in} + \ \mbox{O}\!\left(\frac{1}{n^{4\gamma + 2\alpha - 1}}\right)\!x^4
+ \mbox{O}\!\left(\frac{1}{n^{7\gamma - 1}}\right)\!x^7.
\eea
Given $u \in \R$, we multiply the numerator and denominator of the right side of
the last display by $n^u$, obtaining $nG_{\beta_n, K_n}(x/n^\gamma) = n^u G_n(x)$,
where for any $R > 0$ and all $x \in \R$ satisfying $|x/n^\gamma| < R$
\bea
\label{eqn:c1}
\lefteqn{
G_n(x)} \\ \nonumber
&&  = 
\frac{1}{n^{2\gamma + \alpha - 1 + u}}  \bc(K'(\bc) b - k)(1+\ve_n) x^2
- \frac{1}{n^{4\gamma + \alpha - 1 + u}} 4  c_4 b (1+\ve_n) x^4 \\
\nonumber && \hspace{.2in} + \ \frac{1}{n^{6\gamma -1 + u}} c_6 (1+\ve_n) x^6
+ \mbox{O}\!\left(\frac{1}{n^{2\gamma + 2\alpha - 1 + u}}\right)\!x^2 \\
\nonumber && \hspace{.2in} + \ \mbox{O}\!\left(\frac{1}{n^{4\gamma + 2\alpha - 1 + u}}\right)\!x^4
+ \mbox{O}\!\left(\frac{1}{n^{7\gamma - 1 + u}}\right)\!x^7.
\eea
In this formula $\ve_n \goto 0$ and the big-oh terms are uniform for $x \in (-Rn^\gamma,Rn^\gamma)$.
Since $K'(\bc)b - k \not = 0$ and $c_6 > 0$, the coefficients of $x^2$, $x^4$, and $x^6$ in the first three terms are all nonzero.

In order to obtain the limit of $G_n$, we impose the condition that the powers of $n$
appearing in the first and third terms in the last display equal 0;
i.e., $2\gamma + \alpha - 1 + u = 0 = 6\gamma -1 + u$. 
These two equalities are equivalent to $\gamma = \alpha/4$ and $u = 1 - 6\gamma = 1 - 3\alpha/2$.
With this choice of $\gamma$ and $u$, the powers of $n$ in the second
term and the last three terms in (\ref{eqn:c1}) are positive, and so for all $x \in \R$ 
these four terms converge to 0 as $n \goto \infty$.
It follows that as $n \goto \infty$, we have for all $x \in \R$
\[
G_n(x) = n^{1-u} \gbnkn(x/n^\gamma) \goto g(x) = \bc(K'(\beta) b - k) x^2 + c_6 x^6.
\]
Since the big-oh terms in (\ref{eqn:c1}) are uniform for $x \in (-Rn^\gamma,Rn^\gamma)$, 
the convergence of $G_n(x)$ to $g(x)$ is uniform for $x$ in compact subsets of $\R$.

In the next theorem we derive 
the asymptotic behavior of $m(\bn,\kn)$ for the sequence
$(\bn,\kn)$ defined in (\ref{eqn:newbetanknbc1}) with $K'(\bc) b - k < 0$.
This inequality is equivalent to $(\bn,\kn)$ lying in the phase-coexistence region 
for all sufficiently large $n$. Derived from the general asymptotic
result in Theorem \ref{thm:exactasymptotics}, part (c) of the next theorem expresses the
asymptotic behavior of $\mbnkn \goto 0$ in terms of 
the unique positive, global minimum point $\bar{x}$ of the associated Ginzburg-Landau polynomial $g$.

\begin{thm}
\label{thm:asymptoticsforc1}
For $\alpha > 0$, $b \in \{1,0,-1\}$, and a real number $k \not = K'(\bc) b$, 
define
\[
\beta_n = \bc + {b}/{n^\alpha} \ \mbox{ and } \ K_n = K(\bc) + {k}/{n^\alpha}
\]
as well as $c_6 = 9/40$.  Then $(\bn,\kn)$ converges to the tricritical
point $(\bc,\kc)$. The following conclusions hold.

{\em (a)} For any $\alpha > 0$, $u = 1 - 3\alpha/2$, and $\gamma = \alpha/4$
\[
G_n(x) = n^{1-u} \gbnkn(x/n^\gamma) \goto g(x) = \bc(K'(\bc) b - k) x^2 + c_6 x^6
\]
uniformly for $x$ in compact subsets of $\R$. 

{\em (b)} The Ginzburg-Landau polynomial $g$ has nonzero global minimum points
if and only if $K'(\bc) b - k < 0$.  If this inequality holds, then the global minimum points of $g$ are $\pm \bar{x}$, where
\be
\label{eqn:barxforc1}
\bar{x} = \left({\bc (k - K'(\bc) b)}/[3 c_6]\right)^{1/4}
\ee

{\em (c)} Assume that $K'(\bc) b - k < 0$.  Then for any $\alpha > 0$, $m(\bn,\kn) \goto 0$ and has the asymptotic behavior 
\[
\mbnkn \sim {\bar{x}}/{n^{\alpha/4}}; \ \mbox{ i.e., }
\lim_{n \goto \infty} n^{\alpha/4} \mbnkn = \bar{x}.
\]
When $b \not = 0$, this becomes $\mbnkn \sim \barx|\bc - \bn|^{1/4}$.
\end{thm}

\noi
{\bf Proof.} Part (a) follows from
the discussion leading up to the statement of the
theorem. The first assertion in part (b) is elementary.  
If $K'(\bc) b - k < 0$, then the equation $g'(x) = 6 c_6 x^5 + 2 \bc(K'(\bc) b - k) x = 0$ has solutions
at $\pm \bar{x}$ and at 0, where $\bar{x}$ is defined in (\ref{eqn:barxforc1}).   One easily 
checks that $\pm \bar{x}$ are global minimum points and 0 a local maximum point.  

We now verify the asymptotic behavior of $\mbnkn$ in part (c).
According to Theorem \ref{thm:zngoto0}, $\mbnkn \goto 0$.  
The validity of hypotheses (i) and (ii) of Theorem \ref{thm:exactasymptotics} 
follows from the definition of the sequences $(\bn,\kn)$ and the inequality $K'(\bc) b - k < 0$, 
which by (\ref{eqn:kbetankn}) is equivalent to $K_n > K(\bn)$ for all sufficiently large $n$.
Thus if $K'(\bc) b - k < 0$, then for all sufficiently large $n$, $(\bn,\kn)$ lies in the
phase-coexistence region; $(\bn,\kn)$
is above the spinodal curve if $b = 1$, 
above the second-order curve if $b = -1$, and
above the tricritical point if $b = 0$.
Hypothesis (iii) of Theorem \ref{thm:exactasymptotics} is parts (a) and (b) of the present theorem. 
We now verify hypothesis (iv) of Theorem \ref{thm:exactasymptotics}.
Using (\ref{eqn:c1}) with $\gamma = \alpha/4$ and $u = 1-3\alpha/2$, 
one easily proves that for any $\alpha > 0$ there exists
$R > 0$ such that for all sufficiently large $n \in \N$ and all $x \in \R$ satisfying
$|x/n^\gamma| < R$
\[
G_n(x) = n^{1-u}\gbnkn(x/n^\gamma) \geq H(x) = 
-2\bc|K'(\bc) b - k| x^2 - 8c_4 x^4 + \ts \frac{1}{2} c_6 x^6.
\]
Since $H(x) \goto \infty$ as $|x| \goto \infty$, hypothesis (iv) of Theorem
\ref{thm:exactasymptotics} is satisfied.  The new element is that the term
$-8c_4|b| x^4$ in $H(x)$ must be included in order to bound below the two $x^4$-terms in (\ref{eqn:return})
for all $n$.  This completes the verification of the four hypotheses of Theorem \ref{thm:exactasymptotics}.
We now apply the theorem to conclude that for any $\alpha > 0$, $\mbnkn \sim
\bar{x}/n^\gamma = \barx/n^{\alpha/4}$. Part (c) of the present theorem is proved. \ \ink
 
\skp

We now consider the second choice of sequence $(\bn,\kn)$ converging to 
the tricritical point, which gives a different 
asymptotic behavior of $\mbnkn \goto 0$. Given $\alpha > 0$, $\ell \in \R$, and $\tell \in \R$ we define
\be
\label{eqn:bnkn2a}
\bn = \bc + {1}/{n^\alpha} \ \mbox{ and } \ 
\kn = \kbc + {\kprimebc}/{n^\alpha} + {\ell}/({2n^{2\alpha}}) + {\tell}/({6n^{3\alpha}}).
\ee
The term involving $\tell$ is needed only in the case where $\ell = \ell_c$ 
in order to assure that $(\bn,\kn)$ lies in the phase-coexistence region for all sufficiently large $n$;
see item (ii) after (\ref{eqn:defineellc}). In all other cases the inclusion of the term involving
$\tell$ adds no new features, and we can take $\tell = 0$.

The choice $\ell = \tell = 0$ in (\ref{eqn:bnkn2a}) reduces to the sequence $(\bn,\kn)$ in (\ref{eqn:newbetanknbc1})
when $K'(\bc)b - k = 0$ in that formula. However, as pointed out in the paragraph
after (\ref{eqn:newbetanknbc1}), the analysis given there is valid 
only when $K'(\bc)b - k \not = 0$.

We now consider the behavior of $(\bn,\kn)$ for various choices of $\ell$ and $\tell$.
Since $\bn - \bc = 1/n^\alpha$, we can write 
\[
K_n = \kbc + \kprimebc (\bn - \bc) + \ell (\bn - \bc)^2/2 + \tell (\bn - \bc)^3/6.
\] 
Thus $(\bn,\kn)$ converges to the tricritical point along the curve $(\beta,\tilde{K}(\beta))$, where
for $\beta > \bc$
\[
\tilde{K}(\beta) = \kbc + \kprimebc (\beta - \bc) + \ell (\beta - \bc)^2/2 + \tell (\beta - \bc)^3/6.
\]
This curve is tangent to the spinodal curve at the
tricritical point and satisfies $\tilde{K}''(\bc) = \ell$.  
Thus $\ell > K''(\bc)$ and any $\tell \in \R$ correspond to the sequence $(\bn,\kn)$
converging to the tricritical point from the phase-coexistence region
located above the spinodal curve (see path 4a in Figure 2).
The value $\ell = K''(\bc)$ corresponds to the sequence $(\bn,\kn)$
converging to $(\bc,\kc)$ 
along a curve that coincides with the spinodal curve to order 2
in powers of $\beta - \bc$ in a neighborhood of the tricritical
point.  When $\ell = K''(\bc)$ and $\tell > K'''(\bc)$, 
$(\bn,\kn)$ lies in the phase-coexistence region above the spinodal curve
for all sufficiently large $n$ (see path 4b in Figure 2). Since $K'''(\bc) < 0$ [Lem.\ \ref{lem:kbeta}(d)],
we can take $\tell = 0$. 

The situation for $\ell < K''(\bc)$ is considerably more complicated.  
The discussion is based on three conjectured properties of the function
$K_1(\beta)$, which for $\beta > \bc$ defines the first-order curve.
Since $\lim_{\beta \goto \bc^+} K_1(\beta) = K(\bc)$, by continuity
we extend the definition of $K_1(\beta)$ to $\beta = \bc$ by defining
$K_1(\bc) = K(\bc)$.  In the next paragraph, we assume that the first three right-hand
derivatives of $K_1(\beta)$ exist at $\bc$ and denote them by
$K_1'(\bc)$, $K_1''(\bc)$, and $K_1'''(\bc)$.

We define 
\be 
\label{eqn:defineellc}
\ell_c = -{1}/({4\bc}) - {2}/{\bc^2} + {3}/{\bc^3}.
\ee
In section \ref{section:firstordercurve} we use properties of the appropriate
Ginzburg-Landau polynomials plus numerical evidence to support Conjectures 1, 2, and 3, 
which state the following: (1) $K_1'(\bc) = K'(\bc)$ --- i.e., at
$\bc$ the first-order curve and the spinodal curve have the same right-hand tangent; 
(2) $K_1''(\bc) = \ell_c < 0 < K''(\bc)$; (3) $K_1'''(\bc) > 0$.  If these conjectures
are true, then we have the following picture:
\begin{enumerate}
\item  Assume that $\ell$ satisfies $K''(\bc) > \ell > \ell_c$ and take any $\tell \in \R$.
Then by Conjectures 1 and 2, $(\bn,\kn)$ converges to the tricritical point
from the phase-coexistence region 
along a curve that passes above the first-order curve and below the spinodal curve
in a neighborhood of the tricritical point (see path 4d in Figure 2). 
\item Assume that $\ell = \ell_c$ and take any $\tell > K_1'''(\bc)$.
Then by Conjectures 1 and 2 $(\bn,\kn)$ converges to the tricritical point
along a curve that coincides with the first-order curve to order 2 in powers of $\beta - \bc$
in a neighborhood of the tricritical point (see path 4c in Figure 2).  By Conjecture 3, 
for all sufficiently large $n$, $(\bn,\kn)$ lies in the phase-coexistence region
above the first-order curve and below the spinodal curve.  If we did not include the term
involving $\tell$ in (\ref{eqn:bnkn2a}), then by Conjecture 3 the sequence $(\bn,\kn)$
would lie in the single-phase region below the first-order curve for all sufficiently
large $n$.
\item Assume that $\ell < \ell_c$ and take any $\tell \in \R$.
Then by Conjectures 1 and 2 $(\bn,\kn)$ converges to the tricritical point from the single-phase
region located under the first-order curve.
\end{enumerate}

The structure of the set of global minimum points of the associated Ginzburg-Landau
polynomials $g$ mirrors the phase-transition structure of the region through
which the corresponding sequence $(\bn,\kn)$ passes.  As we will see in items
1--3 before the statement of Theorem \ref{thm:asymptoticsforc2}, the
following three cases arise:
(1) for $\ell > \ell_c$, the global minimum points of $g$ are $\pm \bar{x}(\ell)$, where
$\bar{x}(\ell) > 0$ is defined in (\ref{eqn:barxforc2a}); (2)
for $\ell = \ell_c$, the global minimum points of $g$ are 0 and $\pm \bar{x}(\ell_c) = 
\sqrt{5/3}$; (3) for $\ell < \ell_c$, $g$ has a unique global minimum point at 0. 
These three cases mirror the following features of $\gbnkn$: (1) for $(\bn,\kn)$ above the first-order
curve, the global minimum points of $\gbnkn$ are the symmetric nonzero pair
$\pm m(\bn,\kn)$; (2) for $(\bn,\kn)$ on the first-order
curve, the global minimum points of $\gbnkn$ are 0 and the symmetric nonzero pair
$\pm m(\bn,\kn)$; (3) for $(\bn,\kn)$ below the first-order
curve, $\gbnkn$ has a unique global minimum point at 0.  In addition,
as pointed out in item 4 before the statement of Theorem \ref{thm:asymptoticsforc2},
the set of global minimum points of $g$ undergoes a discontinuous bifurcation at $\ell = \ell_c$.
This mirrors the discontinuous bifurcation that occurs in the set of global
minimum points of $\gbk$ at $K = K_1(\beta)$ for $\beta > \bc$ [Thm.\ \ref{thm:firstorder}(d)].

In order to verify these properties of the Ginzburg-Landau polynomials, we 
must calculate the relevant expansion of $n \gbnkn(x/n^\gamma)$ in 
(\ref{eqn:Taylorforc}).  We first consider the case where $\ell \not = K''(\bc)$; the case where
$\ell = K''(\bc)$ will be discussed later. Since 
\beas
K(\bn) & = & K(\bc + 1/n^\alpha) \\
& = & K(\bc) + K'(\bc)/{n^\alpha} + {K''(\bc)}/({2 n^{2\alpha}}) + K'''(\bc)/({6 n^{3\alpha}}) +
\mbox{O}(1/n^{4\alpha}),
\eeas
we have 
\be
\label{eqn:K''ell}
K(\beta_n) - K_n = 
{(K''(\bc) - \ell)}/({2n^{2\alpha}}) + {(K'''(\bc) - \tell)}/({6n^{3\alpha}}) + \mbox{O}(1/n^{4\alpha})
\ee
and
\[
4 - e^{\bn} = 4(1 - e^{1/n^\alpha}) = -{4}/{n^\alpha} + \mbox{O}(1/n^{2\alpha}).
\]
Substituting the last two formulas into (\ref{eqn:Taylorforc}), we see
that for all $n \in \N$, any $\gamma > 0$, any $R > 0$, and all $x \in \R$ satisfying $|x/n^\gamma| < R$
\bea
\label{eqn:gbnkn52}
\lefteqn{
nG_{\beta_n, K_n}(x/n^\gamma)} \\
\nonumber && = 
\frac{1}{n^{2\gamma + 2\alpha - 1}} \frac{1}{2} \bc(K''(\bc) - \ell)(1+\ve_n) x^2
- \ds\frac{1}{n^{4\gamma + \alpha - 1}} 4 c_4(1+\ve_n)x^4 \\
\nonumber && \hspace{.2in} + \ \frac{1}{n^{6\gamma -1}} c_6 (1+\ve_n)x^6
+ \mbox{O}\!\left(\frac{1}{n^{2\gamma + 3\alpha - 1}}\right)\!x^2 
+ \mbox{O}\!\left(\frac{1}{n^{2\gamma + 4\alpha - 1}}\right)\!x^2\\
\nonumber \nonumber && \hspace{.2in} + \ \mbox{O}\!\left(\frac{1}{n^{4\gamma + 2\alpha - 1}}\right)\!x^4
+ \mbox{O}\!\left(\frac{1}{n^{7\gamma - 1}}\right)\!x^7,
\eea
where $c_4 = 3/16$ and $c_6 = 9/40$. 
Given $u \in \R$, we multiply the numerator and denominator of the right side of
the last display by $n^u$, obtaining $nG_{\beta_n, K_n}(x/n^\gamma) = n^u G_n(x)$,
where for all $n \in \N$, any $\gamma > 0$, any $R > 0$, and all $x \in \R$ satisfying $|x/n^\gamma| < R$
\bea
\label{eqn:c2}
\lefteqn{G_n(x) } \\ 
\nonumber && = 
\frac{1}{n^{2\gamma + 2\alpha - 1 + u}} \frac{1}{2} \bc (K''(\bc) - \ell) (1+\ve_n)x^2
- \ds\frac{1}{n^{4\gamma + \alpha - 1 + u}} 4  c_4 (1+\ve_n) x^4 \\
\nonumber && \hspace{.2in} + \ \frac{1}{n^{6\gamma -1 + u}} c_6 (1+\ve_n) x^6
+ \mbox{O}\!\left(\frac{1}{n^{2\gamma + 3\alpha - 1 + u}}\right)\!x^2 
+ \mbox{O}\!\left(\frac{1}{n^{2\gamma + 4\alpha - 1 + u}}\right)\!x^2\\
\nonumber && \hspace{.2in} + \ \mbox{O}\!\left(\frac{1}{n^{4\gamma + 2\alpha - 1 + u}}\right)\!x^4
+ \mbox{O}\!\left(\frac{1}{n^{7\gamma - 1 + u}}\right)\!x^7.
\eea
In this formula $\ve_n \goto 0$ and the big-oh terms are
uniform for $x \in (-Rn^\gamma, Rn^\gamma)$. 
Since $\ell \not = K''(\bc)$, $c_4 > 0$, and $c_6 > 0$, the coefficients 
of $x^2$, $x^4$, and $x^6$ in the first three terms are all nonzero.  

In order to obtain the limit of $G_n$, we impose the condition that the powers of $n$
appearing in the first three terms in the last display equal 0;
i.e., $2\gamma + 2\alpha - 1 + u = 0 = 4\gamma + \alpha - 1 + u = 6\gamma -1 + u$.  These three 
equalities are equivalent to $\gamma = \alpha/2$ and $u = 1 - 6\gamma = 1 - 3\alpha$.
With this choice of $\gamma$ and $u$, the powers of $n$
in the last four terms in (\ref{eqn:c2}) are positive, and so for all $x \in \R$ these four terms converge to 0
as $n \goto \infty$.  
It follows that as $n \goto \infty$, we have for all $x \in \R$ 
\be
\label{eqn:gell}
G_n(x) = n^{1-u} \gbnkn(x/n^\gamma) \goto g_\ell(x) = \half \bc  (K''(\bc) - \ell) x^2
- 4  c_4 x^4 + c_6 x^6.
\ee
Because the big-oh terms in (\ref{eqn:c2}) are uniform for $x \in (-Rn^\gamma,Rn^\gamma)$, the convergence
of $G_n(x)$ to $g(x)$ is uniform for $x$ in compact subsets of $\R$.
Since $\ell \not = K''(\bc)$, the three
coefficients in $g_\ell$ are all nonzero.  We write the Ginzburg-Landau polynomial as
$g_\ell$ in order to emphasize the dependence on the parameter $\ell$; $g_\ell$ does not
depend on the choice of $\tell$ in (\ref{eqn:bnkn2a}). 

We now briefly consider the case where $\ell = K''(\bc)$.
In this situation we have the right side of (\ref{eqn:K''ell}) with $\ell = K''(\bc)$.
We omit the rest of the calculation showing that when 
$\ell = K''(\bc)$, the limit of $G_n(x) = n^{1-u} \gbnkn(x/n^\gamma)$ 
equals the same Ginzburg-Landau polynomial $g_\ell$.

The Ginzburg-Landau polynomial $g_\ell$ has the form $\atwo x^2 - \afour x^4 + \asix x^6$, 
where 
\be
\label{eqn:a6}
\atwo = \bc  (K''(\bc) - \ell)/2, \afour = 4 c_4 = 3/4 > 0,
\mbox{ and } \asix = c_6 = 9/40 > 0;
\ee
depending on the value of $\ell$, $a_2$ can be positive, 0, or negative.  We are interested
in the structure of the set of global minimum points of $g_\ell$ for variable $\ell$. 
In order to analyze this structure, we define the critical value $a_c = \afour^2/4\asix$.  
According to Theorem \ref{thm:6order}, if $\atwo < a_c$,
then the global minimum points of $\atwo x^2 - \afour x^4 + \asix x^6$
are $\pm \barx(\atwo)$, where $\barx(\atwo) > 0$
is defined in (\ref{eqn:barx}); if $\atwo = a_c$, then the global minimum points of this polynomial are 0 and $\pm (2\atwo/\afour)^{1/2}$; 
and if $\atwo > a_c$, then the unique global minimum point of this polynomial is 0.
Substituting the values of $\atwo$, $\afour$, and $\asix$, we see that 
$a_c = 5/8$.  Defining $\ell_c$ to be the value of $\ell$
for which $\atwo = a_c$, we find that
\be
\label{eqn:ellc}
\ell_c = K''(\bc) - {5}/({4\bc}) = - {1}/({4\bc}) - {2}/{\bc^2} + 
{3}/{\bc^3} = -0.094979.
\ee
The second equality follows from the formula for $K''(\bc)$ given in part (d) of 
Lemma \ref{lem:kbeta}. The structure of the set of global minimum points of the polynomial
$\atwo x^2 - \afour x^4 + \asix x^6$ translates into the following
structure of the set of global minimum points of the Ginzburg-Landau polynomial
\[
g_\ell(x) =  \half \bc (K''(\bc) - \ell) x^2 - 4 c_4 x^4 + c_6 x^6.
\]
The formula for $\bar{x}(\ell)$ in (\ref{eqn:barxforc2a})
is obtained by substituting $a_2$, $a_4$, and $a_6$ from (\ref{eqn:a6})
into the definition (\ref{eqn:barx}) of $\bar{x}(a_2)$.

\begin{enumerate}
\item For $\ell > \ell_c$, which corresponds to $a_2 < a_c$,
the global minimum points of $g_\ell$ are $\pm \bar{x}(\ell)$, where
\be
\label{eqn:barxforc2a}
\bar{x}(\ell) = \ts
\frac{\sqrt{10}}{3} \left(1 + \left(1 - \frac{3\bc}{5}
(K''(\bc) - \ell)\right)^{1/2} \right)^{1/2}.
\ee
The polynomial $g_\ell$ has the same shape as $\gbk$ in Figure 6 in section \ref{section:phasetr}.
\item For $\ell = \ell_c$, which corresponds to $a_2 = a_c$, 
the global minimum points of $g_\ell$ are 0 and $\pm \bar{x}(\ell_c)$, where
$\bar{x}(\ell_c) = \sqrt{5/3}$.  The polynomial $g_\ell$ has the same shape as $\gbk$ in Figure 5 in section 
\ref{section:phasetr}.
\item For $\ell < \ell_c$, which corresponds to $a_2 > a_c$, $g_\ell$ has a unique global minimum point at 0.   
The polynomial $g_\ell$ has the same shape as $\gbk$ in Figure 4 in section \ref{section:phasetr}.
\item The set of global minimum points of $g_\ell$ undergoes a discontinuous bifurcation
at $\ell = \ell_c$, changing discontinuously from $\{0\}$ for $\ell < \ell_c$ to $\{0, \pm \bar{x}(\ell_c)\}$
for $\ell = \ell_c$ to $\{\pm \bar{x}(\ell)\}$ for $\ell > \ell_c$.
This mirrors an analogous property of $\gbk$ given in part (d) of Theorem
\ref{thm:firstorder}.
\end{enumerate}

Assuming Conjectures 1--3 in section \ref{section:firstordercurve},
we contrast properties of the sequence $(\bn,\kn)$ for
$\ell = \ell_c$ and $\tell > K_1'''(\bc)$ with properties of another sequence having the same Ginzburg-Landau polynomial
but a different structure of the set of global minimum points of $\gbnkn$.   
For the first sequence, since $(\bn,\kn)$ lies in the phase-coexistence region
above the first-order curve for all sufficiently large $n$, 
the global minimum points of $\gbnkn$ are the symmetric nonzero pair $\pm \mbnkn$, and as we have just pointed
out, the global
minimum points of $g_{\ell_c}$ are 0 and $\pm\barx(\ell_c)$. 
This is to be contrasted with the sequence $\bn = \bc + 1/n^\alpha$, 
$K_n = K_1(\bn)$.  Since this sequence lies on the first-order curve for all $n$, the global minimum points of 
$\gbnkn$ are 0 and $\pm\mbnkn$. If one replaces $K_1(\bn)$ by the first three terms in its
Taylor expansion plus an error term and uses Conjectures 1 and 2 in section
\ref{section:firstordercurve}, then one sees that the associated Ginzburg-Landau polynomial coincides
with $g_{\ell_c}$.

For the sequence $(\bn,\kn)$ defined in (\ref{eqn:bnkn2a}) 
the asymptotic behavior of $\mbnkn$
is given in the next theorem.  As we verify in the discussion 
after (\ref{eqn:bnkn2a}), $(\bn,\kn)$ lies in the phase-coexistence region for all sufficiently large $n$
under the following conditions on the parameters $\ell$ and $\tell$; in the last two cases these conditions
are supplemented by the appropriate conjectures in section \ref{section:firstordercurve}:
\begin{itemize}
  \item[(i)] $\ell > K''(\bc)$ and $\tell \in \R$, 
  \item[(ii)] $\ell = K''(\bc)$ and $\tell > K'''(\bc)$,
  \item[(iii)] $K''(\bc) > \ell > \ell_c$, $\tell \in \R$, and Conjectures $1$ and $2$,
  \item[(iv)] $\ell = \ell_c$, $\tell > K_1'''(\bc)$, and Conjectures 1--3.
\end{itemize} 

\skp
 
\begin{thm}
\label{thm:asymptoticsforc2}
For $\alpha > 0$, $\ell \in \R$, and $\tell \in \R$, define
\[
\beta_n = \bc + {1}/{n^\alpha} \ \mbox{ and } \ K_n = K(\bc) + K'(\bc)/{n^\alpha}
+  \ell/(2n^{2\alpha}) + \tell/(6n^{3\alpha}) 
\]
as well as $c_4 = 3/16$ and $c_6 = 9/40$. Then $(\bn,\kn)$ converges to the tricritical
point $(\bc,\kc)$.  The following conclusions hold.

{\em (a)} For any $\alpha > 0$, $u = 1 - 3\alpha$, and $\gamma = \alpha/2$
\[
G_n(x) = n^{1-u}\gbnkn(x/n^\gamma) \goto g_\ell(x) = \half \bc (K''(\bc) - \ell) x^2 
 - 4  c_4 x^4 + c_6 x^6
\]
uniformly for $x$ in compact subsets of $\R$.

{\em (b)} The Ginzburg-Landau polynomial $g_\ell$ has nonzero global minimum points if and 
only if $\ell \geq \ell_c = (- \bc^2 - 8\bc + 12)/4 \bc^3$.

\hspace{.25in} {\em (i)} Assume that $\ell > \ell_c$. Then the 
global minimum points of $g_\ell$
are $\pm \bar{x}(\ell)$, where $\bar{x}(\ell)$ is defined in {\em (\ref{eqn:barxforc2a})}.

\hspace{.25in} {\em (ii)} Assume that $\ell = \ell_c$. Then the global minimum points of the Ginzburg-Landau polynomial 
$g_\ell(x)$ are {\em 0} and $\pm \bar{x}(\ell_c)$, where $\bar{x}(\ell_c) = (5/3)^{1/2}$. 

{\em (c)} In each of the cases {\em (i)--(iv)} appearing
before the statement of the theorem and for any $\alpha >0$,
$m(\bn,\kn) \goto 0$ and has the asymptotic behavior 
\[
\mbnkn \sim {\bar{x}(\ell)}/{n^{\alpha/2}} = \barx(\ell)(\bn - \bc)^{1/2}; \ \mbox{ i.e., }
\lim_{n \goto \infty} n^{\alpha/2} \mbnkn = \bar{x}(\ell).
\]
\end{thm}

\noi
{\bf Proof.} Part (a) follows from
the discussion leading up to the statement of the
theorem. Parts (b), (b)(i), and (b)(ii) are consequences of Theorem \ref{thm:6order}. 

We now verify the asymptotic behavior of $\mbnkn$ in part (c).
According to Theorem \ref{thm:zngoto0}, $\mbnkn \goto 0$.  
The validity of hypotheses (i) and (ii) of Theorem \ref{thm:exactasymptotics} 
follows from the definition of the sequence $(\bn,\kn)$
and the discussion leading up to the statement of the present theorem.
\iffalse
In particular, we verify in the first paragraph after (\ref{eqn:bnkn2a}) that
when $\ell \geq K''(\bc)$, $(\bn,\kn)$ lies in the phase-coexistence
region above the spinodal curve for all sufficiently large $n$. 
As we also verify in item 1 in the third paragraph after (\ref{eqn:bnkn2a}),
when $K''(\bc) > \ell > \ell_c$, if Conjectures 1 and 2 in section
\ref{section:firstordercurve} are valid, then for all sufficiently large $n$,
$(\bn,\kn)$ lies in the phase-coexistence region above the first-order curve and below the spinodal curve.
When $\ell = \ell_c$, we verify the same conclusion in item 2 in that paragraph if Conjectures 1, 2, and 3 in section 
\ref{section:firstordercurve} are valid. This completes the verification of hypotheses (i) and (ii)
of Theorem \ref{thm:exactasymptotics}.
\fi
Hypothesis (iii) of Theorem \ref{thm:exactasymptotics}
is parts (a) and (b) of the present theorem. We now verify hypothesis (iv), first in the case
where $\ell \not = K''(\bc)$.  Using (\ref{eqn:c2}) with $\gamma = \alpha/2$ and $u = 1-3\alpha$, 
one easily proves that for any $\alpha > 0$ there exists
$R > 0$ such that for all sufficiently large $n \in \N$ and all $x \in \R$ satisfying
$|x/n^\gamma| < R$
\[
G_n(x) = n^{1-u}\gbnkn(x/n^\gamma) \geq H(x) = - \bc |K''(\bc) - \ell| x^2
-8 c_4 x^4 + \half c_6 x^6.
\]
Since $H(x) \goto \infty$ as $|x| \goto \infty$, hypothesis (iv) of Theorem
\ref{thm:exactasymptotics} is satisfied when $\ell \not = K''(\bc)$.  When $\ell = K''(\bc)$,
we obtain the lower bound in the last display by replacing $H$ there by 
\[
H(x) = - \ts\frac{1}{3} \bc |K'''(\bc) - \tell| x^2 - 8 c_4 x^4 + \half c_6 x^6,
\]
where $K'''(\bc) = -1.024398 < 0$ [Lem.\ \ref{lem:kbeta}(d)].  
This follows from the expansion replacing (\ref{eqn:c2})
when $\ell = K''(\bc)$; the proof is omitted.
The verification of the four hypotheses of Theorem \ref{thm:exactasymptotics} is complete. 
We now apply the theorem to conclude that for any $\alpha > 0$, $\mbnkn \sim
\bar{x}/n^\gamma = \barx/n^{\alpha/2}$. Part (c) 
of the present theorem is proved. \ \ink
 
\skp

We now consider the third choice of sequence $(\bn,\kn)$
converging to the tricritical point, which gives
yet a different asymptotic behavior of $\mbnkn \goto 0$.  Given 
$\alpha > 0$, an integer $p \geq 2$, and $\ell \in \R$, we define
\be
\label{eqn:bnkn3a}
\bn = \bc - {1}/{n^\alpha} \ \mbox{ and } \ 
\kn = \kbc + \sum_{j=1}^{p-1} { K^{(j)}(\bc) (-1)^j}/(j! n^{j\alpha}) + {\ell (-1)^{p}}/(p! n^{p\alpha}).
\ee
In order to simplify the analysis here, we assume that $\ell \not = K^{(p)}(\bc)$. 
The choice $\ell = K^{(p)}(\bc)$ will be discussed in the third paragraph
before Theorem \ref{thm:asymptoticsforc3} and after Theorem \ref{thm:asymptoticsforc4}.

The sequence $(\bn,\kn)$ defined in (\ref{eqn:bnkn3a}) converges to the 
tricritical point $(\bc,\kc)$.
Since $\bn - \bc = -1/n^\alpha$, the convergence takes place along the curve 
\[
\tilde{K}_p(\beta) = \kbc + \sum_{j=1}^{p-1} {K^{(j)}(\bc)}(\beta - \bc)^j/j! + 
{\ell (\beta - \bc)^{p}}/{p!}.
\]
This curve coincides with the second-order curve to order $p-1$ in powers of $\beta - \bc$
in a neighborhood of the
tricritical point and satisfies $\tilde{K}_p^{(p)}(\bc) = \ell$ (see paths 5 and 6 in Figure 2).  
The relationship of the sequence $(\bn,\kn)$ to the second-order curve
depends on the parity of $p$.  We first assume that $p$ is even. 
For all sufficiently large $n$, $\ell > K^{(p)}(\bc)$ corresponds to $(\bn,\kn)$ lying
in the phase-coexistence region located above the second-order curve for $\beta < \bc$
and thus to the free energy functional $\gbnkn$ having its global minimum points at $\pm \mbnkn \not = 0$. On the other hand, for all sufficiently large $n$,
$\ell < K^{(p)}(\bc)$ corresponds to $(\bn,\kn)$ 
lying in the single-phase region under the second-order curve and thus to $\gbnkn$
having a unique global minimum point at 0. 
If $p$ is odd, then the situation is reversed.  
As one can check, in all cases the 
structure of the set of global minimum points 
of the Ginzburg-Landau polynomial mirrors the structure
of the set of global minimum points of $\gbnkn$. 

We now determine the relevant expansion of $n \gbnkn(x/n^\gamma)$ in 
(\ref{eqn:Taylorforc}) where $(\bn,\kn)$ is the sequence in (\ref{eqn:bnkn3a}).  
Since 
\[
K(\bn) = K(\bc + b/n^\alpha) = K(\bc) + \sum_{j=1}^p { K^{(j)}(\bc) (-1)^j}/({j! n^{j\alpha}})
+ \mbox{O}(1/n^{(p+1)\alpha}),
\] 
we have 
\be
\label{eqn:assisi}
K(\beta_n) - K_n = {(K^{(p)}(\bc) - \ell)(-1)^p}/({p!n^{p\alpha}}) + \mbox{O}(1/n^{(p+1)\alpha})
\ee
and
\[
4 - e^{\bn} = 4(1 - e^{-1/n^\alpha}) = {4}/{n^\alpha} + \mbox{O}(1/n^{2\alpha}).
\]
Substituting the last two expressions into (\ref{eqn:Taylorforc}), we see
that for all $n \in \N$, any $\gamma > 0$, any $R > 0$, and all $x \in \R$ satisfying $|x/n^\gamma| < R$
\bea
\label{eqn:gbnkn53}
\lefteqn{nG_{\beta_n, K_n}(x/n^\gamma)} \\ 
\nonumber && = 
\frac{1}{n^{2\gamma + p\alpha - 1}} \frac{1}{p!} \bc (K^{(p)}(\bc) - \ell)(-1)^p (1+\ve_n) x^2 \\
\nonumber && \hspace{.2in} + \ \frac{1}{n^{4\gamma + \alpha - 1}} 4  c_4  (1+\ve_n) x^4 
+ \frac{1}{n^{6\gamma -1}} c_6 (1+\ve_n) x^6  \\
\nonumber && \hspace{.2in} + \ \mbox{O}\!\left(\frac{1}{n^{2\gamma + (p+1)\alpha - 1}}\right)\!x^2
+ \mbox{O}\!\left(\frac{1}{n^{4\gamma + 2\alpha - 1}}\right)\!x^4
+ \mbox{O}\!\left(\frac{1}{n^{7\gamma - 1}}\right)\!x^7,
\eea
where $c_4 = 3/16$ and $c_6 = 9/40$. 
Given $u \in \R$, we multiply the numerator and denominator of the right side of
the last display by $n^u$, obtaining $nG_{\beta_n, K_n}(x/n^\gamma) = n^u G_n(x)$,
where for all $n \in \N$, any $\gamma > 0$, any $R > 0$, and all $x \in \R$ satisfying $|x/n^\gamma| < R$
\bea
\label{eqn:gnxc3}
\lefteqn{G_n(x)} \\ \nonumber 
&& = 
\frac{1}{n^{2\gamma + p\alpha - 1 + u}} \frac{1}{p!} \bc (K^{(p)}(\bc) - \ell)(-1)^p (1+\ve_n) x^2 \\
&& \nonumber \hspace{.2in} + \ \frac{1}{n^{4\gamma + \alpha - 1 + u}} 4  c_4 (1+\ve_n) x^4 
+ \frac{1}{n^{6\gamma -1 + u}} c_6 (1+\ve_n) x^6  \\
&& \nonumber \hspace{.2in} + \ \mbox{O}\!\left(\frac{1}{n^{2\gamma + (p+1)\alpha - 1 + u}}\right)\!x^2
+ \mbox{O}\!\left(\frac{1}{n^{4\gamma + 2\alpha - 1 + u}}\right)\!x^4
+ \mbox{O}\!\left(\frac{1}{n^{7\gamma - 1 + u}}\right)\!x^7.
\eea
In this formula $\ve_n \goto 0$ and the big-oh terms are uniform for $x \in (-Rn^\gamma,Rn^\gamma)$.
Since $\ell \not = K^{(p)}(\bc)$, $b < 0$, $c_4 > 0$, and $c_6 > 0$,
the coefficients of $x^2$, $x^4$, and $x^6$ in the first three terms are all nonzero.

We first consider $p =2$, which gives rise to a different asymptotic
behavior of $\mbnkn \goto 0$ from $p \geq 3$.  We continue to assume
that $\ell \not = K^{''}(\bc)$.  
In order to obtain the limit of $G_n$, we fix $\alpha > 0$ and impose the condition that the three powers of $n$
appearing in the first three terms in (\ref{eqn:gnxc3}) equal 0;
i.e., $2\gamma + 2\alpha - 1 + u = 0 = 4\gamma + \alpha - 1 + u = 6\gamma -1 + u$.  
These three equalities are equivalent to $\gamma = \alpha/2$ and $u = 1 - 6\gamma = 1 - 3\alpha$.
With this choice of $\gamma$ and $u$, the powers of $n$
in the last three terms in (\ref{eqn:gnxc3}) are positive, and so for all $x \in \R$
these three terms converge to 0 as $n \goto \infty$. 
It follows that as $n \goto \infty$, we have for all $x \in \R$
\be
\label{eqn:gofx}
G_n(x) = n^{1-u} \gbnkn(x/n^\gamma) \goto g(x) = \half \bc (K''(\bc) - \ell)x^2 + 4  c_4 x^4 + 
c_6 x^6.
\ee
Since the big-oh terms in (\ref{eqn:gnxc3}) are uniform for
$x \in (-Rn^\gamma,Rn^\gamma)$, the convergence of $G_n$ to $g$ is uniform for
$x$ in compact subsets of $\R$. 

We now briefly consider the case where $\ell = K''(\bc)$.
In this situation the right side of (\ref{eqn:assisi}) must be replaced
by $-K'''(\bc)/(6n^{3\alpha}) + \mbox{O}(1/n^{4\alpha})$.  We omit the rest of the calculation showing that when 
$\ell = K''(\bc)$, the limit of $G_n(x) = n^{1-u} \gbnkn(x/n^\gamma)$ 
equals the same Ginzburg-Landau polynomial $g$ in (\ref{eqn:gofx}).
When $\ell = K''(\bc)$, 
the coefficient of $x^2$ equals 0, and thus $g$ has a unique global minimum point at 0,
an uninteresting case from the viewpoint of the asymptotic behavior of $\mbnkn$.

A major difference between the present situation and that considered in Theorem 
\ref{thm:asymptoticsforc2} arises in the condition guaranteeing
that the Ginzburg-Landau polynomial $g$ in (\ref{eqn:gofx}) has nonzero global minimum points.
It follows from Theorem \ref{thm:a2} 
that since the coefficient of $x^4$ is positive, $g$ has nonzero global minimum points if and only if the coefficient
of $x^2$ is negative; i.e., if and only if $\ell > K''(\bc)$.  If $\ell \leq K''(\bc)$, 
then $g$ has a unique global minimum point at 0, again an uninteresting case from 
the viewpoint of the asymptotic behavior of $\mbnkn$. 

For the sequence $(\bn,\kn)$ defined in 
(\ref{eqn:bnkn3a}) with $p = 2$ and $\ell > K''(\bc)$, the next theorem
describes the asymptotic behavior of $\mbnkn$. The inequality $\ell > K''(\bc)$ in parts
(b) and (c) guarantees that for all sufficiently large $n$, $(\bn,\kn)$ lies in the phase-coexistence region above the 
second-order curve.

\begin{thm}
\label{thm:asymptoticsforc3}
For $\alpha > 0$ and $\ell \in \R$ define
\[
\beta_n = \bc - 1/{n^\alpha} \ \mbox{ and } \ K_n = K(\bc) + K'(\bc)/{n^\alpha}
+ \ell/2n^{2\alpha}
\]
as well as $c_4 = 3/16$ and $c_6 = 9/40$.  Then the sequence
$(\bn,\kn)$ converges to the tricritical point $(\bc,\kc)$. 
The following conclusions hold.

{\em (a)} For any $\alpha > 0$, $u = 1-3\alpha$. and $\gamma = \alpha/2$
\[
G_n(x) = n^{1-u} \gbnkn(x/n^\gamma) \goto g(x) = \half \bc (K''(\bc) - \ell) x^2 
+ 4 c_4 x^4 + c_6 x^6
\]
uniformly for $x$ in compact subsets of $\R$.

{\em (b)} The Ginzburg-Landau polynomial $g$ has nonzero global minimum points
if and only if $\ell > K''(\bc)$.  If this inequality holds, then the global minimum points 
of $g$ are $\pm \bar{x}$, where 
\be
\label{eqn:barxforc3a}
\bar{x} = \ts
\frac{\sqrt{10}}{3}\left(-1 + \left(1 + \frac{3\bc}{5}
(\ell - K''(\bc))\right)^{1/2} \right)^{1/2}.
\ee

{\em (c)} Assume that $\ell > K''(\bc)$.  Then for any $\alpha > 0$, $m(\bn,\kn) \goto 0$ and has the asymptotic behavior 
\[
\mbnkn \sim {\bar{x}}/{n^{\alpha/2}} = \barx(\bc - \bn)^{1/2}; \ \mbox{ i.e., }
\lim_{n \goto \infty} n^{\alpha/2} \mbnkn = \bar{x}.
\]
\end{thm}

\noi
{\bf Proof.} Part (a) follows from
the discussion leading up to the statement of the
theorem.  The first assertion in part (b) is a consequence of Theorem \ref{thm:a2}.  The formula
for $\bar{x}$ comes from the formula for $\bar{x}(a_2)$ in (\ref{eqn:barxa2})
by substituting $a_2 = \half\bc (K''(\bc) - \ell)$, $a_4 = 4 c_4$, and $a_6 = c_6$.

We now verify the asymptotic behavior of $\mbnkn$ in part (c).
According to Theorem \ref{thm:zngoto0}, $\mbnkn \goto 0$.  
The validity of hypotheses (i) and (ii) of Theorem \ref{thm:exactasymptotics} 
follows from the definition of the sequences $(\bn,\kn)$ and the inequality 
$\ell > K''(\bc)$, which by (\ref{eqn:assisi}) is equivalent to $K_n > K(\bn)$
for all sufficiently large $n$. Thus, if $\ell > K''(\bc)$, then $(\bn,\kn)$
lies in the phase-coexistence region for all sufficiently large $n$. 
Hypothesis (iii) of Theorem \ref{thm:exactasymptotics}
is parts (a) and (b) of the present theorem. We now verify hypothesis (iv) of Theorem \ref{thm:exactasymptotics}.
Using (\ref{eqn:gnxc3}) with $\gamma = \alpha/2$ and $u = 1-3\alpha$, one easily proves 
that for any $\alpha > 0$ there exists
$R > 0$ such that for all sufficiently large $n \in \N$ and all $x \in \R$ satisfying
$|x/n^\gamma| < R$
\[
G_n(x) = n^{1-u}\gbnkn(x/n^\gamma) \geq H(x) = - \bc |K''(\bc) - \ell| x^2
+ \ts \frac{1}{2} c_6 x^6.
\]
Since $H(x) \goto \infty$ as $|x| \goto \infty$, hypothesis (iv) in Theorem
\ref{thm:exactasymptotics} is satisfied.  
This completes the verification of the four hypotheses of Theorem \ref{thm:exactasymptotics}.
We now apply the theorem to conclude that for any $\alpha > 0$, $\mbnkn \sim
\bar{x}/n^\gamma = \barx/n^{\alpha/2}$.  Part (c) of the present theorem is proved. \ \ink

\skp
This theorem completes the analysis for $p =2$.  We now continue with 
the analysis for the sequences $(\bn,\kn)$ defined in (\ref{eqn:bnkn3a})
for $p \geq 3$, $\alpha > 0$, and $\ell \not = K^{(p)}(\bc)$.
As we saw in the discussion leading up to Theorem \ref{thm:asymptoticsforc3},
for all $n \in \N$, any $\gamma > 0$, any $R > 0$, all $x \in \R$ 
satisfying $|x/n^\gamma| < R$, we have $nG_{\beta_n, K_n}(x/n^\gamma) = n^u G_n(x)$,
where $G_n$ is given by the expansion (\ref{eqn:gnxc3}):
\bea
\label{eqn:c4}
\lefteqn{G_n(x)} \\ \nonumber 
&& = 
\frac{1}{n^{2\gamma + p\alpha - 1 + u}} \ts\frac{1}{p!} \bc (K^{(p)}(\bc) - \ell)(-1)^p (1+\ve_n) x^2 \\
&& \nonumber \hspace{.2in} + \ \frac{1}{n^{4\gamma + \alpha - 1 + u}} 4  c_4 (1+\ve_n) x^4 
+ \frac{1}{n^{6\gamma -1 + u}} c_6 (1+\ve_n) x^6  \\
&& \nonumber \hspace{.2in} + \ \mbox{O}\!\left(\frac{1}{n^{2\gamma + (p+1)\alpha - 1 + u}}\right)\!x^2
+ \mbox{O}\!\left(\frac{1}{n^{4\gamma + 2\alpha - 1 + u}}\right)\!x^4
+ \mbox{O}\!\left(\frac{1}{n^{7\gamma - 1 + u}}\right)\!x^7.
\eea
In this formula $\ve_n \goto 0$ and the big-oh terms are uniform for $x \in 
(-R n^\gamma, R n^\gamma)$. 

The analysis for $p \geq 3$ is considerably more complicated than in the case $p =2$ just
considered.
In order to obtain the limit of $G_n$, it is useful to denote by $f(\gamma,\alpha,u)$,
$g(\gamma,\alpha,u)$, and $h(\gamma,\alpha,u)$ the respective 
exponents of $n$ in the coefficients
of $x^2$, $x^4$, and $x^6$ in the first three terms in the last display.
Thus, $f(\gamma,\alpha,u) = 2\gamma + p\alpha - 1 + u$,
$g(\gamma,\alpha,u) = 4\gamma + \alpha - 1+ u$, and
$h(\gamma,\alpha,u) = 6\gamma -1 + u$.
In order to obtain a limiting Ginzburg-Landau polynomial $g$ having nonzero global minimum points, 
$g$ must have either three terms or two terms.  The polynomial $g$ has three terms if and only if
there exist $\gamma$, $\alpha$, and $u$ for which the three equalities 
$f(\gamma,\alpha,u) = 0 = g(\gamma,\alpha,u) = h(\gamma,\alpha,u)$ are compatible.
However, a short calculation shows that the three equalities are incompatible. The polynomial
$g$ has two terms if and only if there exist $\gamma$, $\alpha$, 
and $u$ for which at least one of the following sets of two equalities
and one inequality are compatible: 
\begin{enumerate}
\item $f(\gamma,\alpha,u) = 0 = g(\gamma,\alpha,u) <  h(\gamma,\alpha,u)$,
\item  $g(\gamma,\alpha,u) = 0 = h(\gamma,\alpha,u) < f(\gamma,\alpha,u)$,
\item $f(\gamma,\alpha,u) = 0 = h(\gamma,\alpha,u) <  g(\gamma,\alpha,u)$.
\end{enumerate}

Another short calculation shows the two equalities and the one inequality in item 3 are incompatible.
By contrast, the two equalities and the one inequality in item 2 are compatible.
In fact, $g(\gamma,\alpha,u) = 0 = h(\gamma,\alpha,u)$ when $\gamma = \alpha/2$
and $u = 1 - 6\gamma = 1 - 3\alpha$, and $g(\gamma,\alpha,u) < f(\gamma,\alpha,u)$ 
when $\gamma < (p-1)\alpha/2$;
the latter inequality is compatible with $\gamma = \alpha/2$ since $p \geq 3$. 
With this choice of $\gamma$ and $u$, the powers of $n$ in the first term and in the last three terms
in (\ref{eqn:c4}) are positive, and so for all $x \in \R$ these four terms converge
to 0 as $n \goto \infty$.  
It follows that as $n \goto \infty$, we have for all $x \in \R$
\[
G_n(x) = n^{1-u}\gbnkn(x/n^\gamma) \goto g(x)= 4c_4 x^4 + c_6 x^6.
\]
The polynomial $g$ has a unique global minimum point at 0, an uninteresting case from 
the viewpoint of the asymptotic behavior of $\mbnkn$.   

The final case to consider is the compatibility of the two equalities and the one inequality
in item 1.  In fact, $f(\gamma,\alpha,u) = 0 = g(\gamma,\alpha,u)$
when $\gamma = (p-1)\alpha/2$ and $u = 1 - p\alpha - 2\gamma = 1 - (2p-1)\alpha$.
On the other hand, $g(\gamma,\alpha,u)
< h(\gamma,\alpha,u)$ when $\gamma > \alpha/2$, which is compatible with $\gamma = (p-1)\alpha/2$
since $p \geq 3$.  With this choice of $\gamma$ and $u$, the powers of $n$ in the last four terms
in (\ref{eqn:c4}) are positive, and so for all $x \in \R$ these four terms converge
to 0 as $n \goto \infty$.  
It follows that as $n \goto \infty$, we have for all $x \in \R$
\be
\label{eqn:morecareful}
G_n(x) = n^{1-u}\gbnkn(x/n^\gamma) \goto g(x) = \ts \frac{1}{p!}\bc (K^{(p)}(\bc) - \ell)(-1)^p x^2 + 4 c_4 x^4.
\ee
Since the big-oh terms in (\ref{eqn:c4}) are uniform for $x \in (-R n^\gamma,R n^\gamma)$, 
the convergence of $G_n(x)$ to $g(x)$ is uniform for $x$ in compact subsets of $\R$.
The Ginzburg-Landau polynomial $g$ has nonzero global minimum points if and only if
$(K^{(p)}(\bc) - \ell)(-1)^p < 0$. 

For the sequence $(\bn,\kn)$ defined 
in (\ref{eqn:bnkn3a}) with $p \geq 3$ a positive integer and $(K^{(p)}(\bc) - \ell)(-1)^p < 0$,
the asymptotic behavior of $\mbnkn$ is given in the next theorem.
This inequality involving $\ell$ and $K^{(p)}(\bc)$ guarantees
that for all sufficiently large $n$, $(\bn,\kn)$ lies in the phase-coexistence region above
the second-order curve.

\begin{thm}
\label{thm:asymptoticsforc4}
For $p$ a positive integer satisfying $p \geq 3$, $\alpha > 0$, and a real number $\ell \not = K^{(p)}(\bc)$, define
\[
\bn = \bc - 1/{n^\alpha} \ \mbox{ and } \
\kn = \kbc + \sum_{j=1}^{p-1}  K^{(j)}(\bc)(-1)^j/(j! n^{j\alpha}) +  \ell (-1)^p/(p! n^{p\alpha})
\]
as well as $c_4 = 3/16$.  Then $(\bn,\kn)$ converges to the tricritical point $(\bc,\kc)$.
The following conclusions hold.

{\em (a)}  For any $\alpha > 0$, $u = 1 - (2p-1)\alpha$, and $\gamma = (p-1)\alpha/2$
\[
G_n(x) = n^{1-u}\gbnkn(x/n^\gamma) \goto g(x) = \ts \frac{1}{p!} \bc (K^{(p)}(\bc) - \ell) (-1)^p x^2 
+ 4  c_4 x^4 
\]
uniformly for $x$ in compact subsets of $\R$.

{\em (b)} The Ginzburg-Landau polynomial has nonzero global minimum points if and only if
$(K^{(p)}(\bc) - \ell) (-1)^p < 0$. If this inequality holds, then 
the global minimum points of $g$ are $\pm \bar{x}$, where 
\be
\label{eqn:barx4a}
\bar{x} = \left({\bc \! \left(\ell - K^{(p)}(\bc)(-1)^p\right)}/[8  c_4 p!]\right)^{1/2}.
\ee

{\em (c)} Assume that $(K^{(p)}(\bc) - \ell) (-1)^p < 0$. Then
for any $\alpha > 0$, $m(\bn,\kn) \goto 0$ and has the asymptotic behavior 
\[
\mbnkn \sim {\bar{x}}/{n^{(p-1)\alpha/2}} = \barx(\bn - \bc)^{(p-1)/2}; \ \mbox{ i.e., }
\lim_{n \goto \infty} n^{(p-1)\alpha/2} \mbnkn = \bar{x}.
\]
\end{thm}

\noi
{\bf Proof.} Part (a) follows from
the discussion leading up to the statement of the
theorem. The first assertion in part (b) is elementary.  If
$(K^{(p)}(\bc) - \ell) (-1)^p < 0$, then 
$g'(x) = 2 \bc  (K^{(p)}(\bc) - \ell)(-1)^p x/p! + 16 c_4 x^3 = 0$ has solutions $\pm \bar{x}$ 
and 0, where $\bar{x}$ is defined in (\ref{eqn:barx4a}).  
One easily checks that $\pm \bar{x}$ are global minimum points and 0 a local maximum point.  

We now verify the asymptotic behavior of $\mbnkn$ in part (c).
According to Theorem \ref{thm:zngoto0}, $\mbnkn \goto 0$.  
The validity of hypotheses (i) and (ii) of Theorem \ref{thm:exactasymptotics} 
follows from the definition of the sequences $(\bn,\kn)$ and the inequality 
$(K^{(p)}(\bc) - \ell) (-1)^p < 0$, which by (\ref{eqn:assisi}) is equivalent
to $K_n > K(\bn)$ for all sufficiently large $n$. Thus if $(K^{(p)}(\bc) - \ell) (-1)^p < 0$,
then for all sufficiently large $n$, $(\bn,\kn)$ lies in the phase-coexistence region 
above the second-order curve.  
Hypothesis (iii) of Theorem \ref{thm:exactasymptotics}
is parts (a) and (b) of the present theorem. The verification of hypothesis
(iv) of Theorem \ref{thm:exactasymptotics} is more subtle than in the previous theorems.  The limiting
Ginzburg-Landau polynomial is of degree 4, not of degree 6, because for all $x \in \R$
the $x^6$-term in (\ref{eqn:c4}) converges to 0 as $n \goto \infty$. It is much 
more efficient to recalculate this limit by using the formula (\ref{eqn:TaylorG2again})
based on the two-term Taylor expansion for $n\gbnkn(x/n^\gamma)$, rather than
the formula (\ref{eqn:TaylorG6}) based on the three-term Taylor expansion
for $n\gbnkn(x/n^\gamma)$.  Inserting the expressions for $K(\bn) - K_n$ and
for $4 - e^{\bn}$, we obtain in place of (\ref{eqn:c4}) the expansion
\beas
\lefteqn{G_n(x)} \\ \nonumber 
&& = 
\frac{1}{n^{2\gamma + p\alpha - 1 + u}} \frac{1}{p!} \bc 
(K^{(p)}(\bc) - \ell)(-1)^p (1+\ve_n) x^2 \\
&& \nonumber \hspace{.2in} + \ \frac{1}{n^{4\gamma + \alpha - 1 + u}} 4  c_4(1+\ve_n) x^4 
+ \ \mbox{O}\!\left(\frac{1}{n^{2\gamma + (p+1)\alpha - 1 + u}}\right)\!x^2 \\
&& \nonumber \hspace{.2in} \ + 
\mbox{O}\!\left(\frac{1}{n^{4\gamma + 2\alpha - 1 + u}}\right)\!x^4
+ \mbox{O}\!\left(\frac{1}{n^{5\gamma - 1 + u}}\right)\!x^5.
\eeas
Given $\alpha > 0$ and choosing $\gamma = (p-1)\alpha/2$ and $u = 1 - (2p-1)\alpha$,
for all $x \in \R$ we obtain the same $n \goto \infty$ limit as in (\ref{eqn:morecareful}).
Using the last display with these values of $\gamma$ and $u$, one easily proves that for any $\alpha > 0$ there exists
$R > 0$ such that for all sufficiently large $n \in \N$ and all $x \in \R$ satisfying
$|x/n^\gamma| < R$
\[
G_n(x) = n^{1-u}\gbnkn(x/n^\gamma) \geq H(x) =  
-2 \ts \frac{1}{p!}\bc |K^{(p)}(\bc) - \ell| x^2 + 2 c_4 x^4.
\]
Since $H(x) \goto \infty$ as $|x| \goto \infty$, hypothesis (iv) of Theorem
\ref{thm:exactasymptotics} is satisfied.  This completes the verification
of the four hypotheses of that theorem.  We now apply the theorem
to conclude that for any $\alpha > 0$, $\mbnkn \sim
\bar{x}/n^\gamma = \barx/n^{(p-1)\alpha/2}$.  Part (c) 
of the present theorem is proved. \ \ink

\skp
In order to derive the asymptotic behavior of $\mbnkn \goto 0$ for the sequence $(\bn,\kn)$
in the last theorem, we choose $\ell \not = K^{(p)}(\bc)$.  
The choice $\ell = K^{(p)}(\bc)$
corresponds to the sequence $(\bn,\kn)$ lying 
on a curve that coincides with the second-order curve to order $p$ in powers of $\beta - \bc$.
In order to analyze this case, we must know the sign of $K^{(p+1)}(\bc)$. 
Because we are unable to determine this sign analytically, the discussion of this case 
is omitted.

This completes the analysis of the asymptotic behavior of $\mbnkn \goto 0$ for the four sequences
considered in Theorems \ref{thm:asymptoticsforc1}--\ref{thm:asymptoticsforc4}.
In the next section we combine several results derived in the present section 
with other calculations to conjecture a number of properties of the first-order curve.

\section{Properties of the First-Order Curve}
\beginsec
\label{section:firstordercurve}

The starting point of the present section is our analysis in section \ref{section:asymptoticsforc}
of the asymptotics of the magnetization $\mbnkn \goto 0$
for sequences $(\bn,\kn)$ converging to the tricritical point $(\bc,\kc)$.
As in section \ref{section:asymptoticsforb}, the structure of the sets of global minimum points
of the associated Ginzburg-Landau polynomials mirrors features
of the phase transitions of the model.  In the present section we operate differently. 
We shall use the structure of the sets of global minimum points of
the Ginzburg-Landau polynomials not to mirror, but to determine 
features of the phase transitions in the subsets of the phase-coexistence region
through which $(\bn,\kn)$ passes.
These features focus on properties
of the first-order curve, properties that are notoriously difficult
to derive rigorously.  Although the insights into the properties of this curve given by the Ginzburg-Landau
polynomials are not rigorous, we back them up with convincing numerical evidence.

The properties of the first-order curve to be presented in this section are stated in the form of three conjectures
involving the first three right-hand derivatives of $K_1(\beta)$ at $\bc$.  These three conjectures
are used in the proof of part (c) of Theorem \ref{thm:asymptoticsforc2} to verify
the asymptotic behavior of $\mbnkn \goto 0$ given there. 

Before using properties of the Ginzburg-Landau polynomials to determine
properties of the first-order curve, in the next theorem we record properties of the function
$K(\beta) = (e^\beta + 2)/4\beta$.
For $0 < \beta < \bc$, $K(\beta)$ defines the second-order curve, while for $\beta \geq \bc$,
$K(\beta)$ defines the spinodal curve.  Parts (e) and (f) express relationships between $K(\beta)$
and $K_1(\beta)$, which for $\beta > \bc$ defines the first-order curve. $K_1(\beta)$ has the properties
given in Theorem \ref{thm:firstorder}. 

\begin{lemma}
\label{lem:kbeta}
The function $K(\beta)= (e^\beta + 2)/4\beta$ has the following properties.

{\em (a)}  For $\beta > 0,$ $K'(\beta) = ((\beta - 1)e^\beta - 2)/(4\beta^2)$.
There exists
$\beta_1 > \beta_c = \log 4$ such that $K'(\beta) < 0$ for $0 < \beta < \beta_1$.

{\em (b)} $4 K(\beta) - e^\beta = -4\beta K'(\beta)$. 

{\em (c)} For $\beta > 0,$ $K(\beta) > 0$ and 
$K''(\beta) = ((\beta^2 - 2\beta + 2)e^\beta + 4)/(4 \beta^3) > 0$.
Thus $K(\beta)$ is a positive, convex function of $\beta > 0$.

{\em (d)}  $K'(\bc) = (2 \bc - 3)/(2 \bc^2) = -0.059166$, 
$K''(\bc) = (\bc^2 - 2\bc + 3)/{\bc^3} = 0.806706$,
and $K'''(\bc) = (\bc^3 - 3\bc^2 + 6\bc - 9)/\bc^4 = -1.024398$.

{\em (e)} For $\beta > \bc,$ $K(\beta) > K_1(\beta)$. 

{\em (f)}  $\lim_{\beta \goto \bc^+} K_1(\beta) = K(\bc)$. 
\end{lemma}

\noi
{\bf Proof.} (a) For $\beta > 0$ we calculate
$K'(\beta) = ((\beta - 1)e^\beta - 2)/(4\beta^2)$. 
For $0 < \beta < \bc$ the numerator satisfies 
\[
(\beta - 1)e^\beta - 2 < (\bc - 1)e^{\bc} - 2 = 4(\log 4 - 1) - 2 = -0.454823 < 0.
\]
By continuity it follows that there exists
$\beta_1 > \beta_c = \log 4$ such that $K'(\beta) < 0$ for $0 < \beta < \beta_1$.

(b) This follows from the formula for $K'(\beta)$ given in part (a).

(c) Clearly $K(\beta) > 0$ for $\beta > 0$.  For $\beta > 0$, we calculate
\[
K''(\beta) = ((\beta^2 - 2\beta + 2)e^\beta + 4)/(4 \beta^3).
\]
This is positive for all $\beta > 0$ since $\beta^2 - 2\beta + 2 > 0$.
It follows that $K(\beta)$
is a positive, convex function of $\beta > 0$. 

(d)  Since $e^{\bc} = 4$, the formulas for $K'(\bc)$, $K''(\bc)$, and
$K'''(\bc)$ follow from parts (a) and (c). 
 
(e) This is proved in Theorem 3.8 in \cite{EllOttTou}. 

(f) We omit the proof, which is based on a number of technical 
results in sections 3.1 and 3.3 of \cite{EllOttTou}. \ \ink

\skp
The first-order curve is defined by the function $K_1(\beta)$ for $\beta > \bc$.
We use part (f) of the last lemma to extend the definition of this function to $\beta = \bc$
by defining $K_1(\bc) = K(\bc)$.  We next state three properties of the first-order 
curve in the form of three conjectures. They involve the first three right-hand
derivatives of $K_1(\beta)$ at $\bc$.  In doing so, we assume that these derivatives
exist and denote them by $K_1'(\bc)$, $K_1''(\bc)$, and $K_1'''(\bc)$.
In combination with the definition $K_1(\bc) = K(\bc)$, 
the first two conjectures are consistent with the fact that $K_1(\beta) < K(\beta)$
for all $\beta > \bc$ [Lem.\ \ref{lem:kbeta}(e)]. Our main goal in this section
is to use properties of the appropriate Ginzburg-Landau polynomials plus numerical evidence
to support these conjectures.

\skp
\noi
{\bf Conjecture 1.}  The right-hand derivative $K_1'(\bc)$ exists, and $K_1'(\bc) = K'(\bc)$; 
i.e., at $\bc$ the first-order curve and the spinodal curve have the same right-hand tangent. 
Numerically
\[
K_1'(\bc) = K'(\bc) = {1}/{\bc} - {3}/({2\bc^2}) = -0.059166.
\]

\skp
\noi
{\bf Conjecture 2.}  The right-hand derivative $K_1''(\bc)$ exists, and $K_1''(\bc) < 0 < K''(\bc)$; numerically,
\beas
K_1''(\bc)  & = &  K''(\bc) - {5}/({4\bc}) = 
 -{1}/({4\bc}) - {2}/{\bc^2} + {3}/{\bc^3} = -0.094979 \\
& < & K''(\bc) = {1}/{\bc} - {2}/{\bc^2} + {3}/{\bc^3} = 0.806706.
\eeas

\skp
\noi
{\bf Conjecture 3.}  The right-hand derivative $K_1'''(\bc)$ exists, and $K_1'''(\bc) > 0$; numerically,
\[
K_1'''(\bc) = {219}/{(224 \beta _c)} + {3}/{(4 \beta _c^2)} + {6}/{\beta _c^3} - {9}/{\beta _c^4}
= 0.910784.
\]

\skp
Support for Conjecture 1 comes from the form of the Ginzburg-Landau polynomial
when the sequence $(\bn,\kn)$ in Theorem \ref{thm:asymptoticsforc1} 
satisfies $\bn > \bc$ and converges
to the tricritical point $(\bc,\kc)$ along a ray lying under the tangent line to the spinodal
curve $\{(\beta,K(\beta)), \beta > \bc\}$ at the tricritical point. Given $\alpha > 0$
and $k \in \R$, the sequence has the form 
\[
\beta_n = \bc + 1/{n^\alpha} \ \mbox{ and } \ K_n = K(\bc) + {k}/{n^\alpha}.
\]
The requirement that the sequence converges along a ray lying under the tangent line
means that $k < K'(\bc)$.  In this case the Ginzburg-Landau polynomial 
is given by $g(x) = c_6 x^6 + \bc(K'(\bc) - k) x^2$, where $c_6 = 9/40$.  
This has the same form as the 
Ginzburg-Landau polynomial in part (b) of Theorem \ref{thm:asymptoticsforc1} except
that there $K'(\bc) - k < 0$ and here $K'(\bc) - k > 0$.  With this choice of sign
the Ginzburg-Landau polynomial has a unique global minimum point at 0, implying
that the sequence $(\bn,\kn)$ approaches the tricritical point from the single-phase
region. However, as Theorem \ref{thm:firstorder} indicates, when $\beta > \bc$, 
the single-phase region is located under the first-order curve $(\beta,K_1(\beta))$.
It follows that at the tricritical point the right-hand tangent line to the first-order curve
coincides with or lies above the right-hand tangent line to the spinodal curve; i.e.,
$K_1'(\bc) \geq K'(\bc)$. On the other hand, parts (e) and (f) of Lemma \ref{lem:kbeta}
imply that for any $\beta > \bc$
\[
\frac{K_1(\beta) - K_1(\bc)}{\beta - \bc} <  \frac{K(\beta) - K(\bc)}{\beta - \bc}
\]
and thus that $K_1'(\bc) \leq K'(\bc)$.  We conclude that $K_1'(\bc) = K'(\bc)$,
which is the content of Conjecture 1.

By a more detailed argument that involves the uniform convergence of the scaled
free-energy functionals to the Ginzburg-Landau polynomial, we can prove
Conjecture 1 under the assumption that the right-hand derivative $K_1'(\bc)$ exists. Again, because
of parts (e) and (f) of Lemma \ref{lem:kbeta} it suffices to prove that
$K_1'(\bc) \geq K'(\bc)$. We carry this out by showing that $K_1'(\bc) < K'(\bc)$
leads to a contradiction. If this strict inequality holds, then we consider the
same sequence $(\bn,\kn) \goto (\bc,\kc)$ in Theorem \ref{thm:asymptoticsforc1}
that we considered in the preceding paragraph, choosing $k$ to satisfy
$K_1'(\bc) < k < K'(\bc)$.  According to Theorem
\ref{thm:asymptoticsforc1}, for any $\alpha > 0$, $\gamma = \alpha/4$, and $u = 1 - 3\alpha/2$,
$G_n(x) = n^{1-u}\gbnkn(x/n^\gamma)$ converges uniformly to $g(x)$ uniformly for $x$
in compact subsets of $\R$. Since $k > K_1'(\bc)$, for all sufficiently large $n$, $(\bn,\kn)$
lies in the phase-coexistence region located above the first-order curve. Hence the global
minimum points of $\gbnkn$ are the symmetric pair $\pm \mbnkn$ [Thm.\ \ref{thm:firstorder}(c)].  It
follows that the global minimum points of $G_n$ are the symmetric pair $\pm \bar{m}_n = \pm n^\gamma \mbnkn$ [see (\ref{eqn:infgn})],
which converge to $\pm \barx$ as $n \goto \infty$ [Thm.\ \ref{thm:asymptoticsforc1}(b)].

In order to complete the proof, we appeal to several standard results in the theory of analytic functions.  
Since $k < K'(\bc)$, as a function of $z \in \C$ the derivative
$g'(z) = 6 c_6 z^5 + 2\bc(K'(\bc) - k) z$ of the Ginzburg-Landau polynomial has 1 real zero at $z = 0$
and 4 nonreal zeroes at the 4 fourth roots of $\bc(k - K'(\bc))/3c_6$. 
There exists an open set $V$ in the complex plane having the following properties: the boundary of $V$ is a smooth, simple,
closed curve; $V$ contains
the set $\{z \in \C : \Re(z) \in [-2\barx,2\barx], \Im(z) = 0\}$, in which the real zero of $g'$ at $z=0$ lies,
but $V$ does not contain the 4 nonreal zeroes of $g'$; $G_n$ and $g$ are analytic on $V$; as 
$n \goto \infty$, $G_n(z) \goto g(z)$ uniformly for $z \in V$.  
It follows that as $n \goto \infty$, $G_n'(z) \goto g'(z)$ 
uniformly for $z$ in any closed disk contained in $V$
\cite[Thm.\ 3.1.8(i)]{MarHof}. Furthermore, by a corollary of Rouch\'{e}'s
Theorem \cite[p.\ 389]{MarHof}, 
for all sufficiently large $n$, $G_n'$ has the same number of zeroes in $V$ as $g'$, namely 1.
However, this contradicts the fact that for all sufficiently large $n$, $G_n'$ has two zeroes in V
at $\pm \bar{m}_n$.  The contradiction shows that the inequality $K_1'(\bc) < K'(\bc)$ is not valid
and thus that $K_1'(\bc) \geq K'(\bc)$, which is what we want to show.  This completes the proof of Conjecture 1. 

Support for Conjecture 2 comes from Theorem \ref{thm:asymptoticsforc2}.
Given $\alpha > 0$, $\ell \in \R$, and $\tell \in \R$, in that theorem we 
consider the sequence
\[
\beta_n = \bc + {1}/{n^\alpha} \ \mbox{ and } \ K_n = \kbc + K'(\bc)/{n^\alpha}
+ \ell/(2n^{2\alpha}) + \tell/(6n^{3\alpha}).
\]
This sequence converges to the tricritical point along the curve 
$(\beta,\tilde{K}(\beta))$, where
\[
\tilde{K}(\beta) = \kbc + \kprimebc (\beta - \bc) + \ell (\beta - \bc)^2/2 + \tell (\beta - \bc)^3/6.
\]
For $\beta > \bc$, for points $(\beta,K_1(\beta))$ on the first-order curve
the free-energy functional $G_{\beta,K_1(\beta)}$ has three global minimum points
at 0 and $\pm m(\beta,K_1(\beta))$ [Thm.\ \ref{thm:firstorder}(b)].
In addition, as we determine in part (c) of Theorem \ref{thm:asymptoticsforc2}, 
when $\ell = \ell_c = (-\bc^2 - 8 \bc + 12)/4\bc^3$ the limiting
Ginzburg-Landau polynomial $g_\ell$ has three global minimum points at 0 and $\pm (5/3)^{1/2}$. 
This analogous structure of global minimum points both for $G_{\beta,K}$
when $K = K_1(\beta)$ and for $g_\ell$ when $\ell = \ell_c$ suggests the following
conclusion: when $\ell = \ell_c$,
the curve $(\beta,\tilde{K}(\beta))$ along which the given sequence $(\bn,\kn)$
converges to the tricritical point coincides with the first-order curve $(\beta,K_1(\beta))$ 
to order 2 in powers of $\beta - \bc$.  If this is true, then it follows that when $\ell = \ell_c$,
\[
K_1''(\bc) = \tilde{K}''(\bc) = \ell_c.
\]
Since $\ell_c = K''(\bc) - 5/4\bc$ [see (\ref{eqn:ellc})], Conjecture 2 follows.

\iffalse
HERE ARE JON'S COMMENTS ABOUT THE CONJECTURES DATED 1/17/2008.
 
Here are the first three derivatives of the first-order curve at the tricritical point obtained from Mathematica.  The first and second derivatives are correctly obtained from a 6th order expansion of the free energy around zero.  However, the third derivative requires an 8th order expansion of the free energy. I also carried out a 10th order expansion that yielded the same result for all three derivatives.  The full expression for the $K_1(\beta)$ obtained from the 8th order expansion is too unwieldy to display.  It is notable that the correct evaluation of $K_1'''$ is one (and the only?) situation where it was necessary to go beyond 6th order.

\[
K_1'(\bc)=\frac{1}{\beta _c}-\frac{3}{2 \beta _c^2}=\frac{-3+\log (16)}{2 \log ^2(4)}=-0.0591658
\]

\[
K_1''(\bc)=-\frac{1}{4 \beta _c}-\frac{2}{\beta _c^2}+\frac{3}{\beta _c^3}=-\frac{-12+8 \log (4)+\log ^2(4)}{4 \log ^3(4)}=-0.094979
\]

\[
K_1'''(\bc)=\frac{219}{224 \beta _c}+\frac{3}{4 \beta _c^2}+\frac{6}{\beta _c^3}-\frac{9}{\beta _c^4}
=\frac{3 \left(-672+448 \log (4)+56 \log ^2(4)+73 \log ^3(4)\right)}{224 \log ^4(4)}
=0.910784
\]

\skp
\noindent
THIS ENDS JON'S COMMENTS ABOUT THE CONJECTURES.

\fi

We complete this section by citing numerical evidence that supports all three conjectures. Let $c_\beta$
be the function defined in (\ref{eqn:cbeta}). For $\beta > \bc$
the function $K_1(\beta)$ defining the first-order curve is determined by the property that the global minimum points
of $G_{\beta,K_1(\beta)}(x) = \beta K_1(\beta) x^2 - c_\beta(2\beta K_1(\beta) x)$ are 0 and $\pm m$ for
$m = m(\beta,K_1(\beta)) > 0$ [Thm.\ \ref{thm:firstorder}(b)].  This holds if and only if 
\be 
\label{eqn:simultaneous}
G_{\beta,K_1(\beta)}(m) = 0 \ \mbox{ and }
\ G_{\beta,K_1(\beta)}'(m) = 0.  
\ee
Approximating $c_\beta(2\beta K_1(\beta) x)$ by its Taylor expansion to order 6, we solve the equations
in the last display, obtaining an approximation $\bark_1(\beta)$ to $K_1(\beta)$ having the following form:
\be 
\label{eqn:bark1}
\bark_1(\beta) = \frac{2(2 + e^\beta)(64 - 26 e^\beta + e^{2\beta})}{3\beta(144 - 56e^\beta + e^{2\beta})}.
\ee
When $\beta = \bc = \log 4$, we have $\bark_1(\bc) = 3/2\bc = K(\bc)$. 
This is consistent
with the definition of $K_1(\bc) = K(\bc)$ given before Conjecture 1. 
The formula for $\bark_1(\beta)$ is also consistent with the values of $K_1'(\bc)$ and $K_1''(\bc)$
given in Conjectures 1 and 2.  In fact, we calculate
\[
\bark_1'(\bc) = {1}/{\bc} - {3}/(2\bc)^2 = K'(\bc) \ \mbox{ and } \ 
\bark_1''(\bc) = -{1}/(4\bc) - {2}/{\bc^2} + {3}/{\bc^3}.
\]
However, the evidence cited in the next paragraph suggests that $\bark_1'''(\bc)$ is not the correct
value of $K_1'''(\bc)$.

In order to calculate numerically the value of $K_1'''(\bc)$, we approximate 
$c_\beta(2\beta K_1(\beta) x)$ by its Taylor expansion to order 8
and solve equations (\ref{eqn:simultaneous}), obtaining an approximation $\hat{K}_1(\beta)$ 
to $K_1(\beta)$ that is too complicated to display here.  $\hat{K}_1'''(\bc)$ is the value given in Conjecture 3.
Like $\bar{K}_1$ in (\ref{eqn:bark1}), $\hat{K}_1(\bc) = K(\bc)$, $\hat{K}_1'(\bc) = K'(\bc)$, and $\hat{K}_1''(\bc) = K''(\bc)$;
however, $\hat{K}_1'''(\bc) = 0.910784 < 4.53362 = \bar{K}_1'''(\bc)$.  
If we approximate $c_\beta(2\beta K_1(\beta) x)$ by its Taylor expansion to order 10
and solve equations (\ref{eqn:simultaneous}), then the resulting approximation to $K_1(\beta)$ has 
the same value at $\bc$ and the same first three right-hand derivatives at $\bc$ as $\hat{K}_1(\beta)$.
It is reasonable to assume that the same properties hold for any approximation to $K_1(\beta)$
that arises by replacing $c_\beta(2\beta K_1(\beta) x)$ by its Taylor expansion to order 12 or higher.

This completes our discussion of properties of the first-order curve that are consistent 
with properties of the Ginzburg-Landau polynomials plus numerical evidence.
In the next section we relate the results obtained in this paper to the scaling theory of critical phenomena.

\section{Relationship with Scaling Theory of Critical Phenomena}
\beginsec
\label{section:scalingtheory}

The results on the asymptotic behavior of $\mbnkn$ obtained in sections \ref{section:asymptoticsforb} and \ref{section:asymptoticsforc} are related to scaling theory for critical and tricritical points \cite{Rie,Stanley71}.  In this section we review scaling theory and show that its predictions for the magnetization are consistent with the results obtained in those sections.  

Scaling theory is based on the idea that the singular parts of thermodynamic functions near continuous phase transitions are homogeneous functions of the distance to the phase transition.  If there is a single parameter controlling the approach to the phase transition, then the content of scaling theory for a single thermodynamic quantity is simply that its singularities are power laws.  If there is more than one parameter, as is the case here, then scaling theory has a richer content, especially near the tricritical point where the type of phase transition changes in a small neighborhood.

We are interested in the magnetization $m$ as a function of $(\bn,\kn)$, a sequence converging either to a second-order point
$(\beta,K(\beta))$ with $0 < \beta < \bc$ or to the tricritical point $(\beta,K(\beta)) = (\beta_c,K(\beta_c))$. In either case, the relevant parameter-space is two dimensional.  Given any phase-transition point $(\beta,K(\beta))$ with $0 < \beta \leq \beta_c$, the natural coordinate system for scaling theory is a curvilinear system $(\mus,\mut)$ measuring the signed distances from the phase transition point; $\mus$ is the signed distance from the curve of phase transitions and $\mut$ the signed distance from the chosen point along the curve of phase transitions. 
Since we are concerned with the phase-coexistence
region, in all our considerations $\mus\geq 0$; however, $\mut$ may take either sign.  At the tricritical point, $\mut>0$ and $\mus=0$ correspond to the first-order line to the right of the tricritical point while $\mut<0$ and $\mus=0$ correspond to the second order line to the left of the tricritical point. 
At a second-order point, for sufficiently small $|\mut |$, $(0,\mut)$ is also a second-order point. 
Figure 4 shows this coordinate system for the special case of the tricritical point. 

\begin{figure}[h]
\begin{center}
\includegraphics[width=10cm]{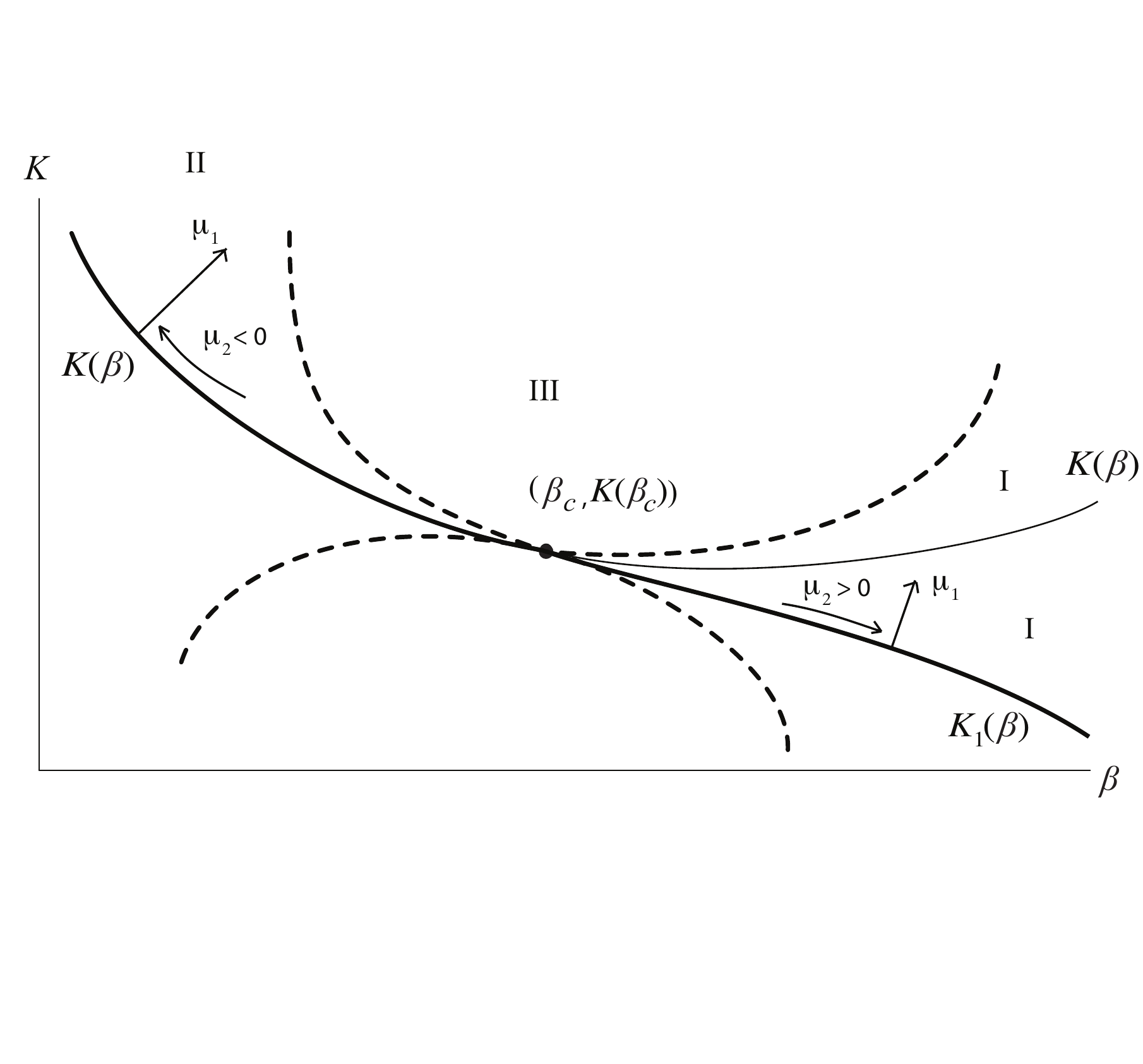}
\vspace{-.9in}
\caption{\footnotesize Curvilinear coordinate system for scaling theory showing the coordinates $\mus$ and $\mut$;
$\mus$ is the signed distance from the phase transition line and $\mut$ the signed distance from the tricritical point along the phase transition line.  Regions I, II, and III are dominated, respectively, by the first-order, second-order, and tricritical 
phase transition.  A similar coordinate system can be defined for any point along the second-order curve.}
\end{center}
\end{figure} 

Scaling theory for the magnetization in a two-dimensional parameter space takes the general form
\begin{equation}
\label{eqn:GenericScaling}
\m(\mus \tll, \mut \tll^\a)  = \tll^\b \m(\mus, \mut),
\end{equation}
where $\tll$ is an arbitrary scale factor and $\a$ and $\b$ are exponents to be determined \cite{Rie}.  The exponents $\a$ and $\b$ are chosen so that the theory is consistent with known exponents for the particular type of phase transition.  In our case, $\a$ and $\b$ depend on whether the phase transition point is a second-order point or the tricritical point.  

We first consider the simpler case of a second-order point.  Then the neighboring points along the phase-transition curve are also second-order points, and there is no singular dependence on $\mut$, implying that $\a=0$.  The singular behavior of the magnetization is controlled 
by $\tbe$, the mean-field magnetization exponent for second-order transitions, which has the value $\tbe=1/2$ \cite{Stanley71}. Choosing 
$\b= \tbe=1/2$, we obtain from (\ref{eqn:GenericScaling})
\be 
\label{eqn:SecondOrderScaling}
\m(\mus \tll, \mut)  = \tll^{\tbe} \m(\mus, \mut) = \tll^{1/2} \m(\mus, \mut).
\ee
Setting $\tll=1/\mus$ yields
\begin{equation}
\label{eqn:SecondOrderScaling2}
\m(\mus , \mut)  = \mus^{\tbe} m(1,\mu_2) = \mus^{1/2} f( \mut);
\end{equation}
$f(\mut)$ is a smooth function of $\mut$ that depends on the chosen point $(\beta,K(\beta))$, and 
the critical amplitude $f(0)$ is presumed to be positive. 
Equation (\ref{eqn:SecondOrderScaling2}) reflects the standard power-law behavior of the magnetization near a critical point.

We now show that (\ref{eqn:SecondOrderScaling2}) is consistent with Theorems \ref{thm:asymptoticsforb} and
\ref{thm:asymptoticsforb2}. These theorems give the exact asymptotic behavior of $m(\bn,\kn)$ for sequences $(\bn,\kn)$ 
converging to a second-order point. For ease of exposition, we refer to the definitions of the sequences according to the labeling in Figure 2
in the introduction, calling them sequences of type 1 and type 2, respectively.  
\iffalse
%Possible paths for these sequences are shown in Figure 2. 
\fi

We first consider the sequence of type 1, which converges to a second-order point $(\beta_0,K(\beta_0))$
along a ray that is above the tangent line to the second-order curve at that point.  Defined in (\ref{eqn:newbetankn}),
this sequence takes the form
\be
\label{eqn:seqtype1} 
\beta_n = \beta_0 + {b}/{n^\alpha} \ \mbox{ and } \ K_n = K(\beta_0) + {k}/{n^\alpha},
\ee
where $b \in \{1,0,-1\}$ and $K'(\beta_0)b - k < 0$. To leading order, the coordinate $\mus$ is given by the distance to the tangent to the second-order curve
at $(\beta_0,K(\beta_0))$; i.e.,
\begin{equation}
\label{eqn:mus}
\mus  \approx (K-K(\beta_0))-K'(\beta_0)(\beta-\beta_0).
\end{equation}
Hence we obtain
\begin{equation}
\label{eqn:mus2}
\mus  \approx (k-K'(\beta_0)b)/n^\alpha.
\end{equation}
The distance $\mut$ is also of order $1/n^\alpha$. However, $f$ is a smooth function of $\mut$ that converges to $f(0) > 0$ as $\mut \goto 0$.
Hence we need only know that $\mu_2 \goto 0$ in order 
to obtain the leading-order behavior of $\m$ from (\ref{eqn:SecondOrderScaling2}) and (\ref{eqn:mus2}), namely,
$
\m \approx (k-K'(\beta_0)b)^{1/2}/n^{\alpha/2}.
$
This asymptotic formula is consistent  with the exact asymptotic behavior of $\mbnkn$ given in Theorem \ref{thm:asymptoticsforb}, 
correctly predicting both the exponent of $n$ and the 
dependence on $k$ and $b$ in the prefactor $\bar{x}$ as given in (\ref{eqn:barxforb}) with $\beta = \beta_0$.

We next consider the sequence of type 2, which converges to a second-order point $(\beta_0,K(\beta_0))$ along a curve lying in the
phase-coexistence region and having the same tangent as the second-order curve at that point.  
Defined in (\ref{eqn:bnknb2}), this sequence takes the form
\be
\label{eqn:seqpscaling}
\bn = \beta_0 + b/{n^\alpha} \ \mbox{ and } \
\kn = K(\beta_0) + \sum_{j=1}^{p-1} K^{(j)}(\beta_0)b^j/(j! n^{j\alpha}) +  \ell b^p/(p! n^{p\alpha}),
\ee
where $b \in \{1,-1\}$, $p \geq 2$, and $(K^{(p)}(\beta_0) - \ell)b^p < 0$.
In this case it is crucial to recall that the scaling variables comprise a curvilinear coordinate system. 
In particular, the coordinate $\mus$ measures the distance from the second-order curve, not the distance from the tangent to this 
curve at $(\beta_0,K(\beta_0))$; as a result
(\ref{eqn:mus}) is not sufficient to determine the asymptotic behavior of $m$. The sequence of type 2 converges to the second-order point along a curve that agrees with the second-order curve 
to order $p-1$ in powers of $\beta - \beta_0$. Hence to leading order $\mus$ is proportional to the difference between the 
last term in the definition of $\kn$ and the term of order $p$ in the Taylor expansion of $K(\bn)$, namely,
$
\mus \approx |\ell-K^{(p)}(\beta_0)|/(p! n^{p\alpha}).
$ 
Substituting this expression into (\ref{eqn:SecondOrderScaling2}) yields
\[
% \label{eqn:SecondOrderScaling2p}
\m \approx (|\ell-K^{(p)}(\beta_0)| / p! n^{p\alpha})^{\tbe} = (|\ell-K^{(p)}(\beta_0)|/p!)^{1/2} / n^{p\alpha/2}.
\]
Again, this asymptotic formula is consistent  with the exact 
asymptotic behavior of $\mbnkn$ given in Theorem \ref{thm:asymptoticsforb2} and correctly captures the square-root dependence
of the prefactor $\barx$ on 
$|\ell-K^{(p)}(\beta_0)|$ given in (\ref{eqn:barx2b}) with $\beta = \beta_0$.

If there is more than one type of phase transition in a neighborhood of a phase-transition point, as is the case near a tricritical point, then scaling theory becomes more complicated \cite{Rie}.  In the case of the tricritical point, the theory involves crossovers between the nearby first-order, second-order, and tricritical phase transitions.  Figure 4 shows the region near the tricritical point. 
\iffalse
%The curvilinear coordinate system with coordinates $\mus$ and $\mut$ measures the distances from the tricritical point; $\mus$ is the signed distance from the second-order curve to the left of the tricritical point and the signed distance from the first-order curve to the right of the tricritical point while $\mut$ is the signed distance along those phase-transition curves. The quantity $\mut$ is positive on the first-order side and negative on the second-order side of the tricritical point.  Since we focus on the phase-coexistence region, $\mus$ is always nonnegative.  
\fi

The three regions I, II, and III separated by dotted lines are controlled by the first-order, the second-order, and the tricritical phase transitions, respectively.  
The mean-field tricritical crossover exponent $\phit$ determines the boundaries of the regions.  In regions I and II we have 
$|\mus | \ll |\mut|^{1/\phit}$ while in region III  $|\mus | \gg |\mut|^{1/\phit}$.  
In region II the magnetization $\m$ is controlled by $\tbe$, the mean-field magnetization exponent for
second-order transitions. In region III the magnetization $\m$ is controlled by $\tbe_t$,
the mean-field magnetization exponent for tricritical transitions, while in region I the magnetization $\m$
approaches a constant value as the first-order line is approached.  These insights are incorporated in the scaling hypothesis  
\begin{equation}
\label{eqn:RiedelScaling}
\m(\mus \tll, \mut \tll^{\phit})  = \tll^{\tbe_t} \m(\mus, \mut),
\end{equation}
where  $\tll$ is an arbitrary scale factor \cite{Rie}. 
This corresponds to (\ref{eqn:GenericScaling}) with $a = \phit$ and $b=\tbe_t$.
Setting $\tll=|\mut|^{-1/\phit}$ yields the alternate form
\begin{equation}
\label{eqn:RiedelScaling2}
\m(\mus, \mut )  =|\mut|^{\tbe_t/\phit} m(\mus/| \mut|^{1/\phit},1) =|\mut|^{\tbe_t/\phit} f_\pm(\mus/ |\mut|^{1/\phit}),
\end{equation}
where $f_+$ is used on the first-order side of the tricritical point ($\mut>0$) and $f_-$ is used 
on the second-order side of the tricritical point ($\mut<0$). 
The values of the three relevant mean-field exponents are $\phit=1/2$, $\tbe=1/2$, and $\tbe_t=1/4$ \cite{RieWeg}. 

We now consider the form taken by the right side of (\ref{eqn:RiedelScaling2}) in each of the three regions.
In region III the arguments of $f_+$ and of $f_-$ are large. 
Hence in order to recover the tricritical power-law behavior of $\m$
we require that $f_+(x) \approx x^{\tbe_t}$ and $f_-(x) \approx x^{\tbe_t}$ as $x \goto \infty$, yielding
\begin{equation}
\label{eqn:TricriticalScaling}
\m(\mus, \mut ) \approx \mus^{\tbe_t} = \mus^{1/4} \ \ [\mbox{region III}].
\end{equation}

In region II with fixed $\mut$ we expect that the scaling is the one given in (\ref{eqn:SecondOrderScaling})
for the second-order curve; i.e., 
\begin{equation}
\label{eqn:SecondOrderScalingagain}
\m(\mus \tll, \mut )  = \tll^{\tbe} \m(\mus, \mut) = \tll^{1/2} \m(\mus, \mut).
\end{equation}
The requirement that the two scaling assumptions (\ref{eqn:RiedelScaling2}) and (\ref{eqn:SecondOrderScalingagain})
are consistent yields an interesting result for the behavior of $\m$ in region II for small  $|\mut|$.  
The asymptotic behavior of $f_-(x)$ as $x \goto 0^+$ must be of the form $x^{\tbe}$ in order that second-order scaling is recovered. 
Thus in region II we find
\begin{equation}
\label{eqn:RegionII}
\m(\mus , \mut )  \approx \mus^{\tbe}|\mut|^{(\tbe_t-\tbe)/\phit} = \mus^{1/2}|\mut|^{-1/2}
\ \ [\mbox{region II}].
\end{equation}

Near the first-order curve in region I, for small positive $\mut$ a similar result can be obtained except that $\m(\mus , \mut )$ must converge to a constant as $\mus \rightarrow 0^+$.  
For (\ref{eqn:RiedelScaling2}) to be consistent with first-order behavior, $f_+(x)$ must also converge to a constant as $x \goto 0^+$.  
Hence along the first-order curve, which is defined by $\mus=0$ and $\mut>0$, we have 
\begin{equation}
\label{eqn:RegionI}
\m(0 , \mut )  \approx \mut^{\tbe_t/\phit} =  \mut^{1/2} \ \ [\mbox{region I}].
\end{equation}

We now show that these results in tricritical scaling theory are consistent with 
Theorems \ref{thm:asymptoticsforc1}--\ref{thm:asymptoticsforc4}.  
\iffalse
%These theorems give the exact asymptotic behavior of $\mbnkn$
%for the sequences $(\bn,\kn)$ in those theorems converging to the tricritical point. For ease of exposition, we refer to the definitions of the sequences in items 3--6 in the introduction, calling them sequences of type 3, 4, 5, and 6.
%Possible paths for these sequences are shown in Figure 2.  
\fi

We first consider the sequence of type 3, which converges to the tricritical point along a ray that is above the tangent line
to the phase-transition curve at the tricritical point. This sequence is defined as in (\ref{eqn:seqtype1}) 
with $\beta_0$ replaced by $\bc$, $b \in \{1,0,-1\}$, and $K'(\bc)b - k < 0$. 
For this sequence (\ref{eqn:mus2}) holds with $\beta_0$ replaced by $\beta_c$.  Thus
$
\mus  \approx (k-K'(\beta_c)b)/n^\alpha,
$
and $\mut$ is of order $1/n^\alpha$. Since this sequence lies in region III, the asymptotic formula 
(\ref{eqn:TricriticalScaling}) predicts $\m \approx \mus^{\tbe_t} \approx (k-K'(\beta_c)b)^{1/4}/n^{\alpha/4}$.  This asymptotic formula is consistent with the exact
asymptotic behavior of $\mbnkn$ given in Theorem \ref{thm:asymptoticsforc1} and 
correctly predicts the 1/4-power dependence on $(k-K'(\beta_c)b)$ given in (\ref{eqn:barxforc1}). 

The sequence of type 4 is defined in (\ref{eqn:bnkn2a}) in terms of real parameters 
$\ell$ and $\tell$.  For $\ell > \ell_c = K''(\bc) - 5/(4\bc)$ and 
\iffalse
\fi
appropriate choices of $\tell$, 
the sequences of type 4a, 4b, and 4d converge to the tricritical point  in the crossover 
region between regions I and III in a neighborhood of the first-order curve.
For these sequences $\mut \approx 1/n^\alpha$ and $\mus \approx (\ell-\ell_c)/n^{2\alpha}$. 
Hence the scaling expression for the magnetization in (\ref{eqn:RiedelScaling2}) becomes
\[
% \label{eqn:4abd}
\m \approx n^{-\alpha \tbe_t/\phit} f_+(\ell-\ell_c) = f_+(\ell-\ell_c)/n^{\alpha /2}. 
\]
We note that $n$ does not appear in the argument of $f_+$ since $1/\phit=2$ and the powers of $n$ cancel. 
This asymptotic
formula is consistent with the exact asymptotic behavior of $\mbnkn$ given in Theorem \ref{thm:asymptoticsforc2}. 
%As we point out in items 4a--4d, this sequence converges to the tricritical point from four different subsets of the phase-coexistence region depending on the values of these parameters.
%Sequences of type 4a, 4b, and 4d are defined by $\ell > \ell_c$, where $\ell_c = K''(\bc) - 5/(4\bc)$
%is conjectured to equal the second derivative of the first-order curve at the
%tricritical point (see Conjecture 2 in section 6 of \cite{EllMacOtt1}). 

The sequence of type 4c is defined in (\ref{eqn:bnkn2a}) with $\ell = \ell_c$ and $\tell > K_1'''(\bc)$.
\iffalse
%If Conjectures 1--3 in section 6 of \cite{EllMacOtt1} are valid, 
%then 
\fi
We conjecture that this sequence converges to the tricritical point along a curve that coincides with the 
first-order curve to order 2 in powers of $\beta - \bc$ and lies in the phase-coexistence region
for all sufficiently large $n$. 
Thus when $\ell=\ell_c$, $\mus \approx 0$ and (\ref{eqn:RegionI}) holds.  Since  $\mut \approx 1/n^\alpha$, we have 
$\m \approx \mut^{1/2} \approx 1/n^{\alpha/2}$.  This result is
consistent with part (c) of Theorem  \ref{thm:asymptoticsforc2}. 

The sequences of type 5 and type 6 approach the tricritical point along a curve that coincides with the second-order
curve to order $p-1$ in powers of $\beta - \bc$.
These sequences are defined in terms of a parameter $\ell$ as in (\ref{eqn:seqpscaling}) with $\beta_0$ replaced by $\bc$ and $b=-1$; 
the sequence of type 5 corresponds to the choice $p = 2$ while
the sequence of type 6 corresponds to $p \geq 3$. Since $1/\varphi_t=2$, the dotted line separating regions II and III deviates quadratically from the second-order curve.  Thus the sequence of type 5, defined for
$\ell > K''(\bc)$, lies in the crossover range between region II and region III. The sequence of type 6 lies within region II since it approaches the second-order curve faster than quadratically.  For a sequence of type 5 we have 
$\mus \approx (\ell - K''(\bc)) / n^{2\alpha}$ and $\mut \approx 1/n^{\alpha}$.  From the general expression (\ref{eqn:RiedelScaling2}) we obtain
\[
% \label{eqn:seq5}
\m \approx n^{-\alpha \tbe_t/\phit} f_-(\ell - K''(\bc)) = n^{-\alpha /2} f_-(\ell - K''(\bc)).
\]
Since $f_-(x) \approx x^{\tbe}$, for small $x$ we find that $\m \approx (\ell - K''(\bc))^{1/2}/n^{\alpha/2}$.
This asymptotic formula is consistent with the exact asymptotic behavior of $\mbnkn$ given in Theorem \ref{thm:asymptoticsforc3}.
It captures the correct dependence of the prefactor $\barx$ on $\ell-K''(\bc)$ for small $\ell-K''(\bc)$ that follows from (\ref{eqn:barxforc3a}).

The sequence of type 6 is defined as in (\ref{eqn:seqpscaling}) with 
$\beta_0$ replaced by $\bc$, $p \geq 3$, $b=-1$, and $(K^{(p)}(\bc) - \ell)(-1)^p < 0$.  Because this sequence
converges to the tricritical point in region II, the scaling expression 
(\ref{eqn:RegionII}) is valid.  In this case $\mus  \approx |\ell - K^{(p)}(\bc)|/n^{p \alpha}$ and $\mut \approx 1/n^\alpha$.  
Substituting these values into (\ref{eqn:RegionII}) yields
\[
% \label{eqn:seq6}
\m \approx( |\ell - K^{(p)}(\bc)|/p!)^{1/2} / n^{(p-1)\alpha/2}.
\]
Once again this asymptotic formula is consistent  with the exact asymptotic behavior of $\mbnkn$
given in Theorem \ref{thm:asymptoticsforc4}.  We note 
that scaling theory predicts the correct square-root dependence of the prefactor $\barx$ on $|\ell - K^{(p)}(\bc)|$ given in
(\ref{eqn:barx4a}).

This completes the discussion of the relationship between the results obtained in 
sections \ref{section:asymptoticsforb} with 
scaling theory for critical and tricritical points \cite{Stanley71}.  We have shown that scaling theory, together with the known mean-field exponents, predicts many of the exact results for $\mbnkn$, capturing both the correct power laws and, in some cases, the dependence on the parameters defining the sequences.    

In the next two appendices we address a number of issues related to several results
contained in the main body of the paper.      

\vspace{.5in}
\noi
\LARGE {\bf Appendix}
\vspace{-.2in}
\normalsize
\appendix
  
\renewcommand{\thesection}{\Alph{section}}          
\renewcommand{\theequation}          
{\Alph{section}.\arabic{equation}}          
\renewcommand{\thedefn}          
{\Alph{section}.\arabic{defn}}          
\renewcommand{\theass}          
{\Alph{section}.\arabic{ass}}   

\section{Properties of Polynomials of Degree 6}
\setcounter{equation}{0}

The main purpose of this section is to analyze the structure of the set of global minimum points
of polynomials of degree 6 having the form $g(x) = a_2 x^2 - a_4 x^4 + a_6 x^6$, 
where $a_2 \in \R$, $a_4 > 0$, and $a_6 > 0$.  We use this information in
section \ref{section:phasetr} in order to motivate the phase-transition
structure of the mean-field B-C model via the Ginzburg-Landau phenomenology.
We also use it in the discussion leading up to Theorem \ref{thm:asymptoticsforc2}
when we show how the predictions of the Ginzburg-Landau phenomenology
can be made rigorous via properties of the Ginzburg-Landau polynomials.  
Specifically, items 1--4 preceding Theorem \ref{thm:asymptoticsforc2}
exhibit the structure of the set of global minimum points
of the associated Ginzburg-Landau polynomial, which precisely mirror
the phase-transition structure of the model. 
We end the section by briefly considering the polynomial $h(x) = a_2 x^2 + a_4 x^4 + a_6 x^6$,
where $a_2 \in \R$, $a_4 > 0$, $a_6 > 0$. 

For variable $a_2 \in \R$ and fixed $a_4 > 0$ and $a_6 > 0$,
the analysis of the set of global minimum points of 
$g(x) = a_2 x^2 - a_4 x^4 + a_6 x^6$ is given
in Theorem \ref{thm:6order} in terms of the quantity
$a_c = a_4^2/4a_6$.  Here are the main features. If $\atwo > a_c$, 
then $g$ has a unique global
minimum point at 0; if $\atwo = a_c$, 
then the global minimum points of $g$ are 0 and nonzero numbers $\pm \barx(a_c)$;
and if $\atwo < a_c$, then the global minimum points of $g$ are 
nonzero numbers $\pm \barx(\atwo)$, where the positive number $\barx(\atwo)$
converges to the positive number $\barx(a_c)$ as $\atwo \goto (a_c)^+$. 
Thus the set of global minimum points of $g$ undergoes
a discontinuous bifurcation at $a_c$, changing 
discontinuously from $\{0\}$ for $\atwo < a_c$ to $\{0,\pm \barx(a_c)\}$ for
$\atwo = a_c$ to $\{\pm \barx(\atwo)\}$ for $\atwo > a_c$.  The discontinuous
bifurcation in the set of global minimum points of 
$g$ is reminiscent of a first-order phase transition, and
the value $a_c$ corresponds to a point on the first-order
curve $K_1(\beta)$. In fact, in section \ref{section:firstordercurve}
we use the analogous behavior
of the set of global minimum points of the 
appropriate Ginzburg-Landau polynomials to deduce properties
of the first-order curve.

In order to highlight the discontinuous bifurcation in the set of global minimum points of $g$
at $a_c$, we give a quick proof of the fact that $g$ has three global minimum
points if and only if $\atwo = a_c = \afour^2/4\asix$.  
Since $g(x) \goto \infty$ as $|x| \goto \infty$, the set of global
minimum points of $g$ is nonempty.  By symmetry,
$g$ has 3 global minimum points at 0 and at $\pm \barx$ for some $\barx > 0$
if and only if $g(\barx) = 0 = g(0)$ and $g'(\barx) = 0$.  Since $\barx > 0$,
these equations reduce to $a_2 - \afour \barx^2 + \asix \barx^4 = 0$ and 
$2 \atwo  - 4 \afour \barx^2 + 6 \asix \barx^4 = 0$. Eliminating $\asix$ from both equations yields
$\barx^2 = 2\atwo/\afour$ or $\barx = \pm (2\atwo/\afour)^{1/2}$.
Substituting this value back into $g(\barx) = 0$ gives $4 \atwo \asix = \afour^2$ or
$\atwo = a_c$.  We conclude that $g$ has three global
minimum points if and only if $\atwo = a_c$, and then the global minimum points
are 0 and $\pm (2\atwo/\afour)^{1/2}$.  This fact is recorded in part (b) of the next
theorem.  

In parts (a), (b), and (c) of the next theorem we 
give information about the global minimum points of $g$ for variable $a_2 \in \R$. 
Part (d) highlights properties of the positive global minimum point of $g$ 
for $a_2 \leq a_c$.  These properties underlie the discontinuous bifurcation in the set of 
global minimum points of $g$ at $a_c$.  For $a_2 > a_c$ [part (a)], $g$ has a similar shape as $\gbk$
in Figure 4 in section \ref{section:phasetr}; for $a_2 = a_c$ [part (b)], $g$ has a similar shape as $\gbk$
in Figure 5 in section \ref{section:phasetr}; and for $a_2 < a_c$ [part (c)], $g$ has a similar shape as $\gbk$
in Figure 6 in section \ref{section:phasetr}.  The elementary proof of the next theorem is omitted.

\begin{thm}
\label{thm:6order}
For variable $\atwo \in \R$ and fixed $\asix > 0$ and $\afour > 0$,
define $g(x) = a_2 x^2 - a_4 x^4 + a_6 x^6$ and $a_c = a_4^2/4 a_6$.  
If $0 \leq \atwo\leq \afour^2/3\asix$, then also define the positive number
\be
\label{eqn:barx}
\barx(\atwo) = \ts \frac{1}{\sqrt{3 a_6}}\left(\afour + (\afour^2 - 3 \atwo \asix)^{1/2}\right)^{\!1/2}.
\ee
The structure of the set of global minimum points of $g$ is as follows.

{\em (a)} If $\atwo > a_c$, then $g$ has a unique global minimum point at $0$. 

{\em (b)} If $\atwo = a_c$, then the global minimum points of $g$ 
are $0$ and $\pm \barx(a_c) = \pm (2\atwo/\afour)^{1/2}$.

{\em (c)} If $a_2 < a_c$, then the global minimum points of $g$ 
are $\pm \barx(\atwo)$.  

{\em (d)}  $\bar{x}(a_2)$ is a positive, decreasing, continuous function
for $a_2 < a_c$, and as $\atwo \goto (a_c)^-$, $\barx(\atwo) 
\goto \barx(a_c)$, the unique positive, global minimum point in part {\em (b)}.
\end{thm}

\skp
We end this appendix by making several observations about the polynomial
$h(x) = a_2 x^2 + a_4 x^4 + a_6 x^6$, where $\atwo \in \R$, $\afour > 0$,
and $\asix > 0$.  Such polynomials arise in Theorem \ref{thm:asymptoticsforc3}.
The elementary proof of the next theorem is omitted.

\begin{thm}
\label{thm:a2}
For variable $a_2 \in \R$ and fixed $\afour > 0$ and $\asix > 0$, define 
$h(x) = a_2 x^2 + a_4 x^4 + a_6 x^6$.  The following conclusions hold.

{\em (a)}  If $\atwo \geq 0$, then $h$ has a unique global minimum point at 0.

{\em (b)} If $\atwo < 0$, then the global minimum points of $h$ are 
$\pm \bar{x}(a_2)$, where
\be
\label{eqn:barxa2}
\barx(\atwo) = \ts\frac{1}{\sqrt{3 a_6}}\left(-\afour + (\afour^2 - 3 \atwo \asix)^{1/2}\right)^{\!1/2}.
\ee
\end{thm}

In the next section we give a second calculation of the asymptotics of $\mbnkn
\goto 0$ in the two cases treated in Theorems \ref{thm:asymptoticsforb}
and \ref{thm:asymptoticsforc1}.

\section{Exact Asymptotics by Another Method}
\beginsec
\label{section:moreasymptotics}

In this appendix we give another proof of the asymptotic behavior of $m(\bn,\kn)$ in two separate cases,
using an argument based directly on the fact that $\mbnkn$ is a positive zero of $\gprimebnkn$.
These two cases correspond to Theorem \ref{thm:asymptoticsforb}, which
deals with sequence $(\bn,\kn)$ converging to a second-order point, 
and to Theorem \ref{thm:asymptoticsforc1},
which deals with sequence $(\bn,\kn)$ having a similar
form as the sequences in Theorem \ref{thm:asymptoticsforb}
but converging to the tricritical point. 
Despite the naturalness of the characterization of $\mbnkn$ as a positive zero of $\gprimebnkn$, 
the proofs of the asymptotic behaviors of $\mbnkn$ given in this
appendix are much more complicated and the respective
theorems give less information than Theorems \ref{thm:asymptoticsforb}
and \ref{thm:asymptoticsforc2}, which are based on properties
of the Ginzburg-Landau polynomials. This emphasizes once again the elegance of the approach
of using properties of these polynomials to deduce the asymptotic behavior of $\mbnkn$.
As in Theorems \ref{thm:asymptoticsforb} and \ref{thm:asymptoticsforc2},
the proofs of the asymptotic behaviors of $\mbnkn$ given in this appendix
rely on the fact that $\mbnkn \goto 0$, which is proved in Theorem \ref{thm:zngoto0}.

We first consider the sequence $(\bn,\kn)$ in Theorem \ref{thm:asymptoticsforb},
which converge to a point on the second-order curve.  
The next theorem expresses the asymptotic behavior of $\mbnkn$ in terms of an explicitly given quantity
$f(b,k)$, which can be seen to agree with the explicit formula for $\bar{x}$ given
in (\ref{eqn:barxforb}) in Theorem \ref{thm:asymptoticsforb}.  
However, in contrast to Theorem \ref{thm:asymptoticsforb}
the next theorem does not identify $f(b,k)$ as the unique, positive, 
global minimum point of the
Ginzburg-Landau polynomial. In this sense, the next theorem gives less information than
Theorem \ref{thm:asymptoticsforb}. 

\begin{theorem}
\label{thm:asymptoticsb}
For $\beta \in (0,\bc)$, $\alpha > 0$,
$b \in \{1,0,-1\}$, and $k \in \R$ satisfying $K'(\beta) b - k < 0$, define
\[
\beta_n = \beta + {b}/{n^\alpha} \ \mbox{ and } \ K_n = K(\beta) + {k}/{n^\alpha}
\]
as well as 
\[
f(b,k) = \left(\frac{96 \beta (k - b K'(\beta)|)}{(e^\beta + 2)^2 \, (4 - e^\beta)}\right)^{\!1/2}.
\]
The sequence $(\bn,\kn)$ converges to the second-order point $(\beta,K(\beta))$.
The following conclusions hold: $m(\bn,\kn) \goto 0$ and has the asymptotic behavior 
\[
\mbnkn \sim {f(b,k)}/{n^{\alpha/2}}; \ \mbox{ i.e., }
\lim_{n \goto \infty} n^{\alpha/2} \mbnkn = f(b,k).
\]
\end{theorem}

\noi {\bf Proof.}  In order to simplify
the notation, we write $\mn$ in place of $m(\bn,\kn)$. For all sufficiently large $n$, $(\bn,\kn)$ lies
above the curve $(\beta,K(\beta))$ for $0 < \beta < \bc$.  Hence for all sufficiently 
large $n$, $\mn$ is a positive global minimum point of $G_{\bn,\kn}$ and thus 
a positive zero of $G_{\bn,\kn}'$.  In fact, $\mn$ is the unique,
positive, global minimum point of $G$ and the unique positive zero of $G_{\bn,\kn}'$,
but this fact is not needed in the proof.  Since
\beas
0 &= & \gbnkn'(\mn) \\
& = & 2\bn \kn \mn - (2 \bn \kn) c'{\bn}(2\bn \kn \mn) \\
& = & 2\bn \kn \mn - 2 \bn \kn \cdot \frac{e^{-\bn} (e^{2\bn\kn \mn} - e^{-2\bn\kn \mn})}
{1 + e^{-\bn} (e^{2\bn\kn \mn} + e^{-2\bn\kn \mn})},
\eeas
it follows that for all sufficiently large $n$, $\mn$ is the unique positive solution of
\be
\label{eqn:taylormn}
\mn = \frac{e^{2\bn\kn \mn} - e^{-2\bn\kn \mn}}{e^{\bn} + e^{2\bn\kn \mn} + 
e^{-2\bn\kn \mn}}.
\ee
We now use Taylor's Theorem to replace the exponentials in 
the numerator by a Taylor expansion to order 3 and the exponentials in
the denominator by a Taylor expansion to order 2.  Since $0 < \beta < \bc$
and $(\bn,\kn) \goto (\beta,K(\beta))$, $\mn \goto 0$ by Theorem \ref{thm:zngoto0}.
Hence there exist error terms $\ve_n = \mbox{O}(\mn^2)\goto 0$ and 
$\delta_n = \mbox{O}(\mn^2) \goto 0$ such that 
\[
\mn = \frac{4\bn\kn \mn + \frac{8}{3}\bn^3\kn^3 \mn^3(1 + \ve_n)}
{e^{\bn} + 2 + 4\bn^2\kn^2 \mn^2(1 + \delta_n)}.
\]
Because $\mn > 0$, this equation can be rewritten in the form
\be
\label{eqn:zn2}
\mn^2 (2\bn\kn)^2 (B_n + \bar{\ve}_n) + C_n = 0,
\ee
where $\bar{\ve}_n = \mbox{O}(\mn^2) \goto 0$ and
\[
B_n = 1 - {2}\bn\kn/3, \ C_n = 2 + e^{\bn} - 4\bn\kn.
\]
Thus
\be
\label{eqn:zn2solved}
\mn =  \left(\frac{-C_n}{(2\bn\kn)^2 (B_n + \bar{\ve}_n)}\right)^{\!1/2}.
\ee

We work out the asymptotic behavior of the terms in this equation.
We have $\bn\kn \goto \beta K(\beta) = (e^\beta + 2)/4$ and thus
\be
\label{eqn:Bn}
(2\bn\kn)^2 (B_n + \bar{\ve}_n) \goto 
\frac{(e^\beta + 2)^2}{4} \left(1 - \frac{(e^\beta + 2)}{6}\right)
= \frac{1}{24} (e^\beta + 2)^2 (4 - e^\beta).
\ee
Since $4 \beta K(\beta) = e^\beta + 2$, $4 K(\beta) - e^\beta = - 4 \beta K'(\beta)$
[Lem.\ \ref{lem:kbeta}(b)], 
and $e^{b/n^\alpha} \sim 1 + b/n^\alpha$,
the asymptotic behavior of $C_n$ is given by
\bea
\label{eqn:cn}
C_n & = & 2 + e^{\bn} - 4\bn\kn \\ \nonumber
& = & 2 + e^\beta e^{b/n^\alpha} - 4 (\beta + b/n^\alpha) (K(\beta) + k/n^\alpha) \\
\nonumber & \sim & - 4\beta (k - b K'(\beta))/n^\alpha.
\eea
Substituting (\ref{eqn:Bn}) and (\ref{eqn:cn}) into (\ref{eqn:zn2}) 
yields the desired asymptotic behavior of $\mn$:
\[
\mn \sim 
\ds \left( \frac{96 \beta(k - b K'(\beta))}{(e^\beta+2)^2 \, (4-e^\beta)}\right)^{\!1/2}
\cdot \frac{1}{n^{\alpha/2}}
\]
The proof of the theorem is complete. \ \ink

\skp
When $\beta = \bc$, the sequence $(\bn,\kn)$ considered in Theorem \ref{thm:asymptoticsb}
converges to the tricritical point, 
and the asymptotic behavior of $\mbnkn$ given in Theorem \ref{thm:asymptoticsb}
fails because in (\ref{eqn:Bn}) $4 - e^{\bc} = 0$.  As we will see in the proof of the next theorem, in this case
the asymptotic behavior of $\mbnkn \goto 0$ is determined by taking higher order terms in the 
Taylor expansions of the exponentials in (\ref{eqn:taylormn}).  

The next theorem expresses the asymptotic behavior of $\mbnkn$ in terms of an explicitly given quantity
$g(b,k)$, which can be seen to agree with the explicit formula for $\bar{x}$ given
in (\ref{eqn:barxforc1}).  However, in contrast to Theorem \ref{thm:asymptoticsforc1}
the next theorem does not identify $g(b,k)$ as the unique, positive, global minimum point of the
Ginzburg-Landau polynomial. In this sense, the next theorem gives less information than
Theorem \ref{thm:asymptoticsforb}.  

\begin{theorem}
\label{thm:asymptoticsc}
For $\alpha > 0$, $b \in \{1,0,-1\}$, and $k \in \R$ satisfying $bK'(\bc) - k < 0$, define
\[
\bn = \beta_c + b/n^\alpha \ \mbox{ and } \ \kn = K(\beta_c) + k/n^\theta
\]
as well as 
\[
g(b,k) = \left({40 \beta_c (k - K'(\bc)b)}/{27}\right)^{\!1/4}.
\]
Then $(\bn,\kn)$ converges to the tricritical point $(\bc,\kc)$.  
The following conclusions hold: $\mbnkn \goto 0$ 
and has the the following asymptotic behavior:
\[
\mbnkn \sim g(k,b)/{n^{\alpha/4}}.
\]
\end{theorem}

\skp
\noi
{\bf Proof.} For all sufficiently large $n$, $(\bn,\kn)$ lies
above the curve $(\beta,K(\beta))$ for $\beta > \bc$.  Hence for all sufficiently 
large $n$, $\mn$ is a positive global minimum point of $G_{\bn,\kn}$
and a positive zero of $\gbnkn'$.  As in the proof of the preceding
theorem, $\mn$ is the unique,
positive, global minimum point of $G$ and the unique positive zero of $G_{\bn,\kn}'$,
but this fact is not needed in the present proof.    Since
\beas
0 & = & \gbnkn'(\mn) \\ & = & 2\bn \kn \mn - (2 \bn \kn) c_{\bn}'(2\bn \kn \mn) \\
& = & 2\bn \kn \mn - 2 \bn \kn \cdot \frac{e^{-\bn} (e^{2\bn\kn\mn} - e^{-2\bn\kn\mn})}
{1 + e^{-\bn} (e^{2\bn\kn\mn} + e^{-2\bn\kn\mn})},
\eeas
it follows that for all sufficiently large $n$, $\mn$ is the unique positive solution of
\be
\label{eqn:taylormnmn}
\mn = \frac{e^{2\bn\kn\mn} - e^{-2\bn\kn\mn}}{e^{\bn} + e^{2\bn\kn\mn} + e^{-2\bn\kn\mn}}.
\ee
We now use Taylor's Theorem to replace the exponentials in the 
numerator by a Taylor expansion to order 5 and 
the exponentials in the denominator by a Taylor expansion to order 4.  
Since $(\bn,\kn) \goto (\bc,\kc)$, $\mn \goto 0$ 
by Theorem \ref{thm:zngoto0}.  
Hence there exist error terms $\ve_n = \mbox{O}(\mn^2) \goto 0$ and 
$\delta_n = \mbox{O}(\mn^2) \goto 0$ such that 
\[
\mn = \frac{2\left( 2\beta_n K_n z_n + 
\frac{(2\beta_n K_n)^3}{3!} z_n^3 + \frac{(2\beta_n K_n)^5}{5!}
z_n^5 (1 + \ve_n) \right)}{e^{\beta_n} + 2 \left(1+ \frac{(2\beta_n K_n)^2}{2} 
z_n^2 + \frac{(2\beta_n K_n)^4}{4!} z_n^4 (1 + \delta_n \right)}. 
\]
Because $\mn > 0$, this equation can be rewritten in the form
\be
\label{eqn:zn4}
\mn^4 \frac{(2\bn\kn)^4}{12} (A_n + \bar{\ve}_n) + \mn^2 (2\bn\kn)^2 B_n + C_n = 0,
\ee
where $\bar{\ve}_n = \mbox{O}(\mn^2) \goto 0$,
\[
A_n = 1 - {2\beta_n K_n}/{5}, \ B_n = 1 - {2}\bn\kn/3, \ C_n = 2 + e^{\bn} - 4\bn\kn.
\]
Without the term involving $\mn^4$, the formula in (\ref{eqn:zn4}) reduces to the quadratic
(\ref{eqn:zn2}) without the error term $\bar{\ve}_n$; this quadratic was used to 
determine the asymptotic behavior of $\mn$ in Theorem \ref{thm:asymptoticsb}. 
It follows from (\ref{eqn:zn4}) that 
\bea
\label{eqn:zn4solved}
z_n^2 & = & \frac{1}{\frac{(2\beta_n K_n)^4}{6} (A_n + \bar{\ve}_n)} \times
\\ \nonumber
&& \hspace{.5in} \left( -(2\beta_n K_n)^2 B_n \pm \left((2\beta_n K_n)^4 B_n^2 - \frac{1}{3} (2\beta_n 
K_n)^4 (A_n + \bar{\ve}_n)C_n\right)^{\!1/2} \right) \\
& = & \frac{6}{(2\beta_n K_n)^2 (A_n + \bar{\ve}_n)} \left( -B_n \pm \left(B_n^2 - \frac{1}{3} 
(A_n + \bar{\ve}_n)C_n \right)^{\!1/2} \right).
\nonumber
\eea

We now work out the asymptotic behavior of the terms in this equation. 
We have $\bn\kn \goto \beta_c K(\beta_c) = 3/2$ and thus
\[
A_n \goto 2/5,
\]
\beas
B_n & = & 1 - 2(\bc + b/n^\alpha)(K(\bc) + k/n^\theta)/3 \\
\nonumber & \sim & -2 b K(\bc)/3n^\alpha - 2\bc k/3n^\alpha,
\eeas
and as in (\ref{eqn:cn}),
\[
C_n \sim - 4\beta (k - b K'(\bc))/n^\alpha.
\]
Since $A_n \goto 2/5$, $\bar{\ve}_n \goto 0$, and $B_n^2/C_n \goto 0$, it follows from (\ref{eqn:zn4solved}) that
\be
\mn^2 \sim \frac{6}{(2\beta_n K_n)^2 A_n} \left(- \frac{1}{3}A_n C_n\right)^{\!1/2}.
\ee
Combining this with the preceding display yields the desired asymptotic behavior of $\mn$:
\[
\mn \sim \left(\frac{40 \bc(k - K'(\beta_c)b)}{27}\right)^{\!1/4} \cdot \frac{1}{n^{\alpha/4}}
\]
The proof of the theorem is complete. \ \ink

\skp

The proofs of the last two theorems make clear that deriving the asymptotic behavior 
of $\mbnkn \goto 0$ in this way requires a detailed asymptotic analysis that is
highly dependent on the specific situation. In the case of Theorem \ref{thm:asymptoticsb}
the starting point is (\ref{eqn:zn2solved}), while in the case of 
Theorem \ref{thm:asymptoticsb} the starting point is (\ref{eqn:zn4solved}). 
The latter formula can also be used to deduce the asymptotic behavior of $\mbnkn \goto 0$
for the sequences appearing in Theorems \ref{thm:asymptoticsforc2}--\ref{thm:asymptoticsforc4}. 
The asymptotic expressions for $B_n$ and $C_n$ 
in (\ref{eqn:zn4solved}) are different for each of these sequences,
 and in each case one must redo the tedious calculation 
leading from (\ref{eqn:zn4solved}) to the answer. By contrast, the proofs of the asymptotic
behavior of $\mbnkn$ given in Theorems \ref{thm:asymptoticsforc1}--\ref{thm:asymptoticsforc4} all
rely on the general result given in Theorem \ref{thm:exactasymptotics}.  
In each of these four cases the asymptotic behavior depends on the unique, positive, global minimum point 
$\bar{x}$ of the associated Ginzburg-Landau polynomial.  One uses the explicit
formula for the sequences only in calculating the form of $\bar{x}$, a much simpler task than
the detailed asymptotics that are required when applying the method in this appendix.

\end{document}